\documentclass[aps,prd,superscriptaddress,groupedaddress,nofootinbib,nobibnotes]{revtex4}

\usepackage{graphicx}
\usepackage{dcolumn}
\usepackage{bm}
\usepackage{amssymb}
\usepackage{epstopdf}
\usepackage{amsmath}
\usepackage{amsfonts}
\usepackage{color}
\usepackage{mathrsfs}
\usepackage[normalem]{ulem}

\newcommand{\be}{\begin{equation}}
\newcommand{\ee}{\end{equation}}
\newcommand{\ba}{\begin{eqnarray}}
\newcommand{\ea}{\end{eqnarray}}
\newcommand{\nn}{\nonumber}
\newcommand{\barr}{\begin{array}}
\newcommand{\earr}{\end{array}}

\newcommand\lsim{\mathrel{\rlap{\lower4pt\hbox{\hskip1pt$\sim$}}
        \raise1pt\hbox{$<$}}}
\newcommand\gsim{\mathrel{\rlap{\lower4pt\hbox{\hskip1pt$\sim$}}
        \raise1pt\hbox{$>$}}}

\def\Var{\mbox{Var}}

\def\th{\bm{\theta}}
\def\l{{\bf l}}
\def\k{{\bf k}}
\def\q{{\bf q}}
\def\x{{\bf x}}

\def\n{{\bf n}}
\def\hE{{\hat {\mathcal E}}}
\def\hT{{\hat T}}
\def\hv{{\hat v}}
\def\hr{{\bf \hat r}}
\def\hdelta{{\hat \delta}}
\def\tT{{\tilde T}}
\def\hphi{{\hat\phi}}
\def\tP{{\widetilde P}}
\def\bP{{\bar P}}
\def\tB{{\widetilde B}}
\def\bB{{\bar B}}

\def\halpha{\hat\alpha}
\def\L{{\mathcal L}}

\begin{document}

\title{KSZ tomography and the bispectrum}

\author{Kendrick M. Smith}
\affiliation{Perimeter Institute for Theoretical Physics, Waterloo, ON N2L 2Y5, Canada}

\author{Mathew~S.~Madhavacheril}
\affiliation{Princeton University, Department of Astrophysical Sciences, Princeton NJ 08540, USA}

\author{Moritz M\"unchmeyer}
\affiliation{Perimeter Institute for Theoretical Physics, Waterloo, ON N2L 2Y5, Canada}

\author{Simone Ferraro}
\affiliation{Berkeley Center for Cosmological Physics, University of California, Berkeley CA 94720, USA}
\affiliation{Miller Institute for Basic Research in Science, University of California, Berkeley CA 94720, USA}

\author{Utkarsh Giri}
\affiliation{Perimeter Institute for Theoretical Physics, Waterloo, ON N2L 2Y5, Canada}

\author{Matthew~C.~Johnson}
\affiliation{Perimeter Institute for Theoretical Physics, Waterloo, ON N2L 2Y5, Canada}
\affiliation{Department of Physics and Astronomy, York University, Toronto, Ontario, M3J 1P3, Canada}

\date{\today}

\begin{abstract}
Several statistics have been proposed for measuring the kSZ effect
by combining the small-scale CMB with galaxy surveys.  We review five
such statistics, and show that they are all mathematically equivalent
to the optimal bispectrum estimator of type $\langle ggT \rangle$.
Reinterpreting these kSZ statistics as special cases of bispectrum estimation 
makes many aspects transparent, for example optimally weighting the estimator, 
or incorporating photometric redshift errors.
We analyze the information content of the bispectrum and show that there
are two observables: the small-scale galaxy-electron power spectrum $P_{ge}(k_S)$,
and the large-scale galaxy-velocity power spectrum $P_{gv}(k)$.
The cosmological constraining power of the kSZ arises from its sensitivity to
fluctuations on large length scales, where its effective noise level
can be much better than galaxy surveys.
\end{abstract}

\maketitle

\tableofcontents
                                   
\section{Introduction}

\begin{figure}[b!]
  \includegraphics[width=0.55\textwidth]{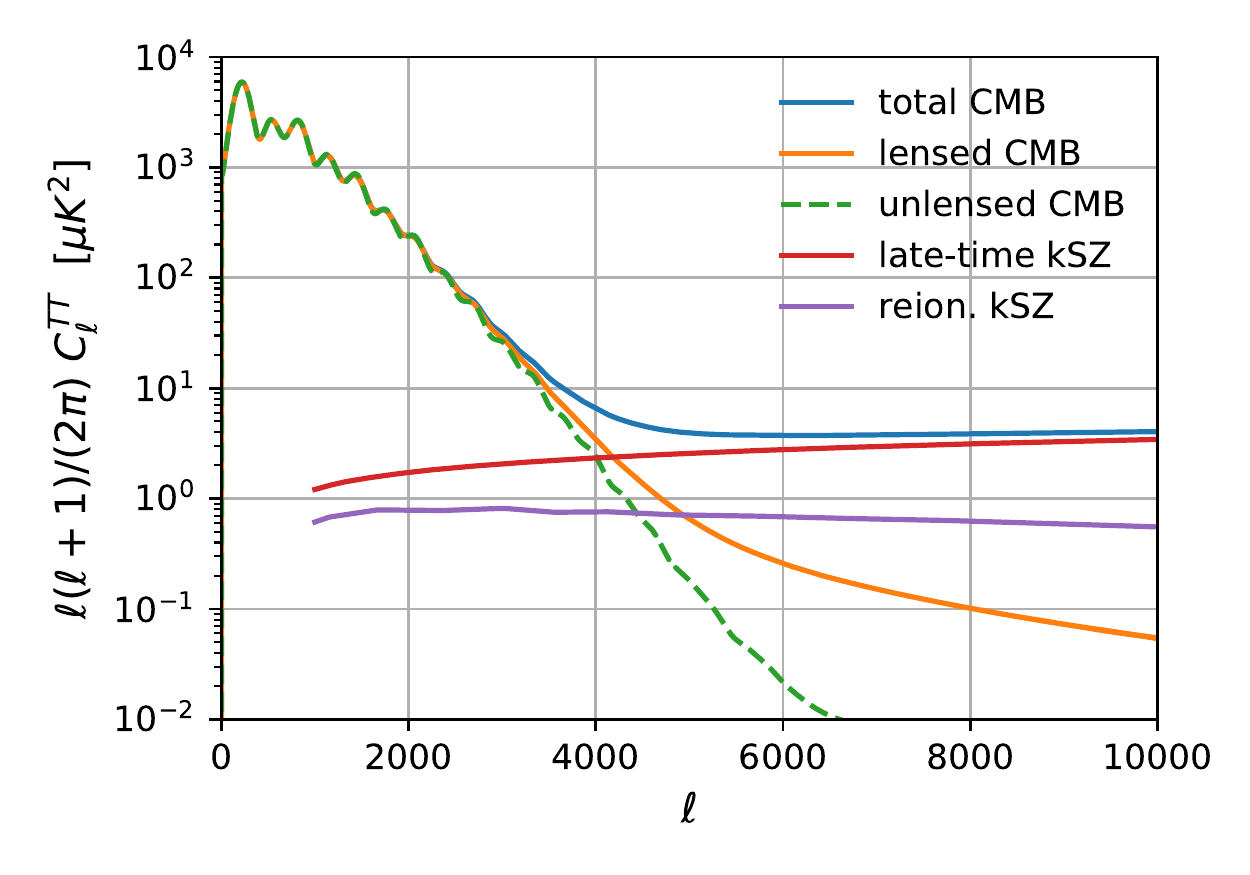}  
  \caption{The CMB power spectrum $C_l^{TT}$ from primary CMB,
    gravitational lensing, late-time kSZ ($z<6$) and reionization kSZ. The late-time
  kSZ was calculated from the halo model (see App.~\ref{app:halo_model}) and
  the reionization kSZ was taken from~\cite{Park:2013mv}.  We have only
  shown contributions with blackbody frequency dependence.  Non-blackbody
  contributions (CIB, tSZ) can be mostly removed using multifrequency
  analysis, but the level of residual contamination will depend on
  experiment-specific details.}
\label{fig:clcmb}
\end{figure}

Over the last 30 years, the CMB temperature power spectrum (Figure~\ref{fig:clcmb})
has been measured with increasing precision.
On large angular scales ($l \lsim 2000$), the CMB is dominated by so-called
``primary'' anisotropy, i.e.~anisotropy which originates on the last scattering
surface at redshift $z \sim 1100$.
On smaller angular scales $2000 \lsim l \lsim 4000$, the CMB receives
large contributions from gravitational lensing: primary anisotropy which has
been lensed by large-scale structure at $0 \lsim z \lsim 5$, shifting
CMB power to smaller scales.
Finally, on the smallest scales $l \gsim 4000$, the CMB becomes dominated
by the kinetic Sunyaev-Zeldovich (kSZ) effect, i.e.~Doppler shifting of CMB
photons by free electrons (Fig.~\ref{fig:clcmb}).  

The primary and lensed CMB have been measured to high precision, and
this has been a gold mine of information for cosmology.
So far, the kSZ effect has been detected at a few sigma
in cross-correlation with large-scale 
structure~\cite{Hand:2012ui,Ade:2015lza,Schaan:2015uaa,Soergel:2016mce,Hill:2016dta,DeBernardis:2016pdv},
but it will be measured much more accurately in the near future.
Qualitatively, it is clear that upcoming kSZ measurements will provide interesting
new information in both astrophysics and cosmology.
On the astrophysics side, the kSZ probes the distribution of electrons in galaxy clusters, 
including cluster outskirts where the gas is too cold to appear in thermal SZ, and not
dense enough to appear in X-rays.
This is a novel observation which can address the ``missing baryon''
problem~\cite{Bregman:2007ac,HernandezMonteagudo:2008jz,Ho:2009iw}.
On the cosmology side, the kSZ is a probe of velocities on large scales.
Potential applications include
dark energy~\cite{DeDeo:2005yr,HernandezMonteagudo:2005ys,Bhattacharya:2007sk},
modified gravity~\cite{Mueller:2014nsa,Bianchini:2015iaa},
neutrino mass~\cite{Mueller:2014dba},
void models~\cite{GarciaBellido:2008gd,Caldwell:2007yu,Clifton:2011sn,Bull:2011wi,Maartens2011,2011CQGra..28p4005Z,Yoo:2010ad},
bulk flows~\cite{Zhang10d,Zhang11b}, and theories predicting significant large-scale inhomogeneity~\cite{Zhang:2015uta,Terrana2016}.

There are well-established statistical frameworks for analyzing the primary and lensed CMB.
The primary CMB is a Gaussian field, and therefore all of the information is contained in the power spectrum.
The lensed CMB is non-Gaussian, but the lens reconstruction quadratic estimator $\hphi$
provides a framework for constructing higher-point statistics~\cite{Hu:2001tn,Hu:2001kj,Okamoto:2003zw}.

In contrast, for the kSZ, many different statistics have been
proposed~\cite{Ho:2009iw,Hand:2012ui,Li:2014mja,Alonso:2016jpy,Deutsch:2017ybc}.
It is not obvious how these statistics relate to each other, how to incorporate
them into larger forecasts involving more datasets, or whether one is more optimal than the others.
One may wonder whether there is a unifying approach.

This paper has three main purposes.
First, we show that if small-scale CMB observations are combined with a galaxy survey on the same patch of sky, 
then the kSZ effect introduces a large three-point correlation function (or bispectrum) involving two 
powers of the galaxy field and one power of the CMB.
Following a standard approach used in other areas of cosmology (for example estimating $f_{NL}$
from the primary CMB), we construct the optimal bispectrum estimator for this signal.
We also construct the bispectrum Fisher matrix, which can be used to forecast total signal-to-noise,
or for more complex multiparameter forecasts.
A crucial property of the kSZ bispectrum is that it is parity-odd under reflections in the radial direction,
and therefore the estimator is not biased by other non-Gaussian signals (CIB, CMB lensing, thermal SZ), which
are parity-even.

A second purpose of the paper is to show that several independently proposed statistics
for analyzing the kSZ effect are mathematically equivalent, if optimally weighted.
These statistics include 
the kSZ template method~\cite{Ho:2009iw},
the pair sum statistic~\cite{Hand:2012ui},
the velocity matched filter from~\cite{Li:2014mja},
the velocity growth method from~\cite{Alonso:2016jpy},
and the velocity reconstruction introduced recently in~\cite{Deutsch:2017ybc}.
We will use the term ``kSZ tomography'' as a catchall term to refer to any of these statistics.

These kSZ tomography statistics have a common property: they are all three-point
estimators involving two powers of the galaxy field and one power of the CMB, as 
we will show explicitly in~\S\ref{sec:equivalence}.\footnote{Not every
 proposed kSZ statistic is a 3-point estimator of the form $\langle ggT \rangle$.
 One of the first kSZ statistics proposed~\cite{Dore:2003ex,DeDeo:2005yr} is a cross correlation 
 between large-scale structure and the squared high-pass filtered CMB.
 Recently this method was used to obtain a 4$\sigma$ measurement of the kSZ~\cite{Hill:2016dta, Ferraro:2016ymw}.
 This is a 3-point estimator of a different type, namely $\langle g T T \rangle$.
 As another example, a four-point estimator $\langle T T T T \rangle$ was recently proposed~\cite{Smith:2016lnt,Ferraro:2018izc}
 which does not use an external large-scale structure dataset and can probe reionization.
 In this paper, we do not consider these statistics, and define ``kSZ tomography'' to mean any kSZ-sensitive
 three-point estimator of type $\langle g g T \rangle$.}
Therefore, the underlying signal is the bispectrum $\langle g g T \rangle$.
In fact, we will show that all of the kSZ tomography statistics (if optimally weighted) are just
different ways of algebraically representing the optimal bispectrum estimator $\hE$.
The estimator $\hE$ is an integral over triples of wavenumbers $\k+\k'+\l/\chi = 0$,
and a double sum over galaxy pairs $(i,j)$.  The ordering of these integrals and sums
can be exchanged, and using different orderings, the estimator $\hE$ can be rewritten
algebraically to take the form of any of the kSZ tomography statistics 
in~\cite{Ho:2009iw,Hand:2012ui,Li:2014mja,Alonso:2016jpy,Deutsch:2017ybc}.

Thus, kSZ tomography is ``bispectrum estimation in disguise'', and a variety
of apparently different statistics are simply different ways of implementing
the optimal bispectrum estimator.
However, the bispectrum perspective has technical advantages.  For example,
it clarifies what kSZ tomography actually measures.  Our calculation (\S\ref{sec:bispectrum})
of the $\langle g g T \rangle$ bispectrum will show that it peaks in the squeezed limit
where the wavenumbers $k_L, k_S$ of the two galaxy modes satisfy 
$k_L \ll k_S$, and is proportional to $P_{gv}(k_L) P_{ge}(k_S)$.
Here, $P_{gv}(k_L)$ is the galaxy-velocity power spectrum on large scales,
and $P_{ge}(k_S)$ is the galaxy-electron power spectrum on small scales.
Thus, kSZ tomography is a measurement of two power spectra $P_{gv}(k_L)$, $P_{ge}(k_S)$,
within a degeneracy which allows an overall constant to be exchanged,
leaving the product $P_{gv}(k_L) P_{ge}(k_S)$ invariant.
This is the well-known optical depth degeneracy in the kSZ~\cite{Battaglia:2016xbi,Flender:2016cjy,Louis:2017hoh,Soergel:2017ahb}.

The third purpose of this paper is to give a simple recipe for incorporating
kSZ tomography into larger analyses (either Fisher matrix forecasts or actual data analysis),
using a quadratic estimator formalism.
Before explaining our recipe, we pause briefly to review CMB lens
reconstruction~\cite{Hu:2001tn,Hu:2001kj,Okamoto:2003zw}, which will turn out to be analogous.

In the case of the lensed CMB, the lens reconstruction quadratic estimator $\hphi(\l)$ 
estimates each Fourier mode $\phi(\l)$ of the CMB lensing potential, using a quadratic
combination of CMB modes.
This naturally leads to higher-point statistics.
If $\hphi$ is correlated with a galaxy survey $g$, the result is is a three-point estimator
involving two powers of the CMB and one power of the galaxies.
The auto power spectrum of $\hphi$ is a four-point estimator in the underlying CMB map.
Furthermore, it is straightforward to incorporate all of these higher-point statistics in larger 
analyses, by including $\phi$ as an additional field with the appropriate noise power spectrum.

Analogously, we propose that kSZ tomography can be included in larger
analyses by including a large-scale {\em radial} velocity reconstruction $\hv_r(\k)$ with
appropriate noise power spectrum.
The quadratic estimator $\hv_r(\k)$ involves one power of the small-scale galaxy field and
one power of the small-scale CMB.  The estimator and its noise power spectrum are given
explicitly in Eqs.~(\ref{eq:hv_final}),~(\ref{eq:Nvr_final}) below.
The quadratic estimator $\hv_r$ was first constructed in~\cite{Deutsch:2017ybc}.

The auto power spectrum of $\hv_r$ is a four-point estimator in the underlying galaxy and
CMB maps, with schematic form $(ggTT)$.
The cross spectrum of $\hv_r$ with a galaxy field is a three-point estimator of schematic
form $(ggT)$.
To incorporate these higher-point statistics into a larger analysis, we simply
include the field $v_r$ with appropriate noise power spectrum.

This is very similar to CMB lensing, but there are some interesting differences
between the kSZ quadratic estimator $\hv_r$ and the lensing estimator $\hphi$,
as we now explain.  The most obvious difference is that $\hv_r(\k)$ is a 3-d reconstruction,
whereas $\hphi(\l)$ is 2-d.  Further differences arise by considering the noise power spectrum
of $\hv_r$, as we explain in the next few paragraphs.

First we note that on large scales, where linear perturbation theory is a good approximation,
the radial velocity $v_r$ is related to the density field $\delta_m$ by:
\be
v_r(\k) = \mu \frac{faH}{k} \, \delta_m(\k)
\ee
where $f=\partial(\log D)/\partial(\log a)$ is the usual RSD parameter, and $\mu=k_r/k$.
Therefore, we can ``convert'' the kSZ-derived radial velocity reconstruction $\hv_r$ to a
reconstruction $\hdelta(\k) = \mu^{-1} (k/faH) \hv_r(\k)$ of the large-scale density field $\delta_m$.
We will show (Eq.~(\ref{eq:Nvr_final}) below) that on large scales, the reconstruction noise
$N_{v_r}(k_L)$ approaches a constant.  Therefore, the kSZ reconstruction noise on the
density field $\delta_m$ has the form:
\be
N_{\delta\delta}^{\rm rec}(k) \propto \mu^{-2} \left( \frac{k}{faH} \right)^2
  \hspace{1cm} \mbox{(as $k\rightarrow 0$)}  \label{eq:noise_intro}
\ee
From this noise power spectrum, we can deduce two qualitative features of the large-scale
kSZ-derived reconstruction.

First, {\em the kSZ cannot reconstruct transverse modes ($\mu \approx 0$), and the kSZ reconstruction cannot be cross-correlated with a 2-d field, such as the CMB lensing potential.}
This is due to the $\mu^{-2}$ prefactor in Eq.~(\ref{eq:noise_intro}), and is easy to understand intuitively:
transverse modes do not contribute to kSZ because the associated radial velocity is zero.
For this reason, in this paper we only consider 3-d large-scale structure fields,
such as galaxy surveys.
A galaxy survey with photometric redshifts is an interesting
intermediate case between 2-d and 3-d~\cite{Keisler:2012eg}.
In this case, there is a signal-to-noise penalty when cross-correlating with the kSZ reconstruction,
but the SNR is still large enough to be interesting.
We work out the details in~\S\ref{sec:photoz_rsd}.

Second, {\em on large scales, kSZ tomography derived from a galaxy survey constrains cosmological modes better than the galaxy survey itself}.
This is because the kSZ reconstruction noise in Eq.~(\ref{eq:noise_intro}) is proportional to $k^2$
on large scales, whereas the Poisson noise of the galaxy survey approaches a constant value $n_g^{-1}$.
We find (Fig.~\ref{fig:noise} below) that the crossover occurs around $k \sim 0.01$ Mpc$^{-1}$,
although the exact value depends on the details of the CMB and galaxy surveys.

Another qualitative feature of the kSZ-derived velocity reconstruction $\hv_r$ is that
it appears with a bias parameter $\langle \hv_r \rangle = b_v v_r^{\rm true}$
which must be marginalized.
This is not initially obvious, but we will show in~\S\ref{sec:forecasts_pheno} that 
this is a consequence of the kSZ optical depth degeneracy, i.e.~astrophysical uncertainty 
in the small-scale galaxy-electron power spectrum $P_{ge}(k_S)$.
Marginalizing $b_v$ fully incorporates the optical depth degeneracy in a larger analysis.

Here is an example to illustrate the power of the velocity reconstruction approach to kSZ tomography.
An interesting recent paper~\cite{Sugiyama:2016rue} showed that the optical depth degeneracy
can be broken using an ``octopolar'' version of the pair sum estimator from~\cite{Hand:2012ui}.
In velocity reconstruction language, the degeneracy breaking can be described as follows.
Consider two fields, the kSZ-derived velocity reconstruction $\hv_r = b_v v_r = b_v \mu (faH/k) \delta_m$,
and a redshift-space galaxy field field $\delta_g = (b_g + f \mu^2) \delta_m$.
If we cross-correlate them, the cross power spectrum has $\mu$ dependence of schematic form
$P_{\hv_r \delta_g} \propto (f a H \sigma_8^2 / k) (b_v b_g \mu + b_v f \mu^3)$.
In this form, we see that a measurement of the $\mu^3$ term breaks the optical depth
degeneracy, in the sense that it pins down the value of $b_v$ (within uncertainty on cosmological
parameters $f, H, \sigma_8$).
This transparent explanation of the degeneracy breaking illustrates the power of the velocity 
reconstruction approach.  A Fisher matrix forecast which includes redshift space distortions
would automatically ``discover'' the degeneracy breaking, without needing to construct 
the octopolar pair sum explicitly (or needing to know that it exists in advance).

Earlier in this introduction, we stated that the kSZ-derived velocity reconstruction $\hv_r$ is mathematically equivalent
to the other kSZ tomography statistics in~\cite{Ho:2009iw,Hand:2012ui,Li:2014mja,Alonso:2016jpy}.
This statement implicitly assumes that we cross-correlate $\hv_r$ with the galaxy
field $g$ on large scales, but do not use it for anything else.
However, in a larger analysis, $\hv_r$ can be correlated with a variety of fields (including itself),
and the $\mu$-dependence of these correlations will lead to extra degeneracy breaking.
In our view, this makes the velocity reconstruction approach more powerful than the
other kSZ tomography statistics, and we advocate using it (at least for cosmology).

Summarizing the results so far, we can give a one-sentence
description of how kSZ tomography fits into the larger picture of cosmological observables.
KSZ tomography reconstructs the largest modes of the universe, with lower noise than
galaxy surveys, up to an overall bias parameter which must be marginalized, and
with the caveat that transverse modes ($\mu \approx 0$) are not reconstructed.

This picture clarifies which cosmological parameters the kSZ can constrain.
The primordial non-Gaussianity parameter $f_{NL}$ is a prime candidate.
The kSZ can be used to reconstruct large-scale density fluctuations with very low noise,
which improves $f_{NL}$ constraints from galaxy surveys by using the sample variance
cancellation idea from~\cite{Seljak:2008xr}.
We present $f_{NL}$ forecasts in the companion paper~\cite{Moritz}.

This paper is organized as follows.
In~\S\ref{sec:bispectrum} we compute the $\langle \delta_g \delta_g T \rangle$ bispectrum,
and construct the optimal bispectrum estimator $\hE$ and its Fisher matrix.
We then show (\S\ref{sec:equivalence}) that the optimal bispectrum estimator $\hE$
can be rewritten algebraically in several different ways, corresponding to
the different kSZ tomography formalisms in~\cite{Ho:2009iw,Hand:2012ui,Li:2014mja,Alonso:2016jpy,Deutsch:2017ybc}.
Armed with this machinery, in~\S\ref{sec:forecasts_pheno} we analyze several 
aspects of kSZ tomography, including 
the velocity reconstruction $\hv_r$ and its formal properties (\S\ref{ssec:cosmology}),
prospects for constraining astrophysics (\S\ref{ssec:astrophysics}).
and 
the optical depth degeneracy (\S\ref{ssec:more_optical_depth}).
In~\S\ref{sec:photoz_rsd}, we show how to incorporate photometric redshifts and
redshift space distortions.  We conclude in~\S\ref{sec:discussion}.

\section{Definitions and notation}
\label{sec:definitions}

Throughout this paper we use the following simplified ``snapshot'' geometry.
We take the universe to be a periodic 3D box with comoving side length $L$ and 
volume $V=L^3$, ``snapshotted'' at some time $t_*$.
We denote the redshift of the snapshot by $z_*$, the comoving distance
to redshift zero by $\chi_*$, etc.

We take the 2D sky to be a periodic square with angular side length $L/\chi_*$,
and define line-of-sight integration by projecting onto the xy-face of the cube,
with a factor $1/\chi_*$ to convert from spatial to angular coordinates.
We denote the transverse coordinates of the box by $(x,y)$, but denote the radial
coordinate by $r$ (not $z$, to avoid notational confusion with the redshift).
We write $(\cdot)_r$ for the radial component of a three-vector, and $\hr$ for a 
unit vector in the radial direction.

With this notation, the kSZ anisotropy is given by the line-of-sight integral:
\be
T_{\rm kSZ}(\th) = K_* \int_0^L \, dr \, q_r(\chi_*\th + r \hr)  \label{eq:ksz_los}
\ee
where $q_i(\x) = \delta_e(\x) v_i(\x)$ is the electron momentum field, 
$K(z)$ is the kSZ radial weight function with units $\mu$K-Mpc$^{-1}$:
\be
K(z) = -T_{\rm CMB} \sigma_T n_{e,0} x_e(z) e^{-\tau(z)} (1+z)^2
\ee
and $\tau(z)$ is the optical depth to redshift $z$.

This simplified geometry neglects lightcone evolution, curved-sky effects,
and survey boundaries, all of which will be nontrivial complications in real
data analysis.
However, it is convenient to ignore these complications when asking questions
such as which KSZ observables we should measure, and how we should interpret them.
When forecasting galaxy surveys, we approximate the true geometry
by our simplified geometry, by matching $z_*$ to the mean redshift of the survey
and matching the box volume $V$ to the comoving volume of the survey.

Our Fourier conventions for a 3D field are:
\be
f(\x) = \int \frac{d^3\k}{(2\pi)^3} f(\k) e^{i\k\cdot x}
  \hspace{1cm}
f(\k) = \int d^3\x \, f(\x) e^{-i\k\cdot x}
\ee
and similarly for a 2D field $T(\th) \leftrightarrow T(\l)$.

In linear theory, the velocity field has zero curl, so we can write $v_j(\k) = (ik_j/k) v(\k)$,
where $v(\k)^* = v(-\k)$.  The linear density field, velocity field, and radial velocity field
are related by
\be
\delta(\k) = \frac{k}{faH} v(\k)  \hspace{1.5cm}  v_r(\k) = \frac{ik_r}{k} v(\k)
\ee
where $f(z) =\partial(\log D)/\partial(\log a)$ is the usual
redshift space distortion parameter.  Sometimes we will also use the notation $\mu = k_r/k$.

We define the galaxy overdensity $\delta_g(\x)$ as a sum of delta functions (or in Fourier
space, a sum of complex exponentials):
\be
\delta_g(\x) = \frac{1}{n_g} \sum_i \delta^3(\x-\x_i)
  \hspace{1cm}
\delta_g(\k) = \frac{1}{n_g} \sum_i e^{-i\k\cdot\x_i} 
\ee
where the sum ranges over 3D galaxy positions $\x_i$, and $n_g$ denotes the
comoving number density of galaxies.  We denote the galaxy bias by $b_g$.

The fiducial cosmological model we assume in our forecasts roughly corresponds to that
determined by Planck with Hubble constant $H_0=67.3$ km/s/Mpc, baryon density $\Omega_b h^2=0.02225$,
cold dark matter density $\Omega_c h^2=0.1198$, scalar spectral index $n_s=0.9645$,
amplitude of scalar fluctuations $A_s=2.2 \times 10^{-9}$, optical depth to reionization $\tau=0.06$,
and minimal sum of neutrino masses $\sum m_{\nu}=0.06$ eV.

For our kSZ forecasts, we will need to know the galaxy auto power spectrum $P_{gg}^{\rm tot}(k)$
and the galaxy-electron cross power spectrum $P_{ge}(k)$.  We model these power spectra using
the halo model.  The main source of modeling uncertainty is the electron halo profile which is assumed,
which affects $P_{ge}(k)$.  In our fiducial model, we use the ``AGN'' electron profile
from~\cite{Battaglia:2016xbi}. We calculate the kSZ power spectrum $C_l$ from
the electron power spectrum $P_{ee}(k)$, calculated self-consistently using the
halo model with the same electron profile.
Details of the halo model and kSZ model are in Appendix~\ref{app:halo_model}.

\section{The $\langle \delta_g \delta_g T \rangle$ bispectrum} 
\label{sec:bispectrum}

The underlying signal for kSZ tomography is a three-point function (or bispectrum) $\langle \delta_g(\k) \delta_g(\k') T(\l) \rangle$
involving two powers of a galaxy field, and one power of the CMB.
There is a standard formalism in cosmology for constructing optimal three-point estimators, and forecasting their
statistical errors (used for example to construct $f_{NL}$ estimators for the CMB).
In this section, we apply this formalism to the kSZ three-point function to construct the optimal bispectrum estimator for kSZ
tomography.  We also derive the bispectrum Fisher matrix, which can be used for forecasting.

The kSZ bispectrum is unusual: it involves two powers of a 3D field $\delta_g(\k)$, and one power of a 2D field $T(\l)$.
This will change some details of the bispectrum formalism.
For example, we will show that the most general bispectrum allowed by symmetry is a function of four variables $B(k,k',l,k_r)$,
rather than the usual function of three variables $B(k,k',k'')$ which arises for three 3D fields (or three 2D fields).

\subsection{Mathematical representation of the bispectrum}

First we write down the most general three-point function $\langle \delta_g \delta_g T \rangle$
allowed by symmetry.
In the simplified geometry used in this paper (\S\ref{sec:definitions}), the statistics of the
fields $\delta_g(\x)$, $T(\th)$ are invariant under the following symmetries.
First, we can rotate both $\delta_g$ and $T$ in the xy-plane.
Second, we can translate in the xy-plane, by applying shifts $(\Delta x, \Delta y)$ to $\delta_g$,
and angular shifts $(\theta_x, \theta_y) = ((\Delta x)/\chi_*, (\Delta y)/\chi_*)$ to $T$.
Third, we can translate $\delta_g$ in the radial direction, leaving $T$ unchanged.

By 3D translation invariance, the three-point function contains the delta function:
\be
\langle \delta_g(\k) \delta_g(\k') T(\l) \rangle 
  = B(\k,\k',\l) \, (2\pi)^3 \delta^3\left( \k + \k' + \frac{\l}{\chi_*} \right)
\ee
Note that the delta function implies that the radial components satisfy $k_r + k'_r = 0$.
Once the radial components are specified, 2D rotation invariance implies that $B(\k,\k',\l)$
only depends on the lengths $k,k',l$.  Therefore we can write
\be
\langle \delta_g(\k) \delta_g(\k') T(\l) \rangle 
  = i B(k,k',l,k_r) (2\pi)^3 \delta^3\left( \k + \k' + \frac{\l}{\chi_*} \right)  \label{eq:B_def}
\ee
where the factor $i$ has been introduced for future convenience.
The permutation symmetry $\k \leftrightarrow \k'$ implies:
\be
B(k,k',l,k_r) = B(k',k,l,-k_r)  \label{eq:Bsym1}
\ee
and by taking the complex conjugate of Eq.~(\ref{eq:B_def}) we get:
\be
B(k,k',l,k_r)^* = -B(k,k',l,-k_r)  \label{eq:Bsym2}
\ee
There is one more symmetry we can use: reflection symmetry in the radial direction.
Under this symmetry, the kSZ temperature transforms with a minus sign, so we get:
\be
B(k,k',l,-k_r) = -B(k,k',l,k_r)  \label{eq:Bsym3}
\ee
Combining Eqs.~(\ref{eq:Bsym1})--(\ref{eq:Bsym3}), we see that $B(k,k',l,k_r)$ is real-valued, antisymmetric in $k,k'$ and odd in $k_r$.

The parity-odd transformation law under radial reflections (Eq.~(\ref{eq:Bsym3}))
has the important consequence that the kSZ bispectrum is orthogonal to bispectra
produced by other secondaries (lensing, Rees-Sciama, residual tSZ, residual CIB).
These secondaries all generate $\delta_g \delta_g T$-bispectra which are parity-even
under radial reflections.\footnote{We have included CMB lensing in this list of parity-even secondaries, even
 though the $\delta_g \delta_g T$-bispectrum produced by lensing is probably very small.
 To see this, we note that if the CMB lensing potential $\phi$ and the {\em unlensed} CMB $T_{\rm unl}$ were statistically independent,
 then lensing would not produce a $\delta_g \delta_g T$-bispectrum, since the statistics of
 the lensed CMB would be invariant under $T \rightarrow (-T)$.
 However, there is a small correlation between $\phi$ and $T_{\rm unl}$ on small scales
 due to the Rees-Sciama effect, and this produces a small, parity-even $\delta_g \delta_g T$-bispectrum.}
As we will show in the next section, this implies that the kSZ tomography estimator
is unbiased by the non-kSZ secondaries.
This property makes kSZ tomography a particularly interesting way of extracting
cosmological information from future CMB experiments, where the main challenge
may be disentangling different contributions, rather than obtaining high SNR.

\subsection{Optimal bispectrum estimator and Fisher matrix}

Given a predicted form of the kSZ bispectrum $B(k,k',l,k_r)$, what is the optimal bispectrum
estimator $\hE$?  To answer this question, we start with the most general three-point estimator
\be
\hE = \int \frac{d^3\k}{(2\pi)^3} \, \frac{d^3\k'}{(2\pi)^3} \, \frac{d^2\l}{(2\pi)^2}
  W(\k,\k',\l) 
     \Big( \delta_g(\k) \delta_g(\k') T(\l) \Big) \,
     (2\pi)^3 \delta^3\left( \k + \k' + \frac{\l}{\chi_*} \right) 
\ee
with weight function $W(\k,\k',\l)$ to be determined by the following constrained optimization
problem.  We minimize the variance $\Var(\hE)$, subject to the constraint that $\hE$ is an unbiased
estimator for the bispectrum amplitude, i.e. $\langle \hE \rangle=1$ if the true bispectrum
is $B$.  When computing $\Var(\hE)$, we assume that the fields $\delta_g,T$ are Gaussian for
simplicity.

Now a short calculation, which we omit since it is similar to bispectrum estimation
in other contexts, gives the following results.  The optimal bispectrum estimator is:
\be
\hE = \frac{1}{2F_{BB}} \int \frac{d^3\k}{(2\pi)^3} \frac{d^3\k'}{(2\pi)^3} \frac{d^2\l}{(2\pi)^2}
     \frac{-iB^*(k,k',l,k_r)}{P_{gg}^{\rm tot}(k) \, P_{gg}^{\rm tot}(k') \, C_l^{TT,\rm tot}}
     \Big( \delta_g(\k) \delta_g(\k') T(\l) \Big) \,
     (2\pi)^3 \delta^3\left( \k + \k' + \frac{\l}{\chi_*} \right)  \label{eq:E_def}
\ee
where $P_{gg}^{\rm tot}$ is the total power spectrum of the galaxy survey including shot noise,
and $C_l^{TT,\rm tot}$ is the total power spectrum of the CMB survey including instrumental noise.
The prefactor $F_{BB}$ is the bispectrum Fisher matrix, which is defined for a pair of bispectra $B,B'$ by:
\be
F_{BB'} = \frac{V}{2} \int \frac{d^3\k}{(2\pi)^3} \frac{d^3\k'}{(2\pi)^3} \frac{d^2\l}{(2\pi)^2}
     \frac{B(k,k',l,k_r)^* \, B'(k,k',l,k_r)}{P_{gg}^{\rm tot}(k) \, P_{gg}^{\rm tot}(k') \, C_l^{TT,\rm tot}} \,
     (2\pi)^3 \delta^3\left( \k + \k' + \frac{\l}{\chi_*} \right)  \label{eq:F_vector}
\ee
where $V$ is the survey volume.
The total signal-to-noise of the kSZ bispectrum is given by $\mbox{SNR} = F_{BB}^{1/2}$.
More generally, given $N$ bispectra to be jointly estimated, their $N$-by-$N$ covariance matrix is the inverse Fisher matrix.
If the bispectrum estimator $\hE$ is constructed assuming bispectrum $B$, and the true bispectrum is $B'$,
then the expectation value of the estimator is $\langle \hE \rangle = F_{BB'} / F_{BB}$.

We can use this last property of the estimator to show that the kSZ bispectrum estimator
is unbiased by parity-even secondaries (CMB lensing, Rees-Sciama, residual tSZ, residual CIB),
as stated without proof in the previous section.
This amounts to showing that $F_{BB'} = 0$, where $B$ is the parity-odd kSZ bispectrum and $B'$ is any parity-even bispectrum.
Writing out the transformation laws explicitly, we have:
\be
B(k,k',l,-k_r) = -B(k,k',l,k_r) 
  \hspace{1cm}
B'(k,k',l,-k_r) = B'(k,k',l,k_r) 
\ee
From the form of the Fisher matrix in Eq.~(\ref{eq:F_vector}), this implies $F_{BB'}=0$.
We note that this argument relies on reflections in the radial direction being an exact symmetry.  
This is true for the simplified snapshot geometry in this paper (\S\ref{sec:definitions}),
but the symmetry is not exact in reality due to evolution along the lightcone, and therefore we expect
some small leakage between kSZ tomography and parity-even secondaries in a more detailed treatment.  
We defer this to future work.

In Eq.~(\ref{eq:F_vector}), we have written the Fisher matrix $F_{BB'}$ as an integral over vector wavenumbers $\k,\k',\l$.
While formally transparent, this is inconvenient for numerical evaluation.
In Appendix~\ref{app:integrals}, we show that $F_{BB'}$ can be written as an integral over scalar wavenumbers:
\be
F_{BB'} = \frac{V}{2} \int dk \, dk' \, dl \, dk_r \, I(k,k',l,k_r) \,
     \frac{B(k,k',l,k_r)^* \, B'(k,k',l,k_r)}{P_{gg}^{\rm tot}(k) \, P_{gg}^{\rm tot}(k') \, C_l^{TT,\rm tot}}  \label{eq:F_scalar}
\ee
where $I(k,k',l,k_r)$ is defined in Eq.~(\ref{eq:I_done}).

\subsection{The tree-level kSZ bispectrum}

In this section we calculate an explicit formula for the kSZ bispectrum $B(k,k',l,k_r)$.
In real space, the kSZ anisotropy $T(\th)$ is given by the line-of-sight integral in Eq.~(\ref{eq:ksz_los}).
Converting to Fourier space and writing $q_r = \delta_e v_r$, this becomes:
\be
T(\l) = \frac{K_*}{\chi_*^2} \int \frac{d^3\q}{(2\pi)^3} \frac{d^3\q'}{(2\pi)^3} \, \Big( \delta_e(\q) v_r(\q') \Big) \, (2\pi)^3 \delta^3\!\left(\q+\q'-\frac{\l}{\chi_*} \right) \label{eq:Tl} 
\ee
Plugging this in, we can write the $\langle \delta_g \delta_g T \rangle$ three-point function as a large-scale structure four-point function:
\be
\big\langle \delta_g(\k) \delta_g(\k') T(\l) \big\rangle
  = \frac{K_*}{\chi_*^2} \int \frac{d^3\q}{(2\pi)^3} \, \frac{d^3\q'}{(2\pi)^3} \, \Big\langle \delta_g(\k) \delta_g(\k') \delta_e(\q) v_r(\q') \Big\rangle \, (2\pi)^3 \delta^3\!\left(\q+\q'-\frac{\l}{\chi_*} \right)
\ee
It would be very difficult to give a complete calculation of this four-point function which
extends to nonlinear scales!  As a starting point, suppose we neglect the connected part of
the four-point function, and compute the disconnected or tree-level part using Wick's theorem:
\ba
\big\langle \delta_g(\k) \delta_g(\k') T(\l) \big\rangle_{\rm tree}
  &=& \frac{K_*}{\chi_*^2} \int \frac{d^3\q}{(2\pi)^3} \, \frac{d^3\q'}{(2\pi)^3} \,
     \Big( \big\langle \delta_g(\k) \delta_e(\q) \big\rangle \big\langle \delta_g(\k') v_r(\q') \big\rangle + (\k \leftrightarrow \k') \Big) 
     \, (2\pi)^3 \delta^3\!\left(\q+\q'-\frac{\l}{\chi_*} \right) \nn \\
  &=& \frac{K_*}{\chi_*^2} \left( P_{ge}(k) \frac{-i k'_r P_{gv}(k')}{k'} + (\k \leftrightarrow \k') \right)
    \, (2\pi)^3 \delta^3\left( \k + \k' + \frac{\l}{\chi_*} \right)
\ea
Comparing with the definition of the bispectrum $B(k,k',l,k_r)$ in Eq.~(\ref{eq:B_def}), we
read off the tree-level kSZ bispectrum in the form:
\be
B(k,k',l,k_r)_{\rm tree} = \frac{K_* k_r}{\chi_*^2} \left( P_{ge}(k) \frac{P_{gv}(k')}{k'} - \frac{P_{gv}(k)}{k} P_{ge}(k') \right)  \label{eq:B_ksz}
\ee
The tree-level kSZ bispectrum is guaranteed to be a good approximation to the true kSZ bispectrum
in the limit where all wavenumbers $k,k',l$ are small (so that loop corrections are small).

However, we are interested in the kSZ bispectrum in a different limit,
namely the ``squeezed'' limit in which one wavenumber $k$ is small (say $\lsim 0.1$ Mpc$^{-1}$), and the
other wavenumbers $k',l$ are large (say $k' \sim l/\chi_* \sim 1$ Mpc$^{-1}$).
To see this intuitively, consider the following argument.
The kSZ is sourced by a real-space product of the form $\delta_e(\x) v_r(\x)$, and almost all of the
power in the velocity field $v_r$ comes from large scales.  Therefore, one of the wavenumbers
must correspond to a large scale, say $k \lsim 0.1$ Mpc$^{-1}$.
On the other hand, the CMB wavenumber must be roughly $l \sim 4000$, since smaller values of $l$ will be
dominated by the primary and lensed CMB, and larger values of $l$ will be noise-dominated.
The triangle condition $\k+\k'+(\l/\chi_*)=0$ then requires the wavenumber $k'$ to correspond to a small scale, 
roughly $k' \sim 1$ Mpc$^{-1}$.

Now we introduce an ansatz which will be of central importance throughout the paper.
We assume that in the squeezed limit ($k \lsim 0.1$ Mpc$^{-1}$ and $k' \sim 1$ Mpc$^{-1}$),
the kSZ bispectrum is accurately approximated by the tree-level expression in Eq.~(\ref{eq:B_ksz}),
but using the {\em nonlinear} small-scale galaxy electron $P_{ge}(k')$ on the RHS.
(Abusing terminology slightly, we will continue to call Eq.~(\ref{eq:B_ksz}) the ``tree-level'' bispectrum,
even though the $P_{ge}$ factor now includes loop and nonperturbative contributions.)

As a direct check that our ansatz is accurate, we have estimated the bispectrum directly from $N$-body simulations,
and compared to the tree-level approximation~(\ref{eq:B_ksz}).  We used the public DarkSky simulation~\cite{Skillman:2014qca}
with box size 1600 $h$ Mpc$^{-1}$ and 4096$^3$ particles, and used dark matter particles instead of electrons,
and halos instead of galaxies.  Empirically, we find in the squeezed limit
($k \le 0.04$ $h$ Mpc$^{-1}$ and $1 \le k' \le 2$ $h$ Mpc$^{-1}$), the tree-level 
bispectrum~(\ref{eq:B_ksz}) is accurate to a few percent or better.
The details of our $N$-body simulation results will be presented in a separate paper~\cite{Utkarsh}.

\begin{figure}[tbh]
  \includegraphics[width=0.45\textwidth]{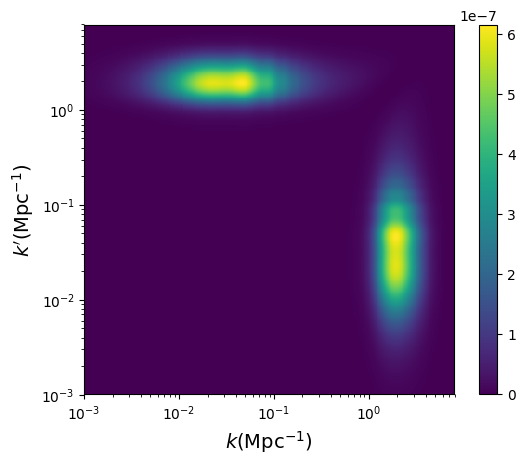}  \includegraphics[width=0.45\textwidth]{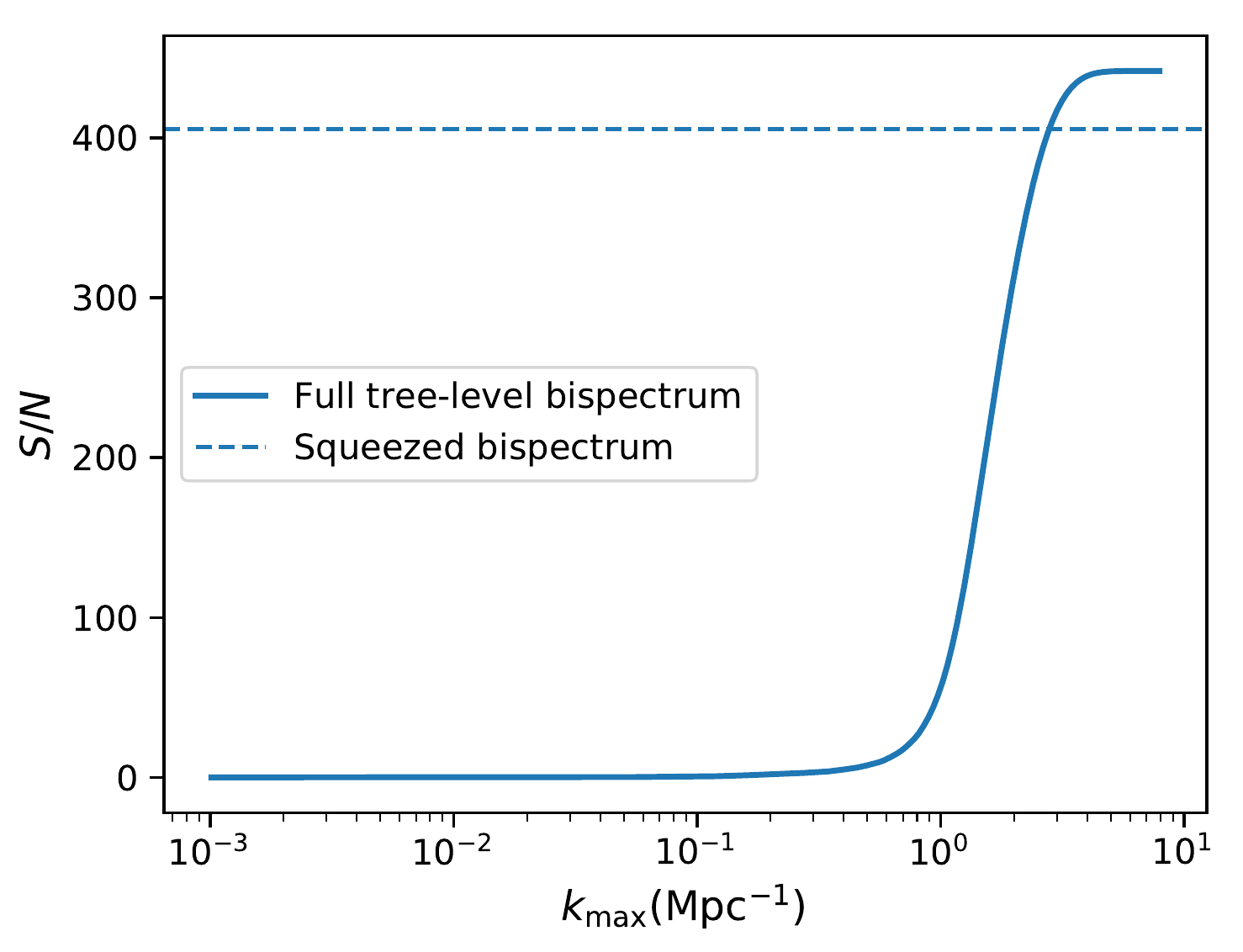}
  \caption{Scale dependence of the kSZ bispectrum.
    {\it Left:} SNR distribution in the $(k,k')$-plane, obtained by plotting the integrand $f(k,k')$ defined in Eq.~(\ref{eq:fkk}).
      As expected, most of the SNR comes from the squeezed limits $k \ll k'$ and $k' \ll k$.
    {\it Right:} The cumulative SNR of the bispectrum as more short-wavelength modes are included in the Fisher integral. 
      The difference between the squeezed limit SNR and the tree-level integral is 11\%.
    In both plots, noise parameters from Simons Observatory and DESI were assumed 
      (see Tables~\ref{tab:galaxy_surveys},~\ref{tab:cmbexp} below).}
  \label{fig:fkk}
\end{figure}

As a check that our approximations are self-consistent, we can show that if we {\em assume} that
the kSZ bispectrum is given by the tree-level expression in Eq.~(\ref{eq:B_ksz}), then the signal-to-noise
is dominated by the squeezed limit $k \ll k'$.
We write the total signal-to-noise of the bispectrum
as $\mbox{SNR}^2 = F_{BB}$, where $F_{BB}$ is the Fisher matrix defined in Eq.~(\ref{eq:F_scalar}).
We then plug in the tree-level kSZ bispectrum in Eq.~(\ref{eq:B_ksz}), and integrate out the variables
$k_r,l$ to write the Fisher matrix as a double integral over $(k,k')$.  After a short calculation we get:
\be
\mbox{SNR}^2 = \frac{V}{2} \int \frac{dk}{k} \frac{dk'}{k'} f(k,k')
\ee
where $f(k,k')$ is defined by:
\be
f(k,k') = \frac{K_*^2}{16\pi^3\chi_*^3} \frac{k^2 (k')^2}{P_{gg}^{\rm tot}(k) P_{gg}^{\rm tot}(k')} 
  \left( P_{ge}(k) \frac{P_{gv}(k')}{k'} - \frac{P_{gv}(k)}{k} P_{ge}(k') \right)^2 
  \int_{|k-k'|\chi_*}^{(k+k')\chi_*} dl \, \frac{\Gamma(k,k',l/\chi_*)^2}{C_l^{TT,\rm tot}}  \label{eq:fkk}
\ee
and $\Gamma$ is defined in Eq.~(\ref{eq:Gamma_def}).
By plotting the integrand $f(k,k')$, we can see which parts of the $(k,k')$-plane contribute to the integral.
We find that almost all of the signal-to-noise comes from the squeezed limit (Figure~\ref{fig:fkk}).

\subsection{Squeezed limit}

Summarizing the previous section, we have argued that the kSZ bispectrum
is dominated by the squeezed limit $k_L \ll k_S$, where $k_L \lsim 0.1$ Mpc$^{-1}$
is a linear scale, and $k_S \sim 1$ Mpc$^{-1}$ is a nonlinear scale.

In the squeezed limit, our previous results simplify.
The bispectrum becomes:
\be
B(k_L,k_S,l,k_{Lr}) = -\frac{K_* k_{Lr}}{\chi_*^2} \frac{P_{gv}(k_L)}{k_L} P_{ge}(k_S)  \label{eq:B_squeezed}
\ee
since the first term in Eq.~(\ref{eq:B_ksz}) is much smaller than the second.
We have omitted the ``tree'' subscript on the LHS since
the tree-level bispectrum is an accurate approximation in the squeezed limit,
provided that the nonlinear power spectrum $P_{ge}(k_S)$ is used on the RHS.
In this paper, we model $P_{ge}(k_S)$ using the halo model (Appendix~\ref{app:halo_model}).
Note that the triangle condition $\k_L + \k_S + (\l/\chi_*) = 0$ implies $l \approx (k_S \chi_*)$ 
and $k_{Sr} = -k_{Lr}$. 

The optimal bispectrum estimator $\hE$ in Eq.~(\ref{eq:E_def}) can be simplified, if we
restrict the integrals to the squeezed limit $k_L \ll k_S$:
\be
\hE = \frac{K_*}{\chi_*^2 F_{BB}} \int \frac{d^3\k_L}{(2\pi)^3} \, \frac{d^3\k_S}{(2\pi)^3} \, \frac{d^2\l}{(2\pi)^2} \,
    \frac{ik_{Lr}}{k_L} \frac{P_{gv}(k_L) P_{ge}(k_S)}{P_{gg}^{\rm tot}(k_L) P_{gg}^{\rm tot}(k_S) C_l^{TT,\rm tot}}
     \Big( \delta_g(\k_L) \delta_g(\k_S) T(\l) \Big) \,
    (2\pi)^3 \delta^3\left( \k_L + \k_S + \frac{\l}{\chi_*} \right)  \label{eq:E_squeezed}
\ee
where the integrals on the RHS should be understood as running over wavenumbers $k_L \ll k_S$
which contribute significantly to the signal-to-noise (as shown in Figure~\ref{fig:fkk}).
Note that there is a factor of two in Eq.~(\ref{eq:E_squeezed}) relative to Eq.~(\ref{eq:E_def}) 
because the $(k,k')$ integral in~(\ref{eq:E_def}) runs over squeezed configurations twice
(for $k \ll k'$ and $k' \ll k$).

The Fisher matrix in Eq.~(\ref{eq:F_vector}) simplifies as follows.
If we make no assumptions on the bispectra other than assuming that the
squeezed limit dominates, then:
\be
F_{BB'} = V \int dk_L \, dk_S \, dk_{Lr} \, \frac{k_L k_S \chi_*^2}{8 \pi^3} 
   \left( \frac{B(k_L,k_S,l,k_{Lr})^* \, B'(k_L,k_S,l,k_{Lr})}{P_{gg}^{\rm tot}(k_L) \, P_{gg}^{\rm tot}(k_S) \, C_l^{TT,\rm tot}} \right)_{l=k_S\chi_*}  \label{eq:F_squeezed}
\ee
where the integral runs over $|k_{Lr}| \le k_L \ll k_S$, and both positive and negative $k_{Lr}$.
This form of the Fisher matrix is nice since the algebraically messy factor $I(k,k',l,k_r)$ does not appear (as in Eq.~(\ref{eq:F_scalar})).
Now suppose we specialize further, by considering bispectra $B,B'$ which are of ``kSZ type'',
meaning that the bispectra are given by the functional form in Eq.~(\ref{eq:B_squeezed}) for different
choices of power spectra $P_{gv}, P_{ge}$:
\be
B(k_L,k_S,l,k_{Lr}) = -\frac{K_* k_{Lr}}{\chi_*^2} \frac{P_{gv}(k_L)}{k_L} P_{ge}(k_S)
  \hspace{1cm}
B'(k_L,k_S,l,k_{Lr}) = -\frac{K_* k_{Lr}}{\chi_*^2} \frac{P'_{gv}(k_L)}{k_L} P'_{ge}(k_S)
\ee
Plugging into Eq.~(\ref{eq:F_squeezed}), the Fisher matrix between kSZ-type bispectra can be written:
\ba
F_{BB'}
&=& V \frac{K_*^2}{8\pi^3 \chi_*^2} \int dk_L \, dk_S \, dk_{Lr} \, k_{Lr}^2 
     \left( \frac{P_{gv}(k_L) P_{gv}'(k_L)}{k_L P_{gg}^{\rm tot}(k_L)} \right)
     \left( \frac{k_S P_{ge}(k_S) P_{ge}'(k_S)}{P_{gg}^{\rm tot}(k_S)} \right)
     \left( \frac{1}{(C_l^{\rm tot})_{l=k_S\chi_*}} \right) \nn \\
&=& V \frac{K_*^2}{12 \pi^3 \chi_*^2}
 \left( \int dk_L \, k_L^2 \frac{P_{gv}(k_L) P'_{gv}(k_L)}{P_{gg}^{\rm tot}(k_L)} \right)
 \left( \int dk_S \, k_S \frac{P_{ge}(k_S) P'_{ge}(k_S)}{P_{gg}^{\rm tot}(k_S)} \frac{1}{(C_l^{\rm tot})_{l=k_S\chi_*}} \right)  \label{eq:F_final}
\ea
where in the second line we have integrated $k_{Lr}$ from $-k_L$ to $k_L$.
The Fisher matrix factorizes as the product of integrals over $k_L$ and $k_S$.
This factorization implies that the measurements of $P_{gv}(k_L)$ and $P_{ge}(k_S)$ 
obtained from kSZ tomography are independent, aside from an overall normalization which is degenerate.
This degeneracy in kSZ tomography is called the ``optical depth degeneracy'' and will be
discussed in more detail in~\S\ref{sec:forecasts_pheno}.

\section{Equivalence with other formalisms}
\label{sec:equivalence}

Summarizing previous sections, we have now shown that the optimal estimator
for kSZ tomography is:
\be
\hE = \frac{K_*}{\chi_*^2 F_{BB}} \int \frac{d^3\k_L}{(2\pi)^3} \, \frac{d^3\k_S}{(2\pi)^3} \, \frac{d^2\l}{(2\pi)^2} \,
    \frac{ik_{Lr}}{k_L} \frac{P_{gv}(k_L) P_{ge}(k_S)}{P_{gg}^{\rm tot}(k_L) P_{gg}^{\rm tot}(k_S) C_l^{TT,\rm tot}}
     \Big( \delta_g(\k_L) \delta_g(\k_S) T(\l) \Big) \,
    (2\pi)^3 \delta^3\left( \k_L + \k_S + \frac{\l}{\chi_*} \right)  \label{eq:hE_equiv1}
\ee
where the integral runs over triangles $\k_L + \k_S + \l/\chi_* = 0$.
The integral should also be restricted to squeezed triangles, in order to ensure 
that Eq.~(\ref{eq:B_squeezed}) for the bispectrum is an accurate approximation.
For example, the integrals could be cut off at $k_{L,\rm max} = 0.1$ Mpc$^{-1}$ and
$k_{S,\rm min} = 1$ Mpc$^{-1}$.  We have shown (Fig.~\ref{fig:fkk}) that almost all
of the signal-to-noise comes from the squeezed limit, so that the value of $\hE$ will
not depend much on the precise choice of cutoffs.

The form of $\hE$ in Eq.~(\ref{eq:hE_equiv1}) is convenient for many calculations, but
less intuitive than the kSZ tomography formalisms proposed 
in~\cite{Ho:2009iw,Hand:2012ui,Li:2014mja,Alonso:2016jpy,Deutsch:2017ybc}.
One may also wonder how best to evaluate $\hE$ algorithmically, given a CMB map and
a galaxy catalog.  There is more than one way to do this, since $\hE$ is a triple integral,
and there is also a double sum over galaxies hidden in the $\delta_g(\k_L) \delta_g(\k_S)$ factor.  
These integrals and sums can be evaluated using different orderings, leading to different 
implementations of the estimator $\hE$.  In this section, we will show that each of the kSZ tomography formalisms
in~\cite{Ho:2009iw,Hand:2012ui,Li:2014mja,Alonso:2016jpy,Deutsch:2017ybc} corresponds to 
a different implementation of $\hE$ as follows:
\begin{itemize}
\item We can write both factors of $\delta_g(\k)$ in Eq.~(\ref{eq:hE_equiv1}) as sums over galaxy positions 
 $\delta_g(\k) = \sum_i e^{-i\k\cdot x_i}$, and bring both sums to the outside, to write
 $\hE$ as a double sum over galaxy positions $\hE = \sum_{ij} W_{ij}$, where
 the pair weighting $W_{ij}$ depends on the CMB temperatures at galaxy positions $i,j$.
 This turns out to be equivalent to the pair sum estimator proposed in~\cite{Hand:2012ui}.
\item We can bring the $\l$-integral to the outside, and write the estimator in the
 schematic form $\hE = \int d^2\l \, T(\l)^* \hT(\l)$, where $\hT$ is a 2D map formed from
 two powers of the galaxy field.  This turns out to be equivalent to the kSZ template formalism from~\cite{Ho:2009iw}.
\item We can write $\delta_g(\k_S) = \sum_i e^{-i\k_S\cdot\x_i}$ as a sum over galaxy positions $\x_i$,
 and bring the sum to the outside, to write $\hE$ as a sum of schematic form $\sum_i \eta_i \tT_i$.
 Here, $\tT_i$ is the high-pass filtered CMB at galaxy location $i$, and 
 $\eta_i$ is an estimate of the radial velocity at galaxy $i$ which is
 derived from the long-wavelength modes of the galaxy survey.
 This turns out to be equivalent to either of the kSZ tomography statistics in~\cite{Li:2014mja,Alonso:2016jpy}.
\item We can bring the $\k_L$-integral to the outside, and write the estimator in the
 schematic form $\hE = \int d^3\k_L \, \delta_g(\k_L)^* \, \hv(\k_L)$, where $\hv(\k_L)$
 is a 3D map formed from one power of the small-scale CMB temperature and one power
 of the small-scale galaxy modes.
 This turns out to be equivalent to the kSZ-derived velocity reconstruction from~\cite{Deutsch:2017ybc}.
\end{itemize}
In the following subsections we will work out the details of this equivalence,
for each of the kSZ tomography statistics in turn.  We will do this in more detail
for the last statistic (the kSZ-derived velocity reconstruction from~\cite{Deutsch:2017ybc}), since some
intermediate results in the derivation will be used in later sections of the paper.

\subsection{Equivalence between the bispectrum and pair sum}
\label{ssec:pair_sum}

In this section we will show that the bispectrum estimator $\hE$ is equivalent to the
pair sum estimator from~\cite{Hand:2012ui}.  We start by writing $\hE$ as:
\be
\hE = \int \frac{d^3\k_L}{(2\pi)^3} \, \frac{d^3\k_S}{(2\pi)^3} \, \frac{d^2\l}{(2\pi)^2} \,
    \frac{ik_{Lr}}{k_L} \frac{P_{gv}(k_L) P_{ge}(l/\chi_*)}{P_{gg}^{\rm tot}(k_L) P_{gg}^{\rm tot}(l/\chi_*) C_l^{TT,\rm tot}}
     \Big( \delta_g(\k_L) \delta_g(\k_S) T(\l) \Big) \,
     (2\pi)^3 \delta^3\left( \k_L + \k_S + \frac{\l}{\chi_*} \right)  \label{eq:hE_equiv_der2}
\ee
where we have replaced $P_{ge}(k_S)$ by $P_{ge}(l/\chi_*)$ in Eq.~(\ref{eq:hE_equiv1}),
and likewise for $P_{gg}^{\rm tot}(k_S)$.
This is an accurate approximation in the squeezed limit $k_L \ll k_S$.
We have also removed a constant prefactor, since the overall normalization of the
estimator will not be important in this section.

In a galaxy catalog, the galaxies are specified as a sequence of 3D locations
$\x_i$ where $i=1,\cdots,N_{\rm gal}$, and the galaxy field $\delta_g(\x)$ is a sum
of delta functions (or in Fourier space, a sum of complex exponentials):
\be
\delta_g(\x) = \frac{1}{n_g} \sum_i \delta^3(\x-\x_i)
  \hspace{1.5cm}
\delta_g(\k) = \frac{1}{n_g} \sum_i e^{-i\k\cdot\x_i} 
\ee
We plug this into Eq.~(\ref{eq:hE_equiv_der2}) and bring the sums to the outside,
to write the result as a sum over galaxy pairs $(i,j)$:
\ba
\hE
 &=& \sum_{ij} \int \frac{d^3\k_L}{(2\pi)^3} \, \frac{d^3\k_S}{(2\pi)^3} \, \frac{d^2\l}{(2\pi)^2} \,
  \frac{ik_{Lr}}{k_L} \frac{P_{gv}(k_L) P_{ge}(l/\chi_*)}{P_{gg}^{\rm tot}(k_L) P_{gg}^{\rm tot}(l/\chi_*) C_l^{TT,\rm tot}}
    T(\l) e^{-i(\k_L\cdot\x_i + \k_S\cdot\x_j)} \,
    (2\pi)^3 \delta^3\left( \k_L + \k_S + \frac{\l}{\chi_*} \right) \nn \\
 &=& \sum_{ij} \int \frac{d^3\k_L}{(2\pi)^3} \, \frac{d^2\l}{(2\pi)^2} \,
    \frac{ik_{Lr}}{k_L} \frac{P_{gv}(k_L) P_{ge}(l/\chi_*)}{P_{gg}^{\rm tot}(k_L) P_{gg}^{\rm tot}(l/\chi_*) C_l^{TT,\rm tot}}
    e^{-i\k_L\cdot(\x_i-\x_j)} e^{i\l\cdot(\x_j^\perp/\chi_*)} T(\l)  \nn \\
 &=& \sum_{ij} \Omega(\x_j-\x_i) \, \tT(\th_j)  \hspace{1.5cm} \mbox{(where $\th_j = \x_j^\perp/\chi_*$)}  \label{eq:pair_sum_der1}
\ea
where we have defined a filtered CMB $\tT(\th)$ and a weight function $\Omega(\x)$ by:
\ba
\tT(\th) &=& \int \frac{d^2\l}{(2\pi)^2} \frac{P_{ge}(l/\chi_*)}{P_{gg}^{\rm tot}(l/\chi_*) C_l^{TT,\rm tot}} T(\l) e^{i\l\cdot\th}  \label{eq:pair_sum_T}  \\
\Omega(\x) &=& \int \frac{d^3\k_L}{(2\pi)^3} \, \frac{ik_{Lr}}{k_L} \frac{P_{gv}(k_L)}{P_{gg}^{\rm tot}(k_L)} e^{i\k_L\cdot\x}
\ea
The quantity $\th_j = \x_j/\chi_*$ defined in Eq.~(\ref{eq:pair_sum_der1}) is just the angular
location of the $j$-th galaxy, in the simplified box geometry used in this paper (\S\ref{sec:definitions}).

To simplify the pair weighting $\Omega(\x_j-\x_i)$ in Eq.~(\ref{eq:pair_sum_der1}),
we note that $\Omega(\x) = \partial_r W(|\x|)$, where:
\be
W(|\x|) = \int \frac{d^3\k}{(2\pi)^3} \frac{P_{gv}(k)}{k P_{gg}^{\rm tot}(k)} e^{i\k\cdot\x}  \label{eq:pair_sum_W}
\ee
and $\partial_r(\cdot)$ denotes the radial derivative.  We evaluate the radial derivative as:
\be
\Omega(\x) = \partial_r W(|\x|) = \frac{\x\cdot\hr}{|\x|} W'(|\x|)
\ee
where $\hr$ is the unit vector in the radial direction.  Plugging into Eq.~(\ref{eq:pair_sum_der1}), we
can write $\hE$ as:
\ba
\hE 
  &=& \sum_{ij} \frac{\x_{ij} \cdot \hr}{|\x_{ij}|} W'(|\x_{ij}|) \tT(\th_j) \nn \\
  &=& \frac{1}{2} \sum_{ij} \frac{\x_{ij} \cdot \hr}{|\x_{ij}|} W'(|\x_{ij}|) (\tT(\th_j) - \tT(\th_i))
\hspace{1cm} \mbox{(where $\x_{ij} = \x_j - \x_i$)}  \label{eq:pair_sum_final}
\ea
In the second line, we have antisymmetrized $\tT(\th_j) \rightarrow (\tT(\th_j) - \tT(\th_i))/2$,
since the remaining factors in the double sum are antisymmetric in $i,j$.

Our final form for $\hE$ in Eq.~(\ref{eq:pair_sum_final}) is a sum over galaxy pairs.
The pair weighting agrees perfectly with the pair sum estimator from~\cite{Hand:2012ui},
including the overall angular dependence $(\x_{ij} \cdot \hr)/|\x_{ij}|$.
Therefore, the bispectrum estimator is equivalent to the pair sum.

This equivalence establishes some interesting properties of the pair sum estimator
which are not obvious in advance.
First, the optimal $l$-weighting of the CMB is given by Eq.~(\ref{eq:pair_sum_T}).
Second, the optimal weighting in the pair separation $r = |\x_{ij}|$ is given by $W'(r)$,
where $W(r)$ is defined in Eq.~(\ref{eq:pair_sum_W}).
Third, the pair sum statistic is an optimal kSZ tomography estimator, if these weightings in $l$ and $r$ are used.
Fourth, the total signal-to-noise of the pair sum can be forecasted as $\mbox{SNR} = F_{BB}^{1/2}$, where
$F_{BB}$ is the bispectrum Fisher matrix defined previously in Eq.~(\ref{eq:F_final}).

\subsection{Equivalence between the bispectrum and kSZ template formalisms}
\label{ssec:ksz_template}

In this section we will show that the bispectrum estimator is equivalent to the kSZ template
formalism from~\cite{Ho:2009iw}.
First we recall the details of the kSZ template formalism.
We start by defining the 3D field:
\be
\eta(\k_L) = \frac{ik_{Lr}}{k_L} \frac{P_{gv}(k_L)}{P_{gg}^{\rm tot}(k_L)} \delta_g(\k_L)  \label{eq:eta_def}
\ee
This can be interpreted as a minimum variance linear reconstruction of the radial velocity $v_r(\k_L)$
from the long-wavelength modes of the galaxy survey.\footnote{We have used the notation $\eta$ (rather than say $\hv_r$) 
 in order to distinguish between two notions of ``large-scale velocity reconstruction'' that will be used throughout
 the paper.  The estimator $\eta$ defined in Eq.~(\ref{eq:eta_def}) is a linear reconstruction of the large-scale radial
 velocity from the large-scale galaxy field.  In the next section we will introduce a kSZ-derived quadratic
 estimator $\hv_r$ which also reconstructs the large-scale radial velocity, using one power of the small-scale
 CMB and one power of the small-scale galaxy field.}
Similarly, we define the 3D field:
\be
\epsilon(\k_S) = \frac{P_{ge}(k_S)}{P_{gg}^{\rm tot}(k_S)} \delta_g(\k_S)  \label{eq:epsilon_def}
\ee
which can be interpreted as a best estimate for the small-scale electron density $\delta_e(\k_S)$, given the galaxy map.
Finally, we define a 2D ``kSZ template'' field $\hT(\th)$, by radially integrating the product of fields $(\eta \epsilon)$:
\be
\hT(\th) = K_* \int_0^L dr \, \eta(\chi_*\th + r \hr) \, \epsilon(\chi_*\th + r \hr) \label{eq:hT_los}
\ee
The kSZ template field $\hT(\th)$ is constructed purely from the galaxy survey,
but we expect it to be highly correlated with the CMB temperature $T(\th)$,
since $\hT$ has been defined using an integral~(\ref{eq:hT_los})
which mimics the line-of-sight integral~(\ref{eq:ksz_los}) for the kSZ.
Ref.~\cite{Ho:2009iw} proposes using the cross power spectrum $C_l^{T\hT}$
as a statistic for kSZ tomography.

To show that $C_l^{T\hT}$ is equivalent to the bispectrum estimator $\hE$,
we proceed as follows.
First, we write $\hT$ in Fourier space, by plugging the definitions of $\eta,\epsilon$
(Eqs.~(\ref{eq:eta_def}),~(\ref{eq:epsilon_def})) into the definition of $\hT$ (Eq.~(\ref{eq:hT_los})):
\be
\hT(\l) = \frac{K_*}{\chi_*^2}  \int \frac{d^3\k_L}{(2\pi)^3} \frac{d^3\k_S}{(2\pi)^3} \frac{ik_{Lr}}{k_L}
    \frac{P_{gv}(k_L)}{P_{gg}^{\rm tot}(k_L)} \frac{P_{ge}(k_S)}{P_{gg}^{\rm tot}(k_S)}
    \Big( \delta_g(\k_L) \delta_g(\k_S) \Big) \, (2\pi)^3 \delta^3\!\left( \k_L + \k_S - \frac{\l}{\chi_*} \right) \label{eq:hT_fourier}
\ee
Then we calculate $C_l^{T\hT}$ as follows:
\ba
C_l^{T\hT} 
 &=& -K_* L \int \frac{d^3\k_L}{(2\pi)^3} \frac{d^3\k_S}{(2\pi)^3} \, \frac{k_{Lr}}{k_L} 
                      \frac{P_{gv}(k_L)}{P_{gg}^{\rm tot}(k_L)}
                      \frac{P_{ge}(k_S)}{P_{gg}^{\rm tot}(k_S)} B(k_L,k_S,l,k_{Lr})  \,
                      (2\pi)^3 \delta^3\!\left( \k_L + \k_S + \frac{\l}{\chi_*} \right) \nn \\
 &=& \frac{K_*^2 L}{\chi_*^2} \int \frac{d^3\k_L}{(2\pi)^3} \frac{d^3\k_S}{(2\pi)^3} \, \frac{k_{Lr}^2}{k_L^2} 
                      \frac{P_{gv}(k_L)^2}{P_{gg}^{\rm tot}(k_L)}
                      \frac{P_{ge}(k_S)^2}{P_{gg}^{\rm tot}(k_S)}  
                      (2\pi)^3 \delta^3\!\left( \k_L + \k_S + \frac{\l}{\chi_*} \right)  \nn \\
 &=& \frac{K_*^2 L}{6 \pi^2 \chi_*^2} 
        \left( \int dk_L \, k_L^2 \frac{P_{gv}(k_L)^2}{P_{gg}^{\rm tot}(k_L)} \right)
        \left( \frac{P_{ge}(k_S)^2}{P_{gg}^{\rm tot}(k_S)} \right)_{k_S=l/\chi_*}   \label{eq:cl_t_ht}
\ea
To get the first line, we have used Eq.~(\ref{eq:hT_fourier}) and the definition~(\ref{eq:B_def}) of the bispectrum.
To get the second line, we have plugged in the bispectrum in the form~(\ref{eq:B_squeezed}).
The third line is a simplification which is valid in the squeezed limit $k_L \ll k_S$.
A similar calculation, omitted for brevity, shows that the auto power spectrum of the template $\hT$ is given by the same expression, i.e.
\be
N_l^{\hT\hT} = C_l^{T\hT}  \label{eq:nlht_equals_cl}
\ee
where we have used the notation $N_l$ since we interpret the auto spectrum of $\hT$ as a noise power spectrum.

Using the results in Eqs.~(\ref{eq:hT_fourier}),~(\ref{eq:cl_t_ht}),~(\ref{eq:nlht_equals_cl}),
a short calculation starting from Eq.~(\ref{eq:hE_equiv1}) now shows that the optimal bispectrum 
estimator $\hE$ can be written in the form:
\be
\hE = \frac{1}{F_{BB}} \int \frac{d^2\l}{(2\pi)^2} \frac{C_l^{T\hT}}{C_l^{TT,\rm tot} N_l^{\hT\hT}} \, \Big( T(\l) \hT(-\l) \Big) \label{eq:E_template}
\ee
(This equation can be simplified using $N_l^{\hT\hT} = C_l^{T\hT}$, but we have written it in a way
which makes equivalence with the kSZ template formalism most transparent.)

This expression for $\hE$ agrees perfectly, including the $l$-weighting,
with the minimum variance estimator for the cross-correlation of two fields $T, \hT$
with auto and cross spectra given by $C_l^{TT,\rm tot}$, $C_l^{T\hT}$, and $N_l^{\hT\hT}$.
This proves that the kSZ tomography statistic $C_l^{T\hT}$ from~\cite{Ho:2009iw}
is equivalent to the optimal bispectrum estimator $\hE$.

\subsection{Equivalence between the bispectrum and velocity matched filter}

In this section, we will show that the bispectrum estimator $\hE$ is equivalent
to the ``velocity matched filter'' statistic from~\cite{Li:2014mja}.

The idea of~\cite{Li:2014mja} is that at the location of each galaxy, the kSZ effect produces a small correlation
between the radial velocity $v_r$ and the high-pass filtered CMB temperature $\tT$.
The correlation is small on a per-object basis, but can be detected by summing over many galaxies.
In our notation, the velocity matched-filter statistic 
is:\footnote{Ref.~\cite{Li:2014mja} uses different notation as follows.  The linear radial velocity $\eta(\x_i)$
 is denoted $v_{\rm rec,i}$, and the high-pass filtered CMB $\tT(\th_i)$ is denoted $K_i$.  The kSZ tomography
 statistic is written $\halpha = \sum_i w_i (K_i / v_{\rm rec,i})$, where $w_i \propto v_{\rm rec,i}^2$,
 or equivalently $\halpha \propto \sum_i v_{\rm rec,i} K_i$.}
\be
\halpha = \sum_i \eta(\x_i) \tT(\th_i)  \hspace{1cm} \mbox{(where $\th_i = \x_i^\perp/\chi_*$)} \label{eq:halpha_def}
\ee
where $\tT(\th)$ is the high-pass filtered CMB defined previously in Eq.~(\ref{eq:pair_sum_T}),
and $\eta(\x)$ is the linear radial velocity reconstruction defined in Eq.~(\ref{eq:eta_def}).

To show that the kSZ tomography statistic $\halpha$ is equivalent to the bispectrum estimator $\hE$,
we start by writing $\hE$ in the form:
\be
\hE = \int \frac{d^3\k_L}{(2\pi)^3} \, \frac{d^3\k_S}{(2\pi)^3} \, \frac{d^2\l}{(2\pi)^2} \,
    \frac{ik_{Lr}}{k_L} \frac{P_{gv}(k_L) P_{ge}(l/\chi_*)}{P_{gg}^{\rm tot}(k_L) P_{gg}^{\rm tot}(l/\chi_*) C_l^{TT,\rm tot}}
     \Big( \delta_g(\k_L) \delta_g(\k_S) T(\l) \Big) \,
     (2\pi)^3 \delta^3\left( \k_L + \k_S + \frac{\l}{\chi_*} \right) 
\ee
where we have replaced $P_{ge}(k_S)$ by $P_{ge}(l/\chi_*)$ in Eq.~(\ref{eq:hE_equiv1}),
and likewise for $P_{gg}^{\rm tot}(k_S)$.  These replacements are valid in the squeezed limit $k_L \ll k_S$.
We then manipulate as follows:
\ba
\hE &=& \frac{1}{n_g} \sum_i 
 \int \frac{d^3\k_L}{(2\pi)^3} \, \frac{d^2\l}{(2\pi)^2} \,
    \frac{ik_{Lr}}{k_L} \frac{P_{gv}(k_L) P_{ge}(l/\chi_*)}{P_{gg}^{\rm tot}(k_L) P_{gg}^{\rm tot}(l/\chi_*) C_l^{TT,\rm tot}}
     \Big( \delta_g(\k_L) T(\l) \Big) e^{i(\k_L + \l/\chi_*) \cdot \x_i} \nn \\
 &=&
\frac{1}{n_g} \sum_i
  \left( \int \frac{d^3\k_L}{(2\pi)^3} \frac{ik_{Lr}}{k_L} \frac{P_{gv}(k_L)}{P_{gg}^{\rm tot}(k_L)} \delta_g(\k_L) e^{i\k_L\cdot\x_i} \right)
  \left( \int \frac{d^2\l}{(2\pi)^2} \, \frac{P_{ge}(l/\chi_*)}{P_{gg}^{\rm tot}(l/\chi_*) C_l^{TT,\rm tot}} T(\l) e^{i\l\cdot(\x_i^\perp / \chi_*)} \right) \nn \\
 &=&
\frac{1}{n_g} \sum_i \eta(\x_i) \tT(\x_i^\perp/\chi_*)
\ea
In the first line, we have plugged in $\delta_g(\k_S) = n_g^{-1} \sum_i e^{-i\k_S\cdot\x_i}$,
and used the delta function to do the $\k_S$-integral.
To get from the second line to the third, we have recognized the factors in parentheses as
the definitions of $\tT(\th)$ and $\eta(\x)$ in Eqs.~(\ref{eq:pair_sum_T}),~(\ref{eq:eta_def}).
The final result is the $\halpha$-statistic in Eq.~(\ref{eq:halpha_def}), completing the proof
that $\halpha$ is equivalent to the bispectrum estimator $\hE$.

\subsection{Equivalence between the bispectrum and the velocity growth method}
\label{ssec:velocity_growth}

In this section, we will show that the optimal bispectrum estimator $\hE$ is
equivalent to the ``velocity growth method'', a kSZ tomography statistic introduced
recently in~\cite{Alonso:2016jpy}.

In~\cite{Alonso:2016jpy}, the large-scale structure catalog is assumed to be a catalog
of galaxy clusters with mass estimates, and a prescription is given for the relative
weighting of mass bins.  Let us first consider the simpler case of a narrow mass bin.
We are also implicitly assuming a narrow redshift bin, by using the ``snapshot'' geometry
from~\S\ref{sec:definitions}.

The kSZ tomography statistic from~\cite{Alonso:2016jpy} is defined by maximizing a likelihood
function $\L(\alpha)$, which in our notation is given by:\footnote{Ref.~\cite{Alonso:2016jpy}
 uses different notation as follows. The linear radial velocity reconstruction $\eta(\x_i)$ is denoted $\hat\beta_r^i$,
 and the high-pass filtered CMB $\hT(\th_i)$ is denoted $a^i_{\rm kSZ}$.  In writing the likelihood in
 Eq.~(\ref{eq:velocity_growth_likelihood}), we have assumed that all clusters are identical, so that 
 the estimated optical depth $\tau_{500}$ is independent of $i$, and so are statistical errors on the
 quantities $\tau_{500}$, $\hT$, $\eta$.  This is a reasonable assumption because we are considering
 narrow mass and redshift bins.}
\be
-\log\L(\alpha) \propto \sum_i \Big( \alpha \eta(\x_i) \tau_{500} - \tT(\th_i) \Big)^2  \label{eq:velocity_growth_likelihood}
\ee
where $\tau_{500}$ is an estimate for the cluster optical depth,
$\tT(\th)$ is the high-pass filtered CMB defined previously in Eq.~(\ref{eq:pair_sum_T}),
and $\eta(\x)$ is the linear radial velocity reconstruction defined in Eq.~(\ref{eq:eta_def}).

Given this likelihood, the maximum-likelihood estimator $\halpha_{\rm ML}$ is:
\be
\halpha_{\rm ML} = \frac{\sum_i \eta(x_i) \tT(\th_i)}{\tau_{500} \sum_i \eta(\x_i)^2}
\ee
We note that statistical fluctuations in the denominator will be small, since there will
be many clusters in the catalog, and the value of $\eta(\x_i)^2$ will be uncorrelated from
one cluster to the next, provided the cluster separation is larger than the correlation length
of the velocity field.  Therefore, to a good approximation, we can replace $\eta(\x_i)^2$ in
the denominator by its expectation value $\langle \eta^2 \rangle$, and write:
\be
\halpha_{\rm ML} \approx \frac{\sum_i \eta(x_i) \tT(\th_i)}{N \tau_{500} \langle \eta^2 \rangle}
\ee
In this form, we see that $\halpha_{\rm ML}$ is proportional to the kSZ tomography statistic
considered in the previous section (in Eq.~(\ref{eq:halpha_def}), also denoted $\halpha$),
where we showed that it is equivalent to the optimal bispectrum estimator $\hE$.

In this analysis, we only considered the case of a narrow mass bin, setting aside
the question of how to optimally weight different mass bins.
Ref.~\cite{Alonso:2016jpy} discusses this optimization in detail, in addition to
the optimal choices of $l$ and $k_L$-weightings which appear in the filtered CMB $\tT$
and velocity reconstruction $\eta$ (Eqs.~(\ref{eq:pair_sum_T}),~(\ref{eq:eta_def})).

In the bispectrum approach used in this paper, these weight optimizations are performed differently.
The optimal $l$ and $k_L$-weightings are part of the optimal bispectrum estimator $\hE$, which
was derived previously in~\S\ref{sec:bispectrum}.
So far, we have not discussed how to optimally weight cluster mass bins.
We will defer this question to~\S\ref{ssec:larger_analyses}
as part of a more general discussion of how to incorporate kSZ tomography into larger analyses.

\subsection{Equivalence between the bispectrum and long-wavelength velocity reconstruction}
\label{ssec:vrec}

In this section, we will show that the optimal bispectrum estimator $\hE$ is equivalent
to the long-wavelength radial velocity reconstruction from~\cite{Deutsch:2017ybc}.

First we recall the idea from~\cite{Deutsch:2017ybc} (see also \cite{Terrana2016,Cayuso:2018lhv} for further related details).
The kSZ induces a squeezed bispectrum of schematic form $\langle v_r(\k_L) \delta_g(\k_S) T(\l) \rangle$.
Therefore, we can build a quadratic estimator for long-wavelength radial velocity modes $v_r(\k_L)$
by summing over pairs $(\delta_g(\k_S) T(\l))$ of short-wavelength modes in the galaxy and CMB maps.
This is analogous to CMB lensing, where there is a squeezed bispectrum of the form
$\langle \phi(\l) T(\l') T(\l'') \rangle$, and consequently the long-wavelength CMB lensing
potential $\phi$ can be reconstructed from small-scale CMB modes.

We next derive the minimum variance estimator $\hv_r(\k_L)$ in the simplified ``snapshot'' geometry
from~\S\ref{sec:definitions}, by solving a constrained optimization problem as follows.
Consider a general quadratic estimator of the form:
\be
\hv_r(\k_L) = \int \frac{d^3\k_S}{(2\pi)^3} \frac{d^2\l}{(2\pi)^2} \,
  W(\k_S,\l) \, \delta_g^*(\k_S) \, T^*(\l) \,
  (2\pi)^3 \delta^3\!\left( \k_L + \k_S + \frac{\l}{\chi_*} \right)  \label{eq:hv_w}
\ee
with weights $W(\k_S,\l)$ to be determined.
We want to find the weights $W(\k_S,\l)$ which minimize the power spectrum of the reconstruction,
subject to the constraint that the reconstruction is unbiased, i.e.~$\langle \hv_r(\k_L) \rangle$
is equal to the true radial velocity $v_r(\k_L)$.  Here, the expectation value $\langle \hv_r(\k_L) \rangle$
is an average over small-scale modes, in a fixed realization of the long-wavelength modes.

A short calculation gives the mean and noise power spectrum of the quadratic estimator in Eq.~(\ref{eq:hv_w}),
for arbitrary weights $W(\k_S,\l)$:
\ba
\langle \hv_r(\k_L) \rangle &=& \frac{K_*}{\chi_*^2} \left[
   \int \frac{d^3\k_S}{(2\pi)^3} \frac{d^2\l}{(2\pi)^2} \, W(\k_S,\l) \, P_{ge}(k_S) 
   (2\pi)^3 \delta^3\!\left( \k_L + \k_S + \frac{\l}{\chi_*} \right)
\right] v_r(\k_L)  \label{eq:expnvr_w}  \\
N_{v_r}(k_L) &=& \int \frac{d^3\k_S}{(2\pi)^3} \frac{d^2\l}{(2\pi)^2} \,
   |W(\k_S,\l)|^2 P_{gg}^{\rm tot}(k_S) C_l^{TT,\rm tot}
   (2\pi)^3 \delta^3\!\left( \k_L + \k_S + \frac{\l}{\chi_*} \right)  \label{eq:nvr_w}
\ea
To get the first line, we have used the identity $\langle \delta_g(\k_S) T(\l) \rangle = (K_* / \chi_*^2) P_{ge}(k_S) v_r(\k_S+\l/\chi_*)$,
which follows from Eq.~(\ref{eq:Tl}).

Now we solve for the weights $W(\k_S,\l)$ which minimize $N_{v_r}(k_L)$, subject to the constraint $\langle \hv_r(\k_L) \rangle = v_r(\k_L)$.
A short calculation using Eqs.~(\ref{eq:expnvr_w}),~(\ref{eq:nvr_w}) shows that $W(\k_S,\l)$ and $N_{v_r}(k_L)$ are related by:
\be
W(\k_S,\l) = N_{v_r}(k_L) \frac{K_*}{\chi_*^2} \frac{P_{ge}(k_S)}{P_{gg}^{\rm tot}(k_S) C_l^{TT,\rm tot}}
\ee
Plugging back into Eqs.~(\ref{eq:hv_w}),~(\ref{eq:nvr_w}), our final expressions for the minimum-variance
quadratic estimator $\hv_r(\k_L)$ and its reconstruction noise power spectrum are:
\ba
\hv_r(\k_L) &=& N_{v_r}(k_L) \frac{K_*}{\chi_*^2} 
 \int \frac{d^3\k_S}{(2\pi)^3} \frac{d^2\l}{(2\pi)^2} \,
    \frac{P_{ge}(k_S)}{P_{gg}^{\rm tot}(k_S) C_l^{TT,\rm tot}} \Big( \delta_g^*(\k_S) T^*(\l) \Big)
    (2\pi)^3 \delta^3\!\left( \k_L + \k_S + \frac{\l}{\chi_*} \right)  \label{eq:hv_final} \\
N_{v_r}(k_L) &=& \frac{\chi_*^4}{K_*^2} \left[ \int \frac{d^3\k_S}{(2\pi)^3} \frac{d^2\l}{(2\pi)^2} \,
    \frac{P_{ge}(k_S)^2}{P_{gg}^{\rm tot}(k_S) C_l^{TT,\rm tot}}
    (2\pi)^3 \delta^3\!\left( \k_L + \k_S + \frac{\l}{\chi_*} \right)
  \right]^{-1} \nn \\
 &=& \frac{\chi_*^2}{K_*^2} \left[ \int \frac{k_S \, dk_S}{2\pi} 
      \left( \frac{P_{ge}(k_S)^2}{P_{gg}^{\rm tot}(k_S) C_l^{TT,\rm tot}} \right)_{l=k_S\chi_*} \right]^{-1} \label{eq:Nvr_final} 
\ea
where we have used $k_L \ll k_S$ in the last line to simplify.

This concludes our description of the quadratic estimator $\hv_r$.
We mention in advance that the expressions for $\hv_r$ and $N_{v_r}$
(Eqs.~(\ref{eq:hv_final}),~(\ref{eq:Nvr_final})) will be used extensively throughout the
rest of the paper.

To construct a kSZ tomography statistic, i.e.~a scalar quantity which is
kSZ-sensitive, we can cross-correlate the radial velocity reconstruction $\hv_r(\k_L)$
with the galaxy field on large scales, with a suitable $\k_L$-weighting.
We next show that this procedure is equivalent to the optimal bispectrum estimator $\hE$.
Starting from Eq.~(\ref{eq:hE_equiv1}) for $\hE$, we plug in Eq.~(\ref{eq:hv_final}) for $\hv_r$,
to write $\hE$ in the form:
\be
\hE = \frac{1}{F_{BB}} \int \frac{d^3\k_L}{(2\pi)^3} \frac{ik_{Lr}}{k_L} \frac{P_{gv}(k_L)}{P_{gg}^{\rm tot}(k_L) N_{v_r}(k_L)} \Big( \delta_g(\k_L) \hv_r(\k_L)^* \Big)  \label{eq:hE_recon}
\ee
Similarly, we start with the Fisher matrix element $F_{BB}$ in the following form:
\be
F_{BB} = V \frac{K_*^2}{\chi_*^4} 
          \int \frac{d^3\k_L}{(2\pi)^3} \frac{d^3\k_S}{(2\pi)^3} \frac{d^2\l}{(2\pi)^2}
          \frac{k_{Lr}^2}{k_L^2}
          \frac{P_{gv}(k_L)^2}{P_{gg}^{\rm tot}(k_L)}
          \frac{P_{ge}(k_S)^2}{P_{gg}^{\rm tot}(k_S)}
          \frac{1}{C_l^{TT,\rm tot}}
    (2\pi)^3 \delta^3\left( \k_L + \k_S + \frac{\l}{\chi_*} \right)
\ee
which follows from Eq.~(\ref{eq:F_vector}), after restricting the integral to squeezed triangles
and plugging in the kSZ bispectrum from Eq.~(\ref{eq:B_squeezed}).
We then plug in Eq.~(\ref{eq:Nvr_final}) for $N_{v_r}$, to write $F_{BB}$ in the form:
\be
F_{BB} = V \int \frac{d^3\k_L}{(2\pi)^3} \frac{k_{Lr}^2}{k_L^2} \frac{P_{gv}(k_L)^2}{P_{gg}^{\rm tot}(k_L) N_{v_r}(k_L)}  \label{eq:F_recon}
\ee
The expressions~(\ref{eq:hE_recon}),~(\ref{eq:F_recon}) for $\hE$ and $F_{BB}$ agree perfectly, including constant factors,
with the minimum variance estimator for the cross-correlation of two fields $\delta_g(\k_L), \hv_r(\k_L)$, with an anisotropic
two-point function of the form $\langle \delta_g(\k')^* \hv_r(\k) \rangle = (ik_r/k) P_{gv}(k) (2\pi)^3 \delta^3(\k-\k')$.
This completes the proof that the optimal bispectrum estimator $\hE$ is equivalent to cross-correlating the
kSZ-derived velocity reconstruction $\hv_r$ with the galaxy field $\delta_g$ on large scales.

The reconstruction $\hv_r$ can be used to build more general statistics as well.
For example, we could consider the auto power spectrum of $\hv_r$ (rather than the cross power spectrum with a galaxy field).
Or we could introduce $\hv_r$ into a larger analysis including many fields which can be cross-correlated with each other.
The $\hv_r$ formalism is also particularly convenient for incorporating redshift-space distortions and photometric redshift errors.
For this reason, the kSZ-derived velocity reconstruction is a particularly powerful approach to kSZ tomography
(at least for cosmology), and we advocate using it.  Most of the rest of the paper is devoted to exploring properties of $\hv_r$
in more detail.

\section{Forecasts and phenomenology}
\label{sec:forecasts_pheno}

So far, we have built up a lot of formal machinery.
We have interpreted kSZ tomography as bispectrum estimation,
constructed the optimal bispectrum estimator, and its Fisher matrix (\S\ref{sec:bispectrum}).
We have shown that the formalisms~\cite{Ho:2009iw,Hand:2012ui,Li:2014mja,Alonso:2016jpy,Deutsch:2017ybc}
for kSZ tomography are equivalent to the bispectrum, and worked out the details of how to translate
between them (\S\ref{sec:equivalence}).

In this section, we will analyze several aspects of kSZ tomography using our machinery.
The ability to translate between different formalisms will be useful, since calculations
which are intuitive in one formalism may not be in others.

In forecasts in this section, we consider galaxy surveys with parameters given in Table~\ref{tab:galaxy_surveys}. 
The parameters for LSST are based on \cite{LSSTSciBook,LSSTSRD}, and those for DESI are based on \cite{DESIReport}. 
We use the LSST Gold sample up to Year 1 (LSST-Y1) and up to Year 10 (LSST-Y10). The DESI sample we consider includes the BGS, LRG, ELG and QSO samples.

For the Planck and CMB-S4 CMB experiments, we model the noise power spectrum in each frequency channel as
\begin{equation}
N_\ell^\nu = N_0^\nu \left(1+\left(\frac{\ell}{\ell_{\mathrm{knee}}}\right)^{\alpha}\right)\rm{exp}\left(\frac{\ell(\ell+1)\theta_{\rm{FWHM}}^2}{8\ln 2}\right).
\end{equation}
with frequencies $\nu$, beamsize $\theta_{\rm{FWHM}}$ and white noise sensitivity $s_w$ given in Table~\ref{tab:cmbexp}. For CMB-S4, we conservatively assume atmospheric noise parameters 
of $\ell_{\mathrm{knee}}=3000$ and $\alpha=-4$ in all frequency channels, and do not include the atmospheric noise term for Planck. We then construct a standard internal linear combination (ILC) noise curve 
from those frequencies in combination with the Planck frequency bandpasses specified in Table~\ref{tab:cmbexp}, with foreground noise from tSZ, 
clustered and point source CIB, and radio point sources \cite{Dunkley2011} in addition to reionization and late-time kSZ. 
For Simons Observatory, we use a parametric fit to the publicly available noise curves~\cite{Ade:2018sbj} for the Goal $f_{\mathrm{sky}}=0.4$ standard ILC cleaned case.

We assume that the CMB experiment overlaps with DESI over $f_{\mathrm{sky}}=0.2$ and with LSST over $f_{\mathrm{sky}}=0.3$. 
For all small-scale power spectra ($P_{ge}(k_S)$, $P_{gg}(k_S)$ and late-time $C_{\ell}^{\rm kSZ}$), we use the halo model 
as described in Appendix~\ref{app:halo_model}, where the stellar mass threshold is chosen such that the predicted number 
density of galaxies is the same as in Table~\ref{tab:galaxy_surveys}.  For large-scale power spectra ($P_{gv}(k_L)$, $P_{gg}(k_L)$), 
we multiply the nonlinear matter power spectrum by the linear galaxy bias in Table~\ref{tab:galaxy_surveys}.

\begin{table}[b!]
\begin{center}
\begin{tabular}{l|llll}
                                              & \textbf{DESI} & \textbf{LSST-Y1} & \textbf{LSST-Y10} \\ \hline
Mean redshift                                 & 0.75          & 0.9          & 1.1          \\
Overlap survey volume ($\mathrm{Gpc}^3$)      & 116            & 113.4           & 180           \\
Overlap $f_{\mathrm{sky}}$              & 0.2            & 0.3           & 0.3           \\
Number density ($\mathrm{Mpc}^{-3}$)    & $1.7 \times 10^{-4}$ & $6.9 \times 10^{-3}$ & $1.2 \times 10^{-2}$          \\
Number density ($\mathrm{arcmin}^{-2}$) & $0.66$   & $18$ & $48$          \\
Galaxy bias                                   & 1.51             & 1.7          & 1.6           \\
Photo-$z$ error $\sigma_z/(1+z)$              & 0             & 0.03         & 0.03         
\end{tabular}
\end{center}
\caption{Galaxy survey parameters used throughout~\S\ref{sec:forecasts_pheno}.
  In the case of LSST, photo-$z$ errors are incorporated into kSZ tomography forecasts
  using machinery which will be developed in~\S\ref{sec:photoz_rsd}.}
\label{tab:galaxy_surveys}
\end{table}

\newcommand{\arcmin}{\mathrm{arcmin}}
\begin{table*}
  \begin{center}

\begin{minipage}{.33\linewidth}
   \begin{tabular}{c|c|c} 
     \hline \hline
      \multicolumn{3}{c}{Planck}   \\
     \hline            
   Frequency & Beam & Noise RMS\\
      (GHz)  & (arcmin)& ($\mu$K-arcmin)   \\
     \hline 
30 & 33 & 145    \\
44 & 23 & 149    \\
70 & 14 & 137    \\
100 & 10 & 65   \\
143 & 7 & 43    \\
217 & 5 & 66    \\
353 & 5 & 200    \\
     \hline
   \end{tabular}
\end{minipage}%
\begin{minipage}{.33\linewidth}
   \begin{tabular}{c|c|c} 
     \hline \hline
        \multicolumn{3}{c}{CMB-S4}\\
     \hline            
   Frequency & Beam & Noise RMS\\
      (GHz)  & (arcmin)& ($\mu$K-arcmin) \\
     \hline 
28 &  7.6 & 20.0  \\
41 &  5.1 & 17.5  \\
90 & 2.4 & 2.0  \\
150 & 1.5 & 1.8 \\
230 & 1.0 & 6.3  \\
     \hline
   \end{tabular}
\end{minipage}

  \end{center}
  \caption{CMB frequency channels, white noise levels, and beam sizes used throughout~\S\ref{sec:forecasts_pheno}. For CMB-S4, we conservatively assume atmospheric noise parameters of $\ell_{\mathrm{knee}}$=3000 and $\alpha=-4$ in all frequency channels. }
  \label{tab:cmbexp}
\end{table*}

\subsection{Total signal-to-noise ratio}
\label{ssec:snr}

The total SNR for kSZ tomography can be computed using any of the three expressions:
\ba
\mbox{SNR}^2
  &=& V \frac{K_*^2}{12 \pi^3 \chi_*^2}
    \left( \int dk_L \, k_L^2 \frac{P_{gv}(k_L)^2}{P_{gg}^{\rm tot}(k_L)} \right)
    \left( \int dk_S \, k_S \frac{P_{ge}(k_S)^2}{P_{gg}^{\rm tot}(k_S)} \frac{1}{(C_l^{\rm tot})_{l=k_S\chi_*}} \right)  \label{eq:total_snr1} \\
  &=& \Omega \int \frac{d^2\l}{(2\pi)^2} \frac{(C_l^{T\hT})^2}{C_l^{TT,\rm tot} N_l^{\hT\hT}}  \label{eq:total_snr2} \\
  &=& V \int \frac{d^3\k_L}{(2\pi)^3} \frac{k_{Lr}^2}{k_L^2} \frac{P_{gv}(k_L)^2}{P_{gg}^{\rm tot}(k_L) N_{v_r}(k_L)}  \label{eq:total_snr3}
\ea
These expressions are mathematically equivalent and correspond to different formalisms introduced previously.
The first expression~(\ref{eq:total_snr1}) is the bispectrum Fisher matrix element $F_{BB}$ from Eq.~(\ref{eq:F_final}).
The second expression~(\ref{eq:total_snr2}) is the total SNR$^2$ for the cross-correlation between the CMB and
the kSZ template $\hT$ from~\S\ref{ssec:ksz_template}.
Here, $\Omega$ is the angular survey area in steradians, and the power spectra $C_l^{T\hT}$, $N_l^{\hT\hT}$ which appear
were given in Eqs.~(\ref{eq:cl_t_ht}),~(\ref{eq:nlht_equals_cl}).
The third expression~(\ref{eq:total_snr3}) is the total SNR$^2$ for the cross-correlation between the large-scale
galaxy field and the kSZ-derived velocity reconstruction $\hv_r$ from~\S\ref{ssec:vrec}.  The reconstruction noise power 
spectrum $N_{v_r}(k_L)$ was given in Eq.~(\ref{eq:Nvr_final}).

\begin{figure}[tbh]
  \includegraphics[width=0.55\textwidth]{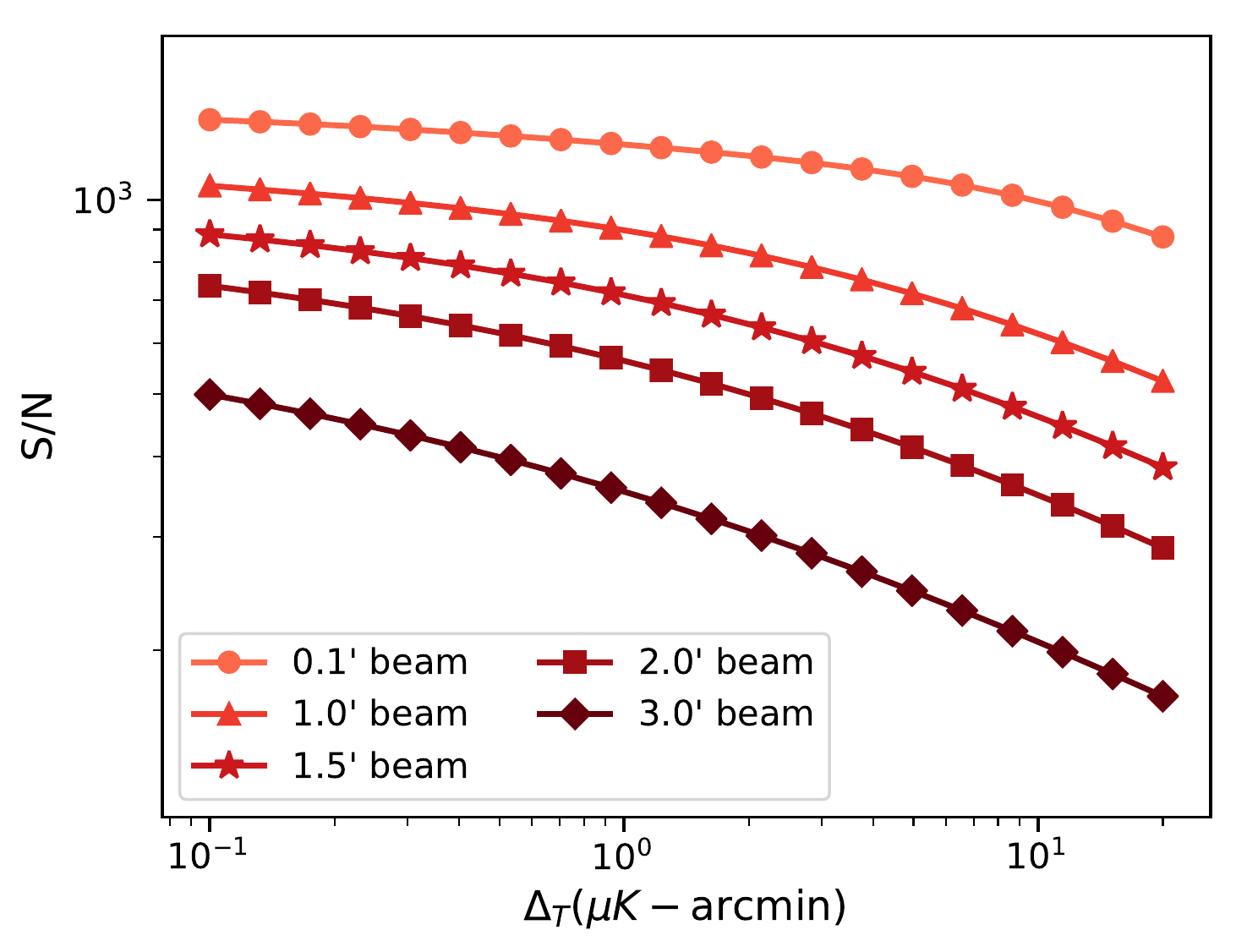}  
  \caption{KSZ tomography signal-to-noise ratio as a function of CMB noise level and beam at 150 GHz, in cross-correlation with DESI. The fiducial CMB experiment configuration is CMB-S4 as described in Table~\ref{tab:cmbexp}, with white noise level and beams in all frequency channels scaled appropriately. The total noise in the CMB includes contributions from the lensed CMB, reionization and late-time kSZ, and the tSZ/CIB/radio residual after standard ILC foreground cleaning.}
\label{fig:sncmb}
\end{figure}

Although kSZ tomography has currently been detected at the few-sigma level, the SNR will rapidly improve in the near future.
We forecast that CMB-S4 will have total $\mbox{SNR} = (653, 333, 366)$, in combination with (DESI, LSST-Y1, LSST-Y10) respectively.
On a shorter timescale, Simons Observatory will have $\mbox{SNR} = (405, 205, 221)$ in combination with the same galaxy surveys.
Note that LSST has lower SNR than DESI, even though its density is higher, due to photo-$z$ errors.
In Figure~\ref{fig:sncmb}, we show more SNR forecasts, for varying CMB parameters in correlation with DESI.

\subsection{What does kSZ tomography actually measure?}
\label{ssec:power_spectra}

In this section we will give a simple answer to the question, ``what does kSZ tomography measure''?
It is convenient to use the bispectrum formalism.
Here, the underlying signal is the squeezed bispectrum:
\be
B(k_L,k_S,l,k_{Lr}) = -\frac{K_* k_{Lr}}{\chi_*^2} \frac{P_{gv}(k_L)}{k_L} P_{ge}(k_S)  \label{eq:B_deg}
\ee
We see that the observables are the large-scale galaxy-velocity power spectrum $P_{gv}(k_L)$
and the small-scale galaxy-electron power spectrum $P_{ge}(k_S)$.

Because the bispectrum in Eq.~(\ref{eq:B_deg}) can be measured as a function of two variables $(k_L,k_S)$,
the power spectra $P_{gv}(k_L)$ and $P_{ge}(k_S)$ can be measured independently, except for one
degeneracy: we have the freedom to multiply $P_{gv}(k_L)$ by a constant $A$, while multiplying $P_{ge}(k_S)$ by $1/A$.
This leaves all kSZ tomography observables unchanged, since the bispectrum~(\ref{eq:B_deg}) is invariant.
This degeneracy is the well-known ``kSZ optical depth 
degeneracy''~\cite{Battaglia:2016xbi,Flender:2016cjy,Louis:2017hoh,Soergel:2017ahb}.\footnote{This is unrelated to another
``optical depth degeneracy'' in the CMB: the cosmological parameters $A_s$ and $\tau$ are constrained
with less precision than the combination $A_s e^{-2\tau}$.}

\begin{figure}[tbh]
  \includegraphics[width=0.45\textwidth]{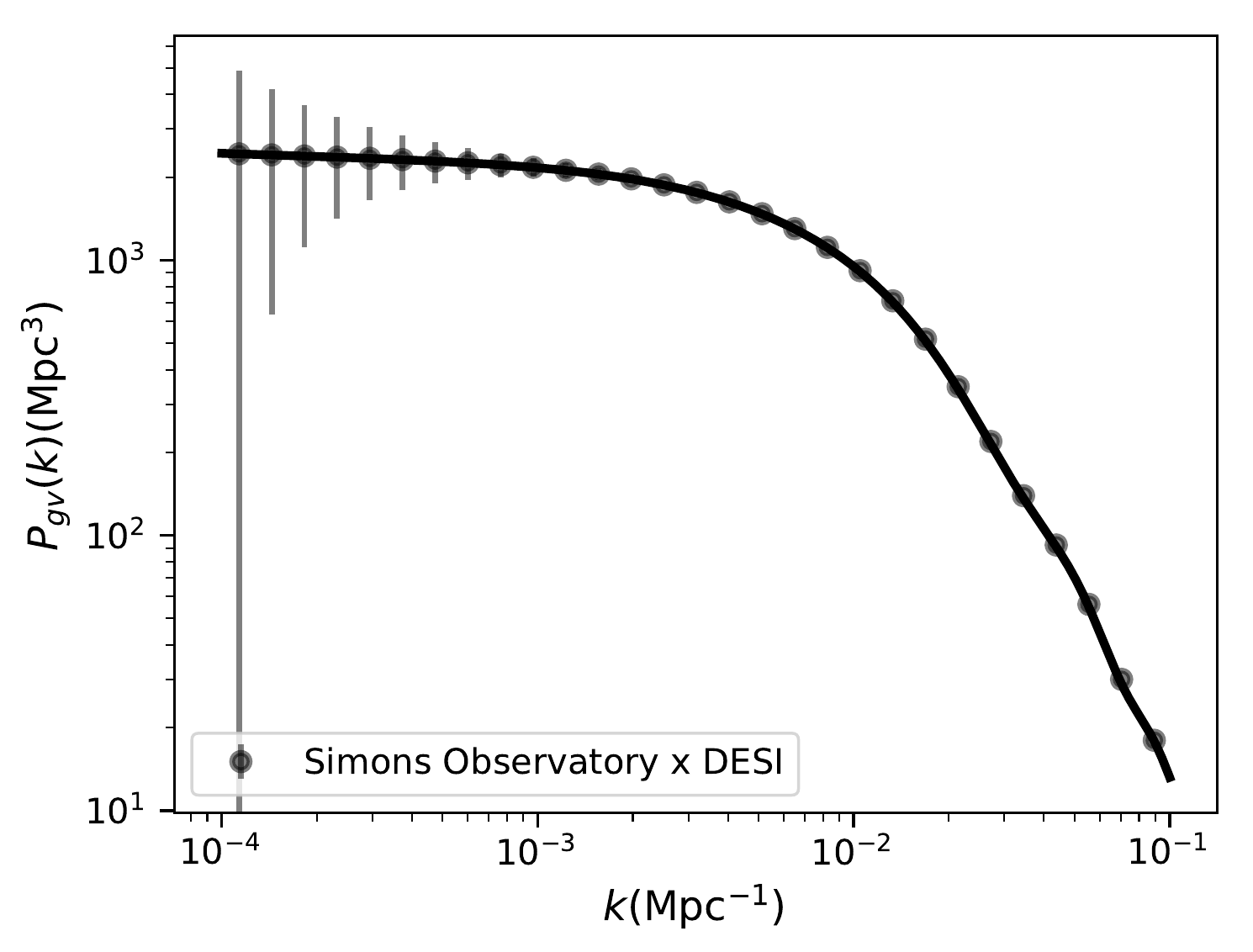}
  \includegraphics[width=0.45\textwidth]{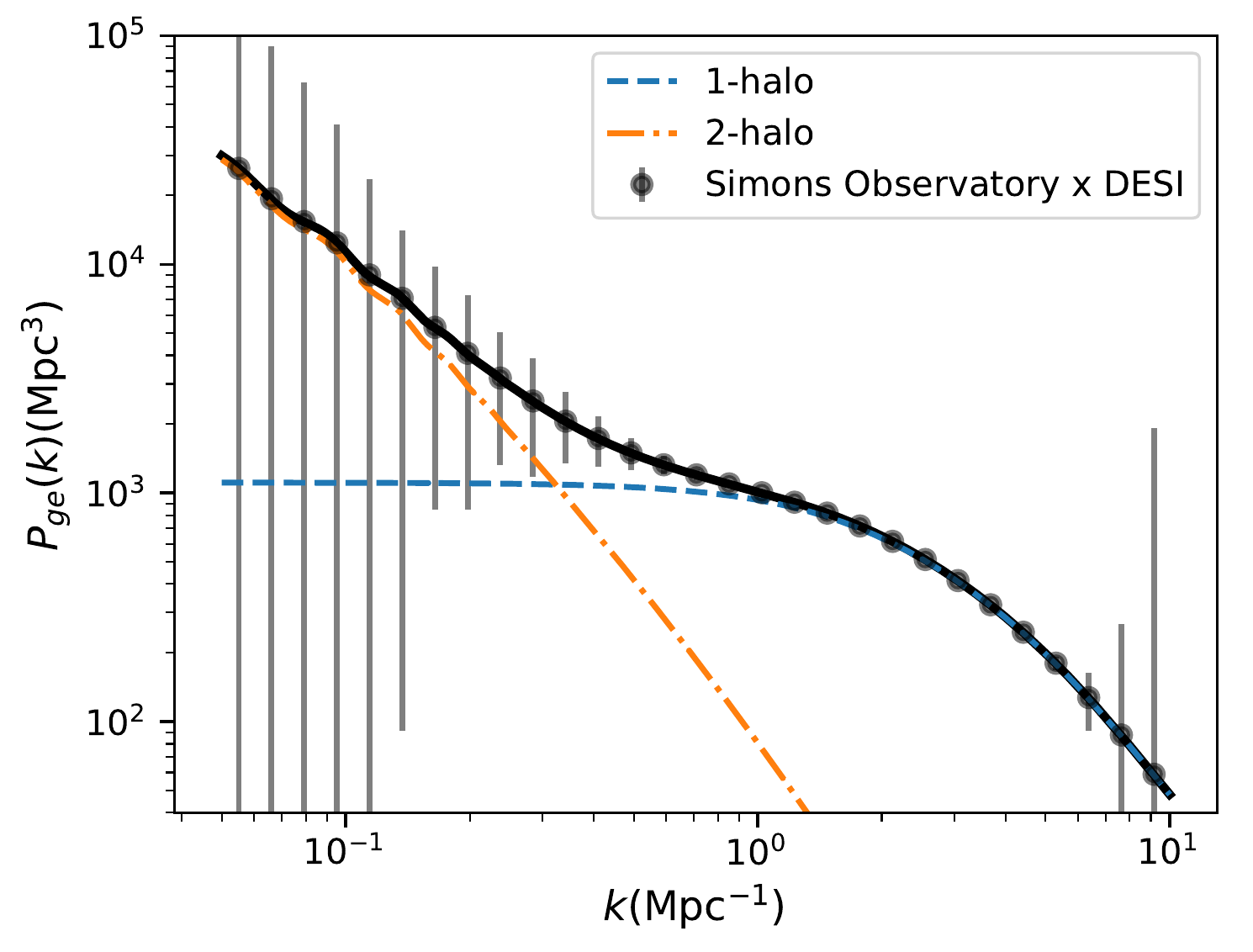}
  \caption{Statistical uncertainties on the galaxy-velocity (left) and galaxy-electron (right) power spectrum from kSZ tomography,
    for Simons Observatory and DESI.  Error bars were computed using Eqs.~(\ref{eq:delta_pgv}),~(\ref{eq:delta_pge}) below.
    The galaxy-electron cross-spectrum includes contributions from
    1-halo (blue dashed) and 2-halo terms (orange dot-dashed). 
    KSZ tomography measures the galaxy-electron cross-spectrum very well in a window of wavenumbers $0.3 \lesssim k \lesssim 5$ Mpc$^{-1}$
    where it is primarily 1-halo dominated.}
  \label{fig:power_spectra}
\end{figure}

Thus, kSZ tomography measures two power spectra $P_{gv}(k_L)$ and $P_{ge}(k_S)$.
The results of a kSZ tomography analysis could be presented as a pair of power spectra
with error bars, as in Figure~\ref{fig:power_spectra}.  When interpreting these plots,
the only subtlety is the optical depth degeneracy, which allows an overall normalization
to be exchanged between $P_{gv}(k_L)$ and $P_{ge}(k_S)$.

\subsection{Constraining cosmology: the large-scale power spectrum $P_{gv}(k_L)$}
\label{ssec:cosmology}

The large-scale galaxy-velocity power spectrum $P_{gv}(k_L)$ can be used to constrain
cosmological parameters.  For simplicity, we first assume that $P_{ge}(k_S)$ is fixed 
to a fiducial value.

For cosmological forecasts, we find it most convenient to use the long-wavelength
velocity reconstruction formalism from~\S\ref{ssec:vrec}.  There we showed that a kSZ-derived
quadratic estimator $\hv_r(\k)$ can reconstruct each mode of the long-wavelength radial velocity
with noise power spectrum:
\be
N_{v_r}(k_L) = \frac{\chi_*^2}{K_*^2} \left[ \int \frac{k_S \, dk_S}{2\pi} 
    \left( \frac{P_{ge}(k_S)^2}{P_{gg}^{\rm tot}(k_S) C_l^{TT,\rm tot}} \right)_{l=k_S\chi_*} \right]^{-1}
\ee
Equivalently, the quadratic estimator can be viewed as a reconstruction of the (non-radial)
velocity $v(\k_L) = \mu^{-1} v_r(\k_L)$ or density $\delta_m(\k_L) = \mu^{-1} (k/faH) v_r(\k_L)$,
with noise power spectra:
\be
N_{vv}^{\rm rec}(k_L,\mu) = \mu^{-2} N_{v_r}(k_L)
  \hspace{1cm}
N_{\delta\delta}^{\rm rec}(k_L,\mu) = \mu^{-2} \left( \frac{k}{faH} \right)^2 N_{v_r}(k_L)  \label{eq:nvv_ndelta}
\ee
Given these noise power spectra, error bars on $P_{gv}(k_L)$ can be computed as:
\ba
\Delta P_{gv} 
  &=& \left( V \int_{\k_L\in b} \frac{d^3\k_L}{(2\pi)^3} 
          \frac{1}{P_{gg}^{\rm tot}(k_L) N_{vv}^{\rm rec}(k_L,\mu)} \right)^{-1/2} \nn \\
  &=& \left( V \int_{k_L^{\rm min}}^{k_L^{\rm max}} \int_{-1}^1 \frac{k_L^2 dk_L \, d\mu}{4\pi^2}
  \frac{1}{P_{gg}^{\rm tot}(k_L) N_{vv}^{\rm rec}(k_L,\mu)} \right)^{-1/2}  \label{eq:delta_pgv}
\ea
where $V$ is the survey volume, and $b = (k_L^{\rm min}, k_L^{\rm max})$ is a $k_L$-bin.
This result was used previously to plot error bars in Figure~\ref{fig:power_spectra}.
We note that Eq.~(\ref{eq:delta_pgv}) can also be derived in the other kSZ tomography formalisms, for example 
by splitting the bispectrum in $k_L$-bins and using the bispectrum Fisher
matrix in Eq.~(\ref{eq:F_final}).

The reconstruction noise power spectrum $N_{\delta\delta}^{\rm rec}(k_L)$ in Eq.~(\ref{eq:nvv_ndelta}) has two novel features.
First, {\em on large scales, kSZ tomography derived from a galaxy survey constrains cosmological modes better than the galaxy survey itself}.
This is because $N_{\delta\delta}^{\rm rec}$ is proportional to $k^2$ on large scales,
whereas the Poisson noise power spectrum of the galaxy survey has a constant value $n_{\rm gal}^{-1}$.
Therefore, on sufficiently large scales, the kSZ-derived noise must be lower.

\begin{figure}[tbh]
  \includegraphics[width=0.45\textwidth]{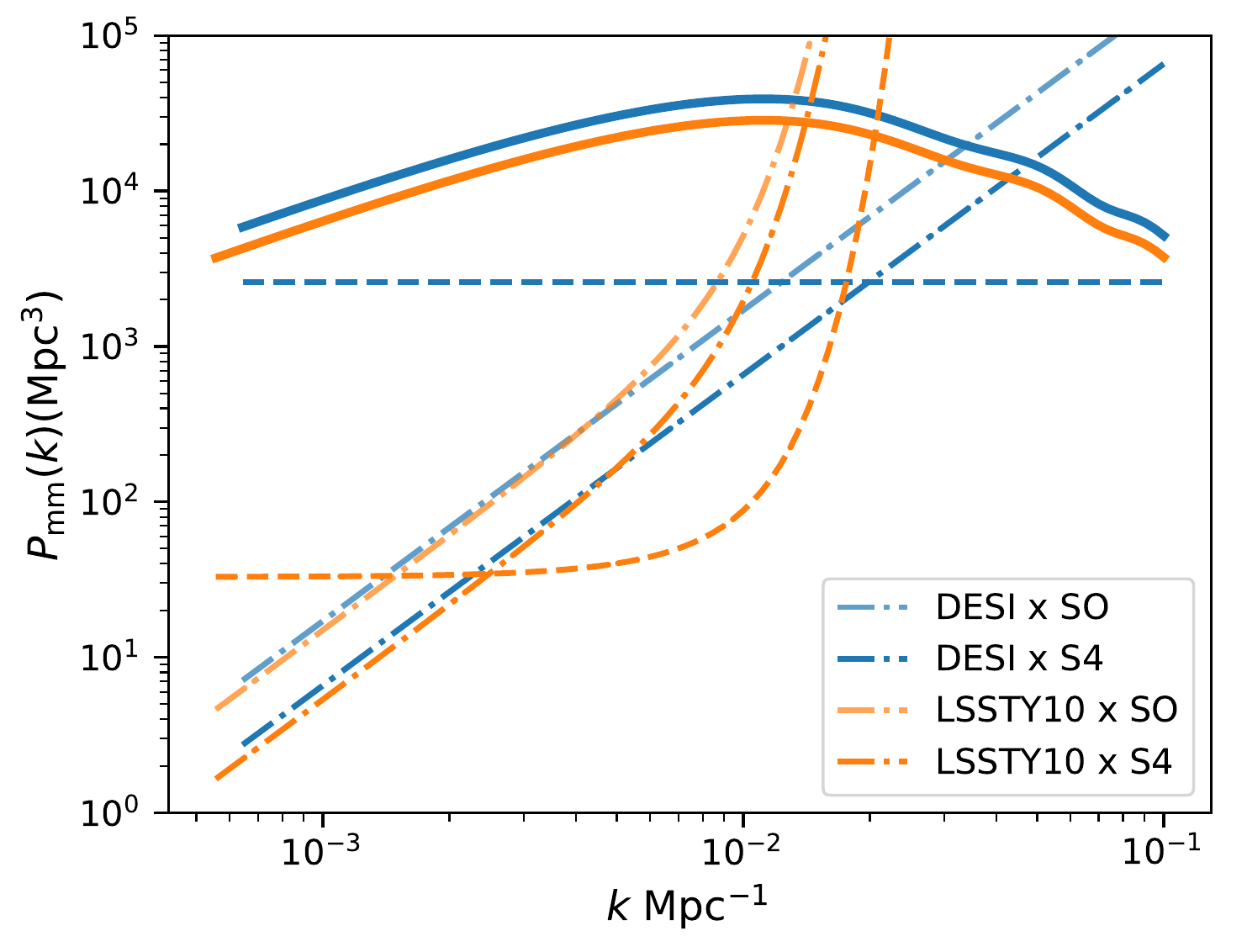}  
  \includegraphics[width=0.45\textwidth]{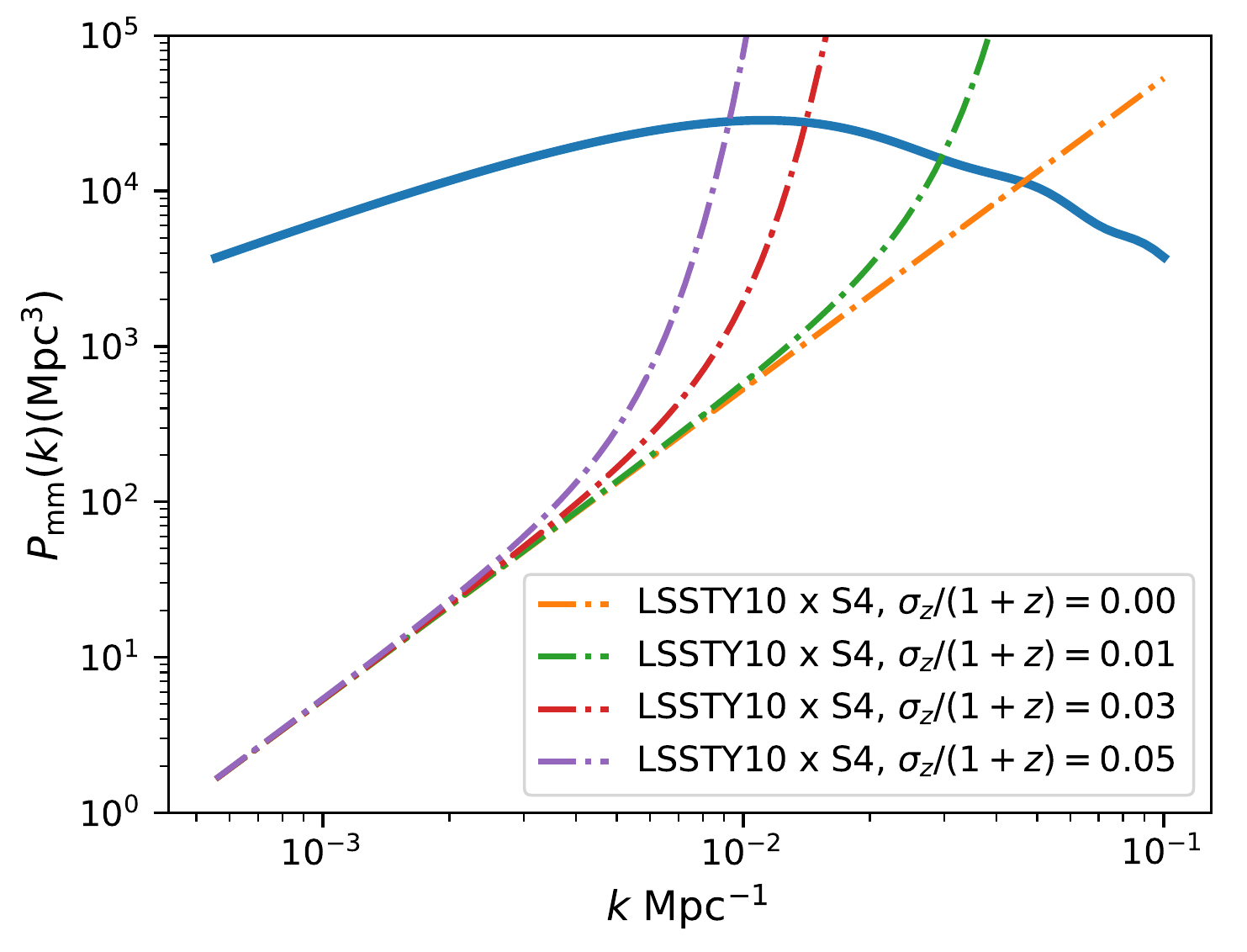}  
  \caption{{\it Left:} Reconstruction noise on large-scale modes using kSZ tomography.
    The solid lines are the total matter power spectrum at the redshifts of DESI and LSST. 
    The dashed lines are the shot noise levels $1/(W^2b^2\bar{n})$ in galaxy clustering (scaled appropriately by the galaxy bias $b$ and photo-$z$ error $W(k)$).
    The dot-dashed lines are reconstruction noise levels $N_{\delta\delta}^{\rm rec}(k_L)$ using kSZ tomography (Eq.~(\ref{eq:nvv_ndelta})),
    for longitudinal modes ($\mu=1$), for various combinations of galaxy and CMB surveys. 
    {\it Right:} The matter power spectrum at the mean redshift of the LSST sample compared with kSZ tomography reconstruction noise, for a few choices of photo-$z$ error $\sigma_z$.}
\label{fig:noise}
\end{figure}

To quantify this, in Figure~\ref{fig:noise}, we compare Poisson noise 
to kSZ-derived noise, for several combinations of galaxy and
CMB surveys.  The crossover occurs around $k_L \sim 0.01$ Mpc$^{-1}$, but
depends on the details of the surveys.

Since future galaxy surveys will generally be sample variance limited on
large scales, one may wonder whether lowering the noise using kSZ tomography
actually gains anything.
In situations where sample variance cancellation is beneficial, the
low-noise measurement from kSZ tomography can be quite helpful.
A prime candidate is constraining $f_{NL}$ using large-scale halo
bias.  This is explored in detail in the companion paper~\cite{Moritz}.

A second novel feature of the reconstruction noise power spectrum in Eq.~(\ref{eq:nvv_ndelta})
is that it is anisotropic, with an overall $\mu^{-2}$ prefactor.
This is easy to understand intuitively.
Since the velocity is curl-free in linear theory, the velocity $v_i$ of a mode points in
a direction parallel to its Fourier wavenumber $k_i$.
In particular, a mode with $\mu=0$ has velocity perpendicular to the line of sight and does
not produce a kSZ signal.
Therefore, its reconstruction noise must be infinite, since the amplitude of the mode cannot
be constrained from kSZ.

The $\mu^{-2}$ dependence has the qualitative consequence that
{\em the kSZ-derived reconstruction of the long-wavelength modes cannot be cross-correlated with a 2D field}, for example the CMB lensing potential $\phi$.
Indeed, in the Limber approximation, only 3D modes with $\mu=0$ will contribute to $\phi$, and these modes have infinite noise in the kSZ reconstruction.
(Because the Limber approximation is not perfect, the cross correlation between $\phi$ and the kSZ reconstruction will not be exactly zero,
but we expect it to be very small.)
For cross-correlations with a 3D field, such as the galaxy-velocity cross correlation $\langle \delta_g v_{\rm rec} \rangle$,
the $\mu^{-2}$ prefactor does not have a qualitative effect, although it does result in an order-one signal-to-noise 
penalty.\footnote{There are other examples of cosmological fields with the property that cross-correlations with 2d fields are always near-zero,
but for different reasons.
The 21-cm brightness temperature $T_b(\n,z)$ has this property, because 21-cm maps must be high-pass filtered in the radial direction in order
to remove Galactic foregrounds.
Similarly, when analyzing Lyman-alpha forest spectra from bright quasars, each spectrum is normalized by dividing by the quasar continuum emission,
which is obtained from the data by some form of low-pass filtering.
This normalization procedure is a radial high-pass filter which removes correlations with 2d fields.}

So far in this section, we have assumed that the small-scale power spectrum $P_{ge}(k_S)$ is known in advance.
Now let us consider the effect of uncertainty in $P_{ge}(k_S)$ when reconstructing long-wavelength modes.
Suppose the quadratic estimator $\hv(\k_L)$ is constructed using a fiducial power spectrum $P_{ge}^{\rm fid}(k_S)$,
but the true power spectrum is $P_{ge}^{\rm true}(k_S) \ne P_{ge}^{\rm fid}(k_S)$.
Then $\hv_r(\k_L)$ will be a biased estimator of $v_r(\k_L)$.
After a short calculation, the bias can be written in the following form:
\be
\langle \hv_r(\k_L) \rangle = b_v v_r(\k_L)
\ee
where the velocity reconstruction bias $b_v$ is given by:
\be
b_v = \frac{\int dk_S \, F(k_S) P_{ge}^{\rm true}(k_S)}{\int dk_S \, F(k_S) P_{ge}^{\rm fid}(k_S)}
  \hspace{1cm} \mbox{where }
F(k_S) = k_S \frac{P_{ge}^{\rm fid}(k_S)}{P_{gg}^{\rm tot}(k_S)} \left( \frac{1}{C_l^{TT,\rm tot}} \right)_{l=k_S\chi_*}  \label{eq:bv_def}
\ee
The details of Eq.~(\ref{eq:bv_def}) are unimportant, except for the crucial property
that the bias $b_v$ is independent of $k_L$.
That is, the kSZ-derived velocity reconstruction actually reconstructs the
velocity (or density) field up to an overall normalization $b_v$ which is not known 
in advance, and therefore must be marginalized.
This is similar to the case of a galaxy field, where the galaxy bias $b_g$ must be marginalized.
In the kSZ context, the bias parameter $b_v$ arises because of the optical depth degeneracy.

\subsection{Constraining astrophysics: the small-scale power spectrum $P_{ge}(k_S)$}
\label{ssec:astrophysics}

KSZ tomography can be used to measure the small-scale galaxy-electron power spectrum $P_{ge}(k_S)$ in $k_S$-bins.
Here, we will neglect the optical depth degeneracy, since $P_{gv}(k_L)$ can be predicted in
advance to a few percent, by combining well-measured cosmological parameters with an external measurement of
galaxy bias (say from cross-correlating with CMB lensing).

First we ask, what is the statistical error $\Delta P_{ge}$ on the power spectrum $P_{ge}(k_S)$ in a $k_S$-bin?
This can be derived in any of our kSZ formalisms, but a quick way to read off the answer using our previous results is as follows.
We start with Eq.~(\ref{eq:total_snr1}) for the total SNR$^2$, which is written in a form where it can be split into $k_S$-bins.
If we restrict to a single bin $(k_S^{\rm min}, k_S^{\rm max})$, and assume that $P_{ge}(k_S)$ has the constant value $P_{ge}^{\rm fid}$
over the bin, then the single-bin SNR is:
\be
\mbox{SNR}^2_{\rm bin}
  = V \frac{K_*^2}{12 \pi^3 \chi_*^2} (P_{ge}^{\rm fid})^2
    \left( \int dk_L \, k_L^2 \frac{P_{gv}(k_L)^2}{P_{gg}^{\rm tot}(k_L)} \right)
    \left( \int_{k_S^{\rm min}}^{k_S^{\rm max}} dk_S \, k_S \frac{1}{P_{gg}^{\rm tot}(k_S)} \frac{1}{(C_l^{\rm tot})_{l=k_S\chi_*}} \right) 
\ee
The single-bin statistical error $\Delta P_{ge}$ and SNR are related by
$\Delta P_{ge} = P_{ge}^{\rm fid} / \mbox{SNR}_{\rm bin}$.  Therefore:
\be
\Delta P_{ge}
  = \left[
       V \frac{K_*^2}{12 \pi^3 \chi_*^2}
       \left( \int dk_L \, k_L^2 \frac{P_{gv}(k_L)^2}{P_{gg}^{\rm tot}(k_L)} \right)
       \left( \int_{k_S^{\rm min}}^{k_S^{\rm max}} dk_S \, k_S \frac{1}{P_{gg}^{\rm tot}(k_S)} \frac{1}{(C_l^{\rm tot})_{l=k_S\chi_*}} \right) 
    \right]^{-1/2}  \label{eq:delta_pge}
\ee
This expression was used previously to show error bars on $P_{ge}(k_S)$ in Figure~\ref{fig:power_spectra}.
One interesting property of this measurement is that the error bars
blow up for both low and small $k_S$.  The power spectrum is constrained in a window of scales (roughly
$0.3 \lesssim k \lesssim 5$ Mpc$^{-1}$) which are mainly 1-halo dominated.

In this paper, our focus is on cosmology, and we will not explore the astrophysical implications
of a precision measurement of $P_{ge}(k_S)$, aside from a few brief comments as follows.

A kSZ-derived measurement of $P_{ge}(k)$ probes the distribution of electrons in halos.
This is similar to galaxy-galaxy lensing, which measures the galaxy-matter power spectrum $P_{gm}(k)$,
and probes the distribution of matter in halos.
In galaxy-galaxy lensing, $P_{gm}(k)$ is usually modelled using the halo model, and we can do the same
for $P_{ge}(k)$, writing it as the sum of 1-halo and 2-halo terms:
\be
P_{ge}(k) = b_g b_e P_{mm}(k) + \frac{1}{n_g \rho_m} \int dm \, m n(M) u_g(k|m) u_e(k|m)  \label{eq:pge_halo_model}
\ee
where $u_g(k|m)$ and $u_e(k|m)$ denote the galaxy and electron profiles respectively.
(For more details on the halo model, see Appendix~\ref{app:halo_model}.)
In particular, ``miscentering'', or the nonzero offset between galaxies and halo centers~\cite{Calafut:2017mzp},
is naturally incorporated by including a galaxy profile $u_g(k|m) \ne 1$ in the model, as is already standard for
galaxy-galaxy lensing.

There is a degeneracy in $P_{ge}(k_S)$ (Eq.~(\ref{eq:pge_halo_model})) between the electron profile $u_e(k|m)$ and the
galaxy profile $u_g(k|m)$.  One way of breaking this degeneracy is to measure the galaxy-matter
power spectrum $P_{gm}(k)$ using galaxy-galaxy lensing with the same galaxy sample.  The dependence on the galaxy profile
largely cancels in the ratio $P_{ge}(k) / P_{gm}(k)$ (but not perfectly, since the galaxy profile 
can depend on halo mass).
Therefore, galaxy-galaxy lensing nicely complements kSZ tomography.

KSZ tomography is also complementary to thermal SZ and X-ray observations, which also probe
the distribution of electrons in halos.
Relative to tSZ and X-ray, kSZ tomography is more sensitive to electrons in the outskirts of halos.
This is because the kSZ profile is proportional to one power of the electron number density $n_e$,
whereas X-ray profiles are proportional to $n_e^2$, and tSZ profiles are proportional to $n_e T$,
where $T$ is the gas temperature.

Normally, a measurement of a power spectrum such as $P_{ge}(k)$ can be converted (by taking a Fourier
transform) to a measurement of the associated correlation function $\zeta_{ge}(r)$.
This is particularly intuitive for kSZ tomography since $\zeta_{ge}(r)$ is the stacked electron profile
around galaxies, which is easy to interpret.
However, kSZ tomography has the unusual property that the error bars on $P_{ge}(k)$ blow up at both
small and large $k$ (Figure~\ref{fig:power_spectra}, right panel).
In real space, this means that if we estimate $\zeta_{ge}(r)$ in $r$-bins, the marginalized error bars 
on each bin will be artificially large and highly correlated.
For this reason, it seems preferable to work in Fourier space, and use the power spectrum 
$P_{ge}(k)$ when visualizing results or performing model fits.

\subsection{More on the optical depth degeneracy}
\label{ssec:more_optical_depth}

As previously described (\S\ref{ssec:power_spectra}), kSZ tomography measures the power
spectra $P_{ge}(k_S)$ and $P_{gv}(k_L)$, up to an overall amplitude which can be exchanged
(the ``optical depth degeneracy'').

If the goal of kSZ tomography is to constrain the galaxy-electron power spectrum $P_{ge}(k_S)$,
then the optical depth degeneracy adds extra uncertainty to the overall amplitude, due to uncertainty in $P_{gv}(k_L)$.
At back-of-the-envelope level, $P_{gv}(k_L)$ can be predicted in advance to a few percent,
since cosmological parameters and galaxy bias can be measured to this accuracy.
Therefore, the optical depth degeneracy should not be an issue if the kSZ tomography measurement 
has total SNR $\lesssim 30$, but should be taken into account above this threshold.

If the goal of kSZ tomography is to constrain cosmological modes on large scales, then
the optical depth degeneracy shows up as a bias parameter $\langle \hv_r \rangle = b_v v_r$
in the velocity reconstruction, which must be marginalized.  For some purposes, for example
the $f_{NL}$ forecasts which we present in the companion paper~\cite{Moritz}, marginalizing $b_v$
turns out to have a minimal effect.
For other purposes, for example if we want to use the overall amplitude of $P_{gv}(k_L)$ to
constrain the cosmological growth rate, then the optical
depth degeneracy is a serious problem, unless it can be broken somehow.

The optical depth degeneracy could be broken (for cosmological purposes) if the galaxy-electron
power spectrum $P_{ge}(k_S)$ can be predicted in advance to better than a few percent.
As previously noted~\cite{Battaglia:2016xbi,Flender:2016cjy,Louis:2017hoh,Soergel:2017ahb},
a {\em necessary} condition for doing this is 
that the mean optical depth $\bar\tau$ of galaxy clusters in the sample must be determined,
since $\bar\tau$ sets the overall amplitude of the 2-halo term in $P_{ge}(k_S)$.
(This is the origin of the term ``optical depth degeneracy''.)

However, we would like to add the observation that predicting $\bar\tau$ is not sufficient
for breaking the optical depth degeneracy, since kSZ tomography is mainly sensitive to $P_{ge}(k_S)$
in the 1-halo dominated regime (Figure~\ref{fig:power_spectra}).
On these scales, $P_{ge}(k_S)$ depends not only on $\bar\tau$, but also on the details of the spatial
profile of the free electrons, including the outskirts of the cluster where the profile is difficult
to measure in X-ray or tSZ.
(``Optical depth degeneracy'' is not really the right term, since $\bar\tau$ is one of several factors 
which determine the small-scale power spectrum $P_{ge}(k_S)$, and we need to know the amplitude
of $P_{ge}(k_S)$ on 1-halo scales to break the degeneracy.)
For this reason, we suspect that breaking the kSZ optical depth degeneracy astrophysically will be very difficult.

Recently,~\cite{Sugiyama:2016rue} proposed breaking the kSZ optical depth degeneracy
in a different way, by using an ``octopolar'' version of the pair sum estimator, rather
than an astrophysical prior on $P_{ge}(k_S)$.  We will study this proposal in the next section.

\subsection{Including kSZ tomography in larger cosmological analyses}
\label{ssec:larger_analyses}

We have shown that the kSZ tomography statistics
in~\cite{Ho:2009iw,Hand:2012ui,Li:2014mja,Alonso:2016jpy,Deutsch:2017ybc} are 
``bispectrum estimation in disguise''
and mathematically equivalent.  In particular, bispectrum estimation can be implemented
by cross-correlating the kSZ-derived velocity reconstruction $\hv_r$ with the galaxy survey $g$
on large scales.

If the kSZ-derived velocity reconstruction $\hv_r$ is included in a larger analysis (either a Fisher matrix
forecast or actual data analysis) with the appropriate noise power spectrum, then additional
higher-point statistics will naturally arise.  For example, consider a forecast with
two fields: the velocity reconstruction $\hv_r$ and the galaxy field $\delta_g$.  The Fisher matrix
would combine contributions from the galaxy auto power spectrum $P_{gg}(k)$, the cross power
spectrum $P_{g\hv}(k)$ (which is really a three-point function $\langle g g T \rangle$), and
the auto power spectrum $P_{\hv\hv}(k)$ (which is really a four-point function $\langle g g T T \rangle$).
This is very similar to CMB lensing, where including the lens reconstruction $\hphi$ in a
larger analysis naturally generates all ``interesting'' three-point and four-point statistics.

Previously, we stated that velocity reconstruction is equivalent to the other kSZ tomography statistics.
This statement implicitly assumes that we cross-correlate $\hv_r$ with the galaxy field $g$ on large scales,
but do not use it for anything else.
However, in a scenario where $\hv_r$ is included in a larger analysis involving more fields,
it automatically captures multiple higher-point statistics and their covariances.
For this reason, we prefer the velocity reconstruction formalism to the other approaches to
kSZ tomography, at least for cosmological purposes.  (For purposes of constraining astrophysics
through measurements of $P_{ge}(k_S)$, the kSZ template method from~\cite{Ho:2009iw} seems simplest.)
Another technical advantage of $\hv_r$ is that it makes the optical depth degeneracy easy
to incorporate, by adding a nuisance parameter $b_v$ and marginalizing it at the end.

As another example, suppose we have $N$ tracer fields, for example corresponding to halos
in different mass bins.  Then we can construct $N$ kSZ-derived velocity reconstructions $\hv_r^{(i)}$,
which can be cross-correlated with tracer field $j$, or with each other.
To find the optimal weighting of all these power spectra, we need to know the
$N$-by-$N$ matrix $N_{v_r}^{(ij)}$ of reconstruction noise power spectra.
Starting from the definition of $\hv_r$ in Eq.~(\ref{eq:hv_final}), a short calculation gives:
\be
N_{v_r}^{(ij)} = \frac{2\pi \chi_*^2}{K_*^2} \frac{A_{ij}}{A_{ii} A_{jj}}
\ee
where we have defined
\be
A_{ij} = \int dk_S \, k_S \frac{P_{ge}^{(i)}(k_S) P_{ge}^{(j)}(k_S) P_{gg}^{(ij)}(k_S)}{P_{gg}^{(i)}(k_S) P_{gg}^{(j)}(k_S)} \left( \frac{1}{C_l^{TT,\rm tot}} \right)_{l=k_S\chi_*}
\ee
In principle, each velocity reconstruction has its own reconstruction bias $b_v^{(i)}$
which must be independently marginalized.
The bias parameters for $i \ne j$ are different because the $k_S$-weighting $F(k_S)$
in Eq.~(\ref{eq:bv_def}) is different for each galaxy field.

As a final illustration of the power of the velocity reconstruction approach, it is straightforward to see how the optical depth
degeneracy gets broken when redshift-space distortions in the galaxy field are included, as shown by~\cite{Sugiyama:2016rue}.
Consider a Fisher matrix forecast with two 3-d fields, a galaxy field $\delta_g$ and its kSZ-derived velocity 
reconstruction $\hv_r$.  On large scales,
\be
\delta_g(\k) = (b_g + f \mu^2) \delta_m(\k) + \mbox{(noise)}
  \hspace{1cm}
\hv_r(\k) = \mu \frac{b_v faH}{k} \delta_m(\k) + \mbox{(noise)}
\ee
where $\mu = k_r/k$ as usual.  Then the cross power spectrum $P_{g\hv_r}(k,\mu)$ is:
\be
P_{g\hv_r}(k,\mu) = \left( \mu b_g b_v + \mu^3 f b_v \right) \frac{faH}{k} P_{mm}(k)
\ee
The two terms have different $\mu$ dependence and their coefficients can be measured
separately.  In particular, the coefficient of the $\mu^3$ term is a measurement of
the parameter combination $b_v f^2 H \sigma_8^2$.  Because $f$, $H$,
and $\sigma_8$ are well-determined cosmological parameters, this pins down $b_v$
and breaks the optical distance degeneracy.

Summarizing, the velocity reconstruction approach is powerful because it fully incorporates kSZ tomography into a larger analysis.
It automatically ``discovers'' subtle effects such as the degeneracy breaking from higher-$\mu$ terms, without needing 
to construct the appropriate statistic (octopole pair sum) explicitly, or even needing to know in advance that it exists.

\section{Photometric redshift errors and redshift space distortions}
\label{sec:photoz_rsd}

As previously explained, kSZ tomography requires a 3D field.
If a 2D field were used, the signal-to-noise would be near-zero.
A galaxy survey with photometric redshifts is an interesting intermediate case between 2D and 3D,
and one may wonder whether photometric surveys are useful for kSZ tomography.

At back-of-the-envelope level, the answer can be worked out as follows.
The effect of photo-$z$ errors is to suppress power in modes of the galaxy survey whose radial
wavenumber $k_r$ is larger than $k_z = H/\sigma_z$, where $\sigma_z$ is the RMS photo-$z$ error.
On the other hand, most of the SNR for kSZ tomography comes from scales $k \sim k_v$, where $k_v \sim 0.02$ $h$ Mpc$^{-1}$
is the velocity correlation length.
Therefore, photo-$z$ errors impose a large SNR penalty in the limit $k_z \ll k_v$, and a small penalty
in the limit $k_z \gg k_v$.
Taking $H \sim 3 \times 10^{-4}$ $h$ Mpc$^{-1}$ and $\sigma_z \sim 0.02$, a typical value of $k_z$ might be
$k_z \sim 0.015$ $h$ Mpc$^{-1}$.
That is, the characteristic scales $k_z$ and $k_v$ are usually comparable, which means that photo-$z$ errors result in an order-one SNR penalty.

To take a concrete example, previously in~\S\ref{ssec:snr}, we found that the total SNR
for kSZ tomography with CMB-S4 and LSST-Y10 was 366.
This forecast includes the effect of photometric redshift errors, using machinery that will be developed in this section.
If we artificially assume that LSST has no photo-$z$ errors, then we find SNR=827.
Thus, photo-$z$ errors reduce total SNR by a factor $\approx$2.3 in this example.

In this section, we will also consider redshift space distortions (RSD),
i.e.~apparent radial displacement of galaxies due to their peculiar velocities.
We will analyze the effect of photo-$z$ errors and RSD on kSZ tomography using a common framework.

\subsection{Modelling photo-z errors and RSD}

In the next few sections, we use a bar $(\bar\cdot)$ to denote ``distorted by photo-z errors and RSD''
and a tilde $(\tilde\cdot)$ to denote ``undistorted''.  We derive expressions for distorted power spectra
such as $\bP_{ge}$, in terms of their undistorted counterparts.

Considering photo-z errors first, we will assume the simplest possible model:
each galaxy has an independent Gaussian redshift error with variance $\sigma_z^2$.
In our halo model (Appendix~\ref{app:halo_model}), this is equivalent to convolving
the real-space galaxy profile $u_g(x)$ by a Gaussian radial kernel with comoving 
width $\Delta x = \sigma_z/H_*$.  In Fourier space, this corresponds to multiplication
by a Gaussian in $k_r$:
\be
u_g(\k) \rightarrow W(k_r) u_g(k) 
  \hspace{1.5cm} \mbox{where } W(k_r) = \exp\left(-\frac{\sigma_z^2}{2 H_*^2} k_r^2 \right) \label{eq:photoz_ug}
\ee
The profile $u_g(\k)$ is now a function of both the length $k=|\k|$
and the radial component $k_r$ of the wavenumber $\k$.

Note that convolving the profile $u_g(\k) \rightarrow W(k_r) u_g(k)$ is not the same thing as convolving the
galaxy field $\delta_g(\k) \rightarrow W(k_r) \delta_g(\k)$ (and the latter would be incorrect).
If we write $\delta_g(\x)$ as a sum of delta functions $n_g^{-1} \sum_i \delta^3(\x-\x_i)$,
then the underlying profile $u_g$ which determines the locations $\x_i$ is convolved with $W(k_r)$,
but the delta functions themselves are not convolved with $W(k_r)$.

Now we analyze the effect of photo-z errors on the total galaxy power spectrum $P_{gg}^{\rm tot}$.
First recall that in the undistorted case, $\tP_{gg}^{\rm tot}$ is the sum of a 2-halo term,
a 1-halo term, and a shot noise term:
\be
\tP_{gg}^{\rm tot}(k) = \tP_{gg}^{2h}(k) + \tP_{gg}^{1h}(k) + \frac{1}{n_g}
\ee
From the explicit formulas for $P_{gg}^{2h}$ and $P_{gg}^{1h}$ in Eqs.~(\ref{eq:pgg_1h}),~(\ref{eq:pgg_2h}),
we see that if we modify the profile $u_g$ as in Eq.~(\ref{eq:photoz_ug}), then the two-halo and
one-halo terms get a factor $W^2$, whereas the shot noise term is unmodified:
\be
\bP_{gg}^{\rm tot}(k,k_r) = W(k_r)^2 \Big( \tP_{gg}^{2h}(k) + \tP_{gg}^{1h}(k) \Big) + \frac{1}{n_g}
  \hspace{1cm} \mbox{(photo-z only)}  \label{eq:photoz_pgg}
\ee
where ``photo-z only'' means that we have included photo-z errors but not redshift-space distortions.
Note that $\bP_{gg}^{\rm tot}$ is anisotropic: it is a function of both $k$ and $k_r$.

We can apply a similar analysis to cross spectra of the form $P_{gX}$,
where $X$ could be the electron field $e$, the matter field $m$, or the velocity field $v$.
We model these cross spectra as 1-halo and 2-halo terms (e.g.~Eqs.~(\ref{eq:pge_1h}),~(\ref{eq:pge_2h}) 
for the $P_{ge}$ case).
Looking at these expressions, we see that both the 1-halo and 2-halo terms get a factor $W(k_r)$.
That is, the effect of photo-z errors on cross spectra is simply:
\be
\bP_{gX}(k,k_r) = W(k_r) \tP_{gX}(k)  \hspace{1cm} \mbox{(photo-z only)}  \label{eq:photoz_pge}
\ee
where $X \in \{ e,m,v \}$.

In real galaxy surveys, modelling photometric errors is more complex than the simple Gaussian
model considered here.  We have assumed a Gaussian error distribution, whereas a real survey
would have a small population of drastic outliers.  We have also assumed that galaxies have
independent photo-z errors, i.e.~a halo with $N$ galaxies would have redshift error $\sigma_z/\sqrt{N}$.
This may be an incorrect assumption if the errors have systematic dependence, e.g.~on metallicity.
Exploring these issues further is outside the scope of this paper.

Now we consider redshift-space distortions.
On large scales, the effect of RSD is given by the Kaiser formula, which states that the galaxy profile $u_g$ is modified as:
\be
u_g(\k) \rightarrow \left( 1 + \beta \frac{k_{Lr}^2}{k_L^2} \right) u_g(k) 
\ee
where $\beta = f/b_g$ and $f = \partial(\log D)/\partial(\log a)$.
As in the photo-z case, convolving the profile $u_g$ is not equivalent to convolving the galaxy field $\delta_g$.
Large-scale power spectra are modified as:
\begin{align}
\bP_{gg}^{\rm tot}(k_L,k_{Lr}) &= \left( 1 + \beta \frac{k_{Lr}^2}{k_L^2} \right)^2 \Big( \tP_{gg}^{2h}(k_L) + \tP_{gg}^{1h}(k_L) \Big) + \frac{1}{n_g}
  & \mbox{(RSD only)} \nn \\
\bP_{gX}(k_L,k_{Lr}) &= \left( 1 + \beta \frac{k_{Lr}^2}{k_L^2} \right) \tP_{gX}(k_L)  & \mbox{(RSD only)}  \label{eq:rsd_large_scale}
\end{align}
where $X \in \{ e,m,v \}$ and a large scale has been assumed.

On small scales, redshift space distortions (``Fingers of God'') are nonlinear and difficult to model.
However, we will be interested in ``near-transverse'' small-scale modes where $k_{Sr}$ is small, even though $k_S$ is large.
This is because kSZ tomography always involves a delta function of the form $\delta^3(\k_L + \k_S + \l/\chi_*)$ which
implies $k_{Sr} = -k_{Lr}$.
For near-transverse small-scale modes, we have checked with simulations that the effect of redshift space distortions is small,
and we will neglect it in this paper.
Details of the simulations will be presented separately in~\cite{Utkarsh}.
Thus, on small scales we will assume:
\be
\bP_{gg}^{\rm tot}(k_S,k_{Sr}) = \tP_{gg}^{\rm tot}(k_S)
  \hspace{1cm}
\bP_{gX}(k_S,k_{Sr}) = \tP_{gg}^{\rm tot}(k_S)
  \hspace{1cm}
\mbox{(RSD only)}  \label{eq:rsd_small_scale}
\ee
where $X \in \{ e,m,v \}$ and a near-transverse small-scale mode has been assumed.

Summarizing this section, our model for photo-z errors and RSD's on large and small scales is defined by 
Eqs.~(\ref{eq:photoz_pgg}),~(\ref{eq:photoz_pge}),~(\ref{eq:rsd_large_scale}),~(\ref{eq:rsd_small_scale}) above.
Combining these results, our ``bottom-line'' model including both effects is:
\ba
\bP_{gg}^{\rm tot}(k_L,k_{Lr}) 
  &=& W(k_{Lr})^2 \left( 1 + \beta \frac{k_{Lr}^2}{k_L^2} \right)^2 \Big( \tP_{gg}^{2h}(k_L) + \tP_{gg}^{1h}(k_L) \Big) + \frac{1}{n_g}  \label{eq:bPgg_kl} \\
\bP_{gg}^{\rm tot}(k_S,k_{Sr}) 
  &=& W(k_{Sr})^2 \Big( \tP_{gg}^{2h}(k_S) + \tP_{gg}^{1h}(k_S) \Big) + \frac{1}{n_g}  \label{eq:bPgg_ks}  \\
\bP_{gX}(k_L,k_{Lr})
  &=& W(k_{Lr}) \left( 1 + \beta \frac{k_{Lr}^2}{k_L^2} \right) \tP_{gX}(k_L) \label{eq:bPgx_kl} \\
\bP_{gX}(k_S,k_{Sr})
  &=& W(k_{Sr}) \, \tP_{gX}(k_S)  \label{eq:bPgx_ks}
\ea
where $X \in \{ e, m, v \}$, and $W(k_r)$ is the Fourier transformed
photo-$z$ error distribution defined in Eq.~(\ref{eq:photoz_ug}).

\subsection{The kSZ bispectrum with photo-$z$'s and RSD}

Next we consider the combined effect of photo-z errors and RSD on previous results in the paper.
In some cases, the derivations involve repeating analysis from previous sections, which we do in streamlined form.

First we consider the kSZ bispectrum.
In the squeezed limit $k_L \ll k_S$, the distorted bispectrum $\bB$ can be written in any of the following forms:
\ba
\bB(k_L,k_S,l,k_{Lr}) 
  &=& -\frac{K_* k_{Lr}}{\chi_*^2} \frac{\bP_{gv}(k_L,k_{Lr})}{k_L} \bP_{ge}(k_S,k_{Lr}) \nn \\
  &=& -\frac{K_* k_{Lr}}{\chi_*^2} W(k_{Lr})^2 \left( 1 + \beta \frac{k_{Lr}^2}{k_L^2} \right) \frac{\tP_{gv}(k_L)}{k_L} \tP_{ge}(k_S) \nn \\
  &=& W(k_{Lr})^2 \left( 1 + \beta \frac{k_{Lr}^2}{k_L^2} \right) \tB(k_L,k_S,l,k_{Lr})  \label{eq:Bbar_squeezed}
\ea
generalizing Eq.~(\ref{eq:B_squeezed}) in the undistorted case.

Note that in the first line of Eq.~(\ref{eq:Bbar_squeezed}), we have written $\bP_{ge}(k_S,k_{Lr})$ on the RHS instead of $\bP_{ge}(k_S,k_{Sr})$.
We have used the relation $k_{Sr} = -k_{Lr}$ to eliminate $k_{Sr}$ in favor of $k_{Lr}$,
for notational consistency with the LHS, where only $k_{Lr}$ appears.
We will do the same elsewhere in this section without commenting on it explicitly.

The minimum-variance bispectrum estimator $\hE$ is given by:
\ba
\hE &=& \frac{K_*}{\chi_*^2 F_{BB}} \int \frac{d^3\k_L}{(2\pi)^3} \, \frac{d^3\k_S}{(2\pi)^3} \, \frac{d^2\l}{(2\pi)^2} \,
    \frac{ik_{Lr}}{k_L} \frac{\bP_{gv}(k_L,k_{Lr}) \bP_{ge}(k_S,k_{Lr})}{\bP_{gg}^{\rm tot}(k_L,k_{Lr}) \bP_{gg}^{\rm tot}(k_S,k_{Lr}) C_l^{TT,\rm tot}} \nn \\
&& \hspace{1.5cm} \times
     \Big( \delta_g(\k_L) \delta_g(\k_S) T(\l) \Big) \,
    (2\pi)^3 \delta^3\left( \k_L + \k_S + \frac{\l}{\chi_*} \right)  \label{eq:Ebar_squeezed}
\ea
generalizing Eq.~(\ref{eq:E_squeezed}) in the undistorted case.

As in the undistorted case, the integrals in Eq.~(\ref{eq:Ebar_squeezed}) 
should be understood as running over wavenumbers $k_L \ll k_S$
which contribute significantly to the signal-to-noise (Figure~\ref{fig:fkk}).
The barred power spectra on the RHS of Eq.~(\ref{eq:Ebar_squeezed}) are
given by Eqs.~(\ref{eq:bPgg_kl})--(\ref{eq:bPgx_ks}).

The Fisher matrix $F_{BB}$ is given by any of the following forms:
\ba
F_{BB} 
  &=& V \int \frac{d^3\k_L}{(2\pi)^3} \frac{d^3\k_S}{(2\pi)^3} \frac{d^2\l}{(2\pi)^2}
     \frac{\bB(k_L,k_S,l,k_{Lr})^2}{\bP_{gg}^{\rm tot}(k_L,k_{Lr}) \, \bP_{gg}^{\rm tot}(k_S,k_{Lr}) \, C_l^{TT,\rm tot}} \,
     (2\pi)^3 \delta^3\left( \k_L + \k_S + \frac{\l}{\chi_*} \right) \nn \\
  &=& V \int dk_L \, dk_S \, dk_{Lr} \, \frac{k_L k_S \chi_*^2}{8 \pi^3} 
   \left( \frac{\bB(k_L,k_S,l,k_{Lr})^2}{\bP_{gg}^{\rm tot}(k_L,k_{Lr}) \, \bP_{gg}^{\rm tot}(k_S,k_{Lr}) \, C_l^{TT,\rm tot}} \right)_{l=k_S\chi_*}  \nn \\
 &=& V \frac{K_*^2}{8\pi^3 \chi_*^2} \int dk_L \, dk_S \, dk_{Lr} \, 
     \left( \frac{k_{Lr}^2}{k_L} \frac{\bP_{gv}(k_L,k_{Lr})^2}{\bP_{gg}^{\rm tot}(k_L,k_{Lr})} \right)
     \left( \frac{k_S \bP_{ge}(k_S,k_{Lr})^2}{\bP_{gg}^{\rm tot}(k_S,k_{Lr})} \right)
     \left( \frac{1}{(C_l^{\rm tot})_{l=k_S\chi_*}} \right)  \label{eq:F_distorted}
\ea
generalizing Eqs.~(\ref{eq:F_vector}),~(\ref{eq:F_squeezed}),~(\ref{eq:F_final}) in the undistorted case.
In the second and third lines, the integral runs over $|k_{Lr}| \le k_L \ll k_S$, with positive or negative $k_{Lr}$.

Recall that in the undistorted case, $F_{BB}$ could be simplified further by doing the $k_{Lr}$ integral,
leading to the simple form of the Fisher matrix in Eq.~(\ref{eq:F_final}).  This does not work in the
distorted case because the $k_{Lr}$ dependence of the integrand is more complicated, due to the 
$\bP_{gg}^{\rm tot}$ denominators.

\subsection{Constraining astrophysics using $P_{ge}(k_S)$, with photo-z's and RSD}

Previously in \S\ref{ssec:astrophysics}, we forecasted error bars on the galaxy-electron power spectrum
$\tP_{ge}(k_S)$ in the undistorted case.  In this section we will generalize to include photo-z errors and RSD.

For simplicity, we will assume that $\bP_{gv}(k_L,k_{Lr})$ is known in advance.
For example, it could be given by Eq.~(\ref{eq:bPgx_kl}) above, with $\beta$ and $W(k_{Lr})$ assumed known.
We will forecast constraints on the small-scale galaxy-electron power spectrum $P_{ge}$ twice,
with different levels of generality.

First, a general ``two-variable'' forecast: suppose $\bP_{ge}(k_S,k_{Sr})$ is a free function of two
variables $(k_S, k_{Sr})$, which we want to measure using kSZ tomography.
We will derive an expression for the statistical error $\Delta \bP_{ge}$ over a ``band'' $\beta$,
which can be an arbitrary subset of the $(k_S, k_{Sr})$ plane (including positive and negative values of $k_{Sr}$).
Following the derivation in~\S\ref{ssec:astrophysics}, we start with Eq.~(\ref{eq:F_distorted}) for the
total SNR$^2$ and restrict the integrals to the band $\beta$, to obtain the SNR$^2$ in the band, and
the bandpower error $\Delta \bP_{ge} = \bP_{ge} / \mbox{SNR}_\beta$.  The result is:
\ba
\mbox{SNR}^2_\beta &=& V \frac{K_*^2}{\chi_*^2} \int_{(k_S,k_{Sr}) \in \beta} \frac{k_S \, dk_S \, dk_{Sr}}{2\pi}
   \frac{A(k_{Sr}) \bP_{ge}(k_S,k_{Sr})^2}{\bP_{gg}^{\rm tot}(k_S,k_{Sr})} \left( \frac{1}{C_l^{TT,\rm tot}} \right)_{l=k_S\chi_*} \label{eq:snr2_beta} \\
\Delta \bP_{ge} &=& \left[ V \frac{K_*^2}{\chi_*^2} \int_{(k_S,k_{Sr}) \in \beta} \frac{k_S \, dk_S \, dk_{Sr}}{2\pi}
   \frac{A(k_{Sr})}{\bP_{gg}^{\rm tot}(k_S,k_{Sr})} \left( \frac{1}{C_l^{TT,\rm tot}} \right)_{l=k_S\chi_*} \right]^{-1/2}
\ea
where we have defined:
\be
A(k_{Lr}) = \int_{|k_{Lr}|}^\infty \frac{k_L dk_L}{4\pi^2} \frac{k_{Lr}^2}{k_L^2} 
   \frac{\bar P_{gv}(k_L,k_{Lr})^2}{\bar P_{gg}^{\rm tot}(k_L,k_{Lr})}
\ee
Second, we do a ``one-variable'' forecast, where we make the extra assumptions that $\bP_{ge}(k_S,k_{Sr}) = W(k_{Sr}) \tP_{ge}(k_S)$
as in Eq.~(\ref{eq:bPgx_ks}), and the photo-z error distribution $W(k_{Sr})$ is known.  In the one-variable
forecast, we want to measure the {\em undistorted} galaxy-electron power spectrum $\tP_{ge}(k_S)$ in a $k_S$-bin $(k_S^{\rm min}, k_S^{\rm max})$.
We specialize Eq.~(\ref{eq:snr2_beta}) by setting $\bP_{ge}(k_S,k_{Sr}) = W(k_{Sr}) \tP_{ge}(k_S)$ and integrate out $k_{Sr}$,
to obtain the SNR$^2$ in the $k_S$-bin, and the bandpower error $\Delta \tP_{ge} = \tP_{ge} / \mbox{SNR}_{\rm bin}$:
\ba
\mbox{SNR}^2_{\rm bin} &=& V \frac{K_*^2}{\chi_*^2} \int_{k_S^{\rm min}}^{k_S^{\rm max}} \frac{k_S \, dk_S}{2\pi} B(k_S) 
  \tP_{ge}(k_S)^2 \left( \frac{1}{C_l^{TT,\rm tot}} \right)_{l=k_S\chi_*} \\
\Delta \tP_{ge} &=& \left[ V \frac{K_*^2}{\chi_*^2} \int_{k_S^{\rm min}}^{k_S^{\rm max}} \frac{k_S \, dk_S}{2\pi} B(k_S) 
   \left( \frac{1}{C_l^{TT,\rm tot}} \right)_{l=k_S\chi_*} \right]^{-1/2}
\ea
Here, we have defined:
\be
B(k_S) = \int_{-k_S}^{k_S} dk_{Sr} \, \frac{W(k_{Sr})^2 A(k_{Sr})}{\bP_{gg}^{\rm tot}(k_S,k_{Sr})}
\ee

\subsection{Constraining cosmology with photo-z's and RSD}

Previously in~\S\ref{sec:forecasts_pheno}, we argued that for cosmological applications,
kSZ tomography is best formulated as a quadratic estimator $\hv_r(\k_L)$ which reconstructs
long-wavelength modes of the radial velocity field.
In this section, we revisit this analysis in the presence of RSD and photo-$z$ errors.
We mention in advance that we will construct two different quadratic estimators, 
a ``minimum variance'' estimator $\hv_r^{\rm mv}(\k_L)$ and a ``robust'' estimator 
$\hv_r^{\rm rob}(\k_L)$.

The minimum variance estimator $\hv_r^{\rm mv}(\k_L)$ has the best possible
reconstruction noise power spectrum, but has the drawback that if $\bP_{ge}(k_S, k_{Sr})$
is not known perfectly, then the reconstruction bias is a function $b_v^{\rm mv}(k_{Lr})$.
This is in contrast to the undistorted case, where a single bias parameter $b_v$
must be marginalized.
The robust estimator $\hv_r^{\rm rob}$ has higher noise, but its velocity bias $b_v^{\rm rob}$ 
is constant on large scales, under certain assumptions which we will state explicitly.

To construct the minimum variance estimator $\hv_r^{\rm mv}$, we repeat the logic from~\S\ref{ssec:vrec}, allowing
power spectra to be anisotropic.
We consider a general quadratic estimator of the form:
\be
\hv_r^{\rm mv}(\k_L) = \int \frac{d^3\k_S}{(2\pi)^3} \, \frac{d^2\l}{(2\pi)^2} \,
  W(\k_S,\l) \, \delta_g^*(\k_S) \, T^*(\l) \, (2\pi)^3 \delta^3\left( \k_L + \k_S + \frac{\l}{\chi_*} \right)
\ee
and solve for the weights $W(\k_S,\l)$ which minimize the noise power spectrum
\be
N_{v_r}^{\rm mv}(k_L, k_{Lr}) = \int \frac{d^3\k_S}{(2\pi)^3} \, \frac{d^2\l}{(2\pi)^2} \,
  |W(\k_S,\l)|^2 \bP_{gg}^{\rm tot}(k_S,k_{Sr}) C_l^{TT,\rm tot} \,
  (2\pi)^3 \delta^3\left( \k_L + \k_S + \frac{\l}{\chi_*} \right)  \label{eq:hv_dist_noise}
\ee
subject to the constraint $\langle \hv_r^{\rm mv}(\k_L) \rangle = v_r(\k_L)$, which is equivalent to:
\be
1 = \frac{K_*}{\chi_*^2} \int \frac{d^3\k_S}{(2\pi)^3} \, \frac{d^2\l}{(2\pi)^2} \,
 W(\k_S,\l) \bP_{ge}(k_S,k_{Sr}) \, (2\pi)^3 \delta^3\left( \k_L + \k_S + \frac{\l}{\chi_*} \right)  \label{eq:hv_dist_constraint}
\ee
This constrained minimization problem can be solved by a short calculation involving Lagrange multipliers.
The minimum variance estimator $\hv_r^{\rm mv}(\k_L)$ and its noise power spectrum $N_{v_r}^{\rm mv}(k_L,k_{Lr})$
are found to be:
\ba
\hv_r^{\rm mv}(\k_L) &=& \frac{K_*}{\chi_*^2} N_{v_r}^{\rm mv}(k_L,k_{Lr}) 
  \int \frac{d^3\k_S}{(2\pi)^3} \, \frac{d^2\l}{(2\pi)^2} \,
    \frac{\bP_{ge}(k_S,k_{Sr})}{\bP_{gg}^{\rm tot}(k_S,k_{Sr}) C_l^{\rm tot}} \delta_g(\k_S)^* T(\l)^*
    \, (2\pi)^3 \delta^3\left( \k_L + \k_S + \frac{\l}{\chi_*} \right)  \\
N_{v_r}^{\rm mv}(k_L,k_{Lr}) &=& \frac{\chi_*^4}{K_*^2} 
   \left[ \int \frac{d^3\k_S}{(2\pi)^3} \, \frac{d^2\l}{(2\pi)^2} \, 
       \frac{\bP_{ge}(k_S,k_{Sr})^2}{\bP_{gg}^{\rm tot}(k_S,k_{Sr}) C_l^{TT, \rm tot}} \,
       (2\pi)^3 \delta^3\left( \k_L + \k_S + \frac{\l}{\chi_*} \right)
   \right]^{-1} \nn \\
 &=& \frac{\chi_*^2}{K_*^2} 
   \left[ \int \frac{k_S \, dk_S}{2\pi} \,
       \left( \frac{\bP_{ge}(k_S,k_{Lr})^2}{\bP_{gg}^{\rm tot}(k_S,k_{Lr}) C_l^{TT, \rm tot}} \right)_{l=k_S\chi_*}
   \right]^{-1}   \label{eq:nv_mv}
\ea
The final result is very similar to the quadratic estimator derived previously in the undistorted case
in Eq.~(\ref{eq:hv_final}) above.
Note that the reconstruction noise power spectrum is anisotropic in the presence of photo-$z$ errors.
We have written it as $N_{v_r}(k_L,k_{Lr})$, but we note that it only depends on $k_{Lr}$.

Now we analyze the effect of the optical depth degeneracy, by assuming that the estimator $\hv_r^{\rm mv}$ is defined
using fiducial power spectrum $\bP_{ge}^{\rm fid}(k_S,k_{Sr})$, which may differ from the true power spectrum $\bP_{ge}^{\rm true}(k_S,k_{Sr})$.
A short calculation shows that $\langle \hv_r^{\rm mv}(\k_L) \rangle = b_v^{\rm mv}(k_{Lr}) v_r(\k_L)$, where the
velocity reconstruction bias $b_v^{\rm mv}(k_{Lr})$ is given by:
\be
b_v^{\rm mv}(k_{Lr}) = \frac{\int dk_S \, F(k_S,k_{Lr}) \bP_{ge}^{\rm true}(k_S,k_{Lr})}{\int dk_S \, F(k_S,k_{Lr}) \bP_{ge}^{\rm fid}(k_S,k_{Lr})}
  \hspace{1cm}
\mbox{where }
F(k_S,k_{Lr}) = \left( \frac{k_S \bP_{ge}^{\rm fid}(k_S,k_{Lr})}{\bP_{gg}^{\rm tot}(k_S,k_{Lr}) C_l^{TT,\rm tot}} \right)_{l=k_S\chi_*}  \label{eq:bvmv1}
\ee
We see that the velocity bias is not constant on large scales: it depends on the radial component $k_{Lr}$ of
the wavenumber $\k_L$.  This is a potential problem for cosmological parameter constraints, since it may require
introducing many nuisance parameters in order to parameterize the velocity bias $b_v^{\rm mv}(k_{Lr})$.

\subsection{A quadratic estimator which is robust to photo-$z$ errors}

We now construct a ``robust'' velocity reconstruction estimator $\hv_r^{\rm rob}$ 
whose reconstruction bias $b_v$ is constant on large scales, as in the undistorted case.
The construction is simple: we define
\be
\hv_r^{\rm rob}(\k_L) = W(k_{Lr})^{-1} \hv_r^{\rm und}(\k_L)  \label{eq:hvr_rob}
\ee
where $\hv_r^{\rm und}(\k_L)$ is the {\em undistorted} quadratic estimator, defined by:
\ba
\hv_r^{\rm und}(\k_L) &=& N_{v_r}^{\rm und}(k_L) \frac{K_*}{\chi_*^2} 
 \int \frac{d^3\k_S}{(2\pi)^3} \frac{d^2\l}{(2\pi)^2} \,
    \frac{\tP_{ge}(k_S)}{\tP_{gg}^{\rm tot}(k_S) C_l^{TT,\rm tot}} \Big( \delta_g^*(\k_S) T^*(\l) \Big)
    (2\pi)^3 \delta^3\!\left( \k_L + \k_S + \frac{\l}{\chi_*} \right)  \\
N_{v_r}^{\rm und}(k_L) &=& \frac{\chi_*^2}{K_*^2} \left[ \int \frac{k_S \, dk_S}{2\pi} 
      \left( \frac{\tP_{ge}(k_S)^2}{\tP_{gg}^{\rm tot}(k_S) C_l^{TT,\rm tot}} \right)_{l=k_S\chi_*} \right]^{-1}  \label{eq:nv_und}
\ea
This is the same as the previous definition in Eq.~(\ref{eq:hv_final}), but we have rewritten it
to emphasize that it is defined using undistorted power spectra $\tP_{gg}^{\rm tot}$, $\tP_{ge}$
throughout.

With the prefactor $W(k_{Lr})^{-1}$ in Eq.~(\ref{eq:hvr_rob}), the robust estimator $\hv_r^{\rm rob}$
is an unbiased reconstruction in the distorted case, i.e.~$\langle \hv_r^{\rm rob}(\k_L) \rangle = v_r(\k_L)$.
This statement is not obvious, but follows from a short calculation using 
Eqs.~(\ref{eq:bPgg_ks}),~(\ref{eq:bPgx_ks}),~(\ref{eq:hv_dist_constraint}).

Another short calculation shows that the reconstruction noise of the robust estimator is:
\be
N_{v_r}^{\rm rob}(k_L, k_{Lr}) = \left( \frac{N_{v_r}^{\rm und}(k_L)}{W(k_{Lr})} \right)^2 
   \frac{K_*^2}{\chi_*^2} \int \frac{k_S \, dk_S}{2\pi} 
      \left( \frac{\tP_{ge}(k_S)^2 \bP_{gg}^{\rm tot}(k_S,k_{Lr})}{\tP_{gg}^{\rm tot}(k_S)^2 C_l^{TT,\rm tot}} \right)_{l=k_S\chi_*}  \label{eq:nv_rob}
\ee
where $N_{v_r}^{\rm und}(k_L)$ is the undistorted reconstruction noise in Eq.~(\ref{eq:nv_und}).

Next we compute the reconstruction bias $b_v$ for the robust estimator.
Suppose the velocity reconstruction $\hv_r^{\rm rob}(\k_L)$
is defined using fiducial galaxy-electron power spectrum $\bP_{ge}^{\rm fid}(k_S, k_{Sr}) = W_{\rm fid}(k_{Sr}) \tP_{ge}^{\rm fid}(k_S)$,
and the true power spectrum is $\bP_{ge}^{\rm true}(k_S, k_{Sr}) = W_{\rm true}(k_{Sr}) \tP_{ge}^{\rm true}(k_S)$.
Then a short calculation shows that $\langle \hv_r^{\rm rob}(\k_L) \rangle = b_v^{\rm rob}(k_{Lr}) v_r(\k_L)$, where the
velocity reconstruction bias $b_v^{\rm rob}(k_{Lr})$ is given by:
\be
b_v^{\rm rob}(k_{Lr}) = \frac{W_{\rm true}(k_{Lr})}{W_{\rm fid}(k_{Lr})} \, b_v^{\rm und}  \label{eq:bv_rob}
\ee
where $b_v^{\rm und}$ is the undistorted bias parameter, defined previously in Eq.~(\ref{eq:bv_def})
and independent of $\k_L$.

From Eq.~(\ref{eq:bv_rob}), we see that the velocity reconstruction bias $b_v^{\rm rob}$ is independent of $\k_L$
if the photometric error distribution $W(k_{Lr})$ is well-characterized, so that $W_{\rm fid}(k_{Lr}) = W_{\rm true}(k_{Lr})$
to a good approximation.
If $W(k_{Lr})$ is poorly characterized, then more nuisance parameters would be necessary, to model uncertainty
in the photometric error distribution.

This situation is qualitatively similar to weak gravitational lensing,
where photo-$z$ errors must be well-characterized to avoid introducing extra nuisance parameters.
Because weak lensing is of central importance for upcoming large-scale structure surveys,
photometric redshift errors are expected to be precisely characterized.
Therefore, it seems reasonable to assume that in the kSZ context,
photometric redshift errors will also be characterized well enough that the bias $\hv_r^{\rm rob}$
is constant on large scales.
In the context of a real photometric survey such as LSST, this assumption should probably be 
revisited using detailed survey-specific modeling, but this is outside the scope of this paper.
Our analysis here is simply to show that there is no ``showstopper'' problem
in doing kSZ tomography using photometric catalogs.

In summary, we have now shown how to modify the minimum-variance velocity reconstruction $\hv_r^{\rm mv}(\k_L)$,
obtaining a ``robust'' reconstruction $\hv_r^{\rm rob}(\k_L)$ whose bias $b_v^{\rm rob}$ is constant on large
scales.
This construction depends on the following assumptions.  First, the distorted and undistorted galaxy-electron 
power spectra must be related by $\bP(k_S, k_{Sr}) = W(k_{Sr}) \tP(k_S)$.  Second, the photometric redshift error
distribution $W(k_{Sr})$ must be well-characterized.

In principle, the robust estimator has higher reconstruction noise than the minimum-variance estimator.
However, the two are nearly equal in practice.
For example, for LSSTY10 $\times$ S4, the noise curves are identical at large scales and at most 3\% different on small scales. 
In this paper, we have used minimum-variance noise curves in forecasts (since it makes no practical difference),
but in real data analysis we recommend using the robust estimator.

Throughout this section, we have constructed reconstruction estimators $\hv_r(\k_L)$ for the radial velocity.
As in the undistorted case, a radial velocity reconstruction $\hv_r$ can be converted
to either a reconstruction of the full velocity field $\hv(\k_L) = -i\mu^{-1} \hv_r(\k_L)$,
or the density field $\hdelta(\k_L) = \mu^{-1} (k_L/faH) \hv_r(\k_L)$, with noise power spectra
\be
N_{vv}(\k_L) = \mu^{-2} N_{v_r}(k_L)
  \hspace{1.5cm}
N_{\delta\delta}(\k_L) = \mu^{-2} \left( \frac{k_L}{faH} \right)^2 N_{v_r}(k_L)
\ee

\section{Discussion}
\label{sec:discussion}

We have shown that several proposed kSZ tomography statistics
are ``bispectrum estimation in disguise'' and mathematically equivalent.
Among these statistics, the kSZ-derived radial velocity quadratic estimator $\hv_r$ is
particularly convenient, since it naturally generates additional higher-point
statistics.  For example, an auto correlation of the form $\langle \hv_r(\k)^* \hv_r(\k) \rangle$
is a four-point estimator in the underlying CMB and galaxy fields.

This perspective puts kSZ tomography
on the same footing as more familiar higher-point estimators in cosmology,
making its properties more transparent.
For example, the degeneracy breaking mechanism recently proposed in~\cite{Sugiyama:2016rue} 
appears ``automatically'' when $\hv_r$ is included in a Fisher matrix forecast which also includes
a galaxy survey with redshift-space distortions.

There are two kSZ tomography observables.
First, kSZ tomography measures the small-scale galaxy-electron power spectrum $P_{ge}(k_S)$
on 1-halo dominated scales.  This measurement probes the distribution of electrons in halos
and will be interesting to combine with other probes, especially galaxy-galaxy lensing,
thermal SZ, and X-ray observations.

Second, kSZ tomography measures 3-d cosmological modes on large scales, with lower
noise than can be achieved with galaxy surveys.  Thus, even though the kSZ appears
on small scales in the CMB, its cosmological constraining power arises from its
ability to constrain large-scale physics.

In this paper, we have sometimes made simplifications or approximations which could
be explored in more detail in future work:
\begin{itemize}
\item The simplified ``snapshot'' geometry from~\S\ref{sec:definitions} neglects evolution
  along the lightcone, and makes the flat-sky approximation.
\item We have assumed that the kSZ anisotropy is sourced by the large-scale velocity
  field $v_r$.  This is an approximation to a gauge-invariant quantity, namely the
  CMB dipole in the electron rest frame.  On Hubble scales this approximation may
  become inaccurate.
\item We used symmetry arguments to show that the kSZ bispectrum is unbiased by
  contributions from other CMB secondaries.  These symmetry arguments break down
  in the presence of sky cuts or evolution along the lightcone.
\item We have neglected terms which are subleading in the squeezed
 limit $k_L \ll k_S$, but such terms may become important at high SNR.
\item We have not included all non-Gaussian contributions to higher $N$-point functions.
 For example, our forecasts assume that the quadratic estimator $\hv_r$ has auto correlations
 of the form $\langle \hv_r(\k)^* \hv_r(\k) \rangle \propto (P_{v_r}(k) + N_{v_r}(k))$.
 This is an approximation to a four-point function of type $\langle \delta_g \delta_g T T \rangle$.
 Similar approximations are often made in the context of CMB lens reconstruction.
 We are in the process of using $N$-body simulations to study this issue systematically~\cite{Utkarsh}.
\item Our model for photometric errors assumes that the error distribution is known perfectly,
 that drastic outliers are negligible, and that every galaxy has an independent photo-$z$ error.
\end{itemize}

\section*{Acknowledgements}

Research at Perimeter Institute is supported by the Government of Canada
through Industry Canada and by the Province of Ontario through the Ministry of Research \& Innovation.
MSM is grateful to Perimeter for supporting visits during which this work was carried out. 
KMS was supported by an NSERC Discovery Grant and an Ontario Early Researcher Award. SF was supported by Miller and BCCP fellowships at 
the University of California, Berkeley.
MCJ is supported by the National Science and Engineering Research Council through a Discovery grant.
We thank Nick Battaglia, Neal Dalal, Jo Dunkley, Emmanuel Schaan, Marcel Schmittfull, Uro\v{s} Seljak, David Spergel and 
Martin White for useful discussions and 
Emanuela Dimastrogiovanni for collaboration during the early stages of this work.

\bibliographystyle{h-physrev}
\bibliography{ksz_bispectrum}

\begin{thebibliography}{10}

\bibitem{Park:2013mv}
H.~Park {\em et~al.},
\newblock Astrophys. J. {\bf 769}, 93 (2013), 1301.3607.

\bibitem{Hand:2012ui}
N.~Hand {\em et~al.},
\newblock Phys. Rev. Lett. {\bf 109}, 041101 (2012), 1203.4219.

\bibitem{Ade:2015lza}
{Planck Collaboration} {\em et~al.},
\newblock Astron. Astrophys. {\bf 586}, A140 (2016), 1504.03339.

\bibitem{Schaan:2015uaa}
E.~Schaan {\em et~al.},
\newblock Phys. Rev. {\bf D93}, 082002 (2016), 1510.06442.

\bibitem{Soergel:2016mce}
{DES/SPT Collaboration} {\em et~al.},
\newblock Mon. Not. Roy. Astron. Soc. {\bf 461}, 3172 (2016), 1603.03904.

\bibitem{Hill:2016dta}
J.~C. Hill, S.~Ferraro, N.~Battaglia, J.~Liu, and D.~N. Spergel,
\newblock Phys. Rev. Lett. {\bf 117}, 051301 (2016), 1603.01608.

\bibitem{DeBernardis:2016pdv}
F.~De~Bernardis {\em et~al.},
\newblock JCAP {\bf 1703}, 008 (2017), 1607.02139.

\bibitem{Bregman:2007ac}
J.~N. Bregman,
\newblock Ann. Rev. Astron. Astrophys. {\bf 45}, 221 (2007), 0706.1787.

\bibitem{HernandezMonteagudo:2008jz}
C.~Hernandez-Monteagudo and R.~A. Sunyaev,
\newblock Astron. Astrophys. {\bf 490}, 25 (2008), 0805.3702.

\bibitem{Ho:2009iw}
S.~Ho, S.~Dedeo, and D.~Spergel,
\newblock (2009), 0903.2845.

\bibitem{DeDeo:2005yr}
S.~DeDeo, D.~N. Spergel, and H.~Trac,
\newblock (2005), astro-ph/0511060.

\bibitem{HernandezMonteagudo:2005ys}
C.~Hernandez-Monteagudo, L.~Verde, R.~Jimenez, and D.~N. Spergel,
\newblock Astrophys. J. {\bf 643}, 598 (2006), astro-ph/0511061.

\bibitem{Bhattacharya:2007sk}
S.~Bhattacharya and A.~Kosowsky,
\newblock Phys. Rev. {\bf D77}, 083004 (2008), 0712.0034.

\bibitem{Mueller:2014nsa}
E.-M. Mueller, F.~de~Bernardis, R.~Bean, and M.~D. Niemack,
\newblock Astrophys. J. {\bf 808}, 47 (2015), 1408.6248.

\bibitem{Bianchini:2015iaa}
F.~Bianchini and A.~Silvestri,
\newblock Phys. Rev. {\bf D93}, 064026 (2016), 1510.08844.

\bibitem{Mueller:2014dba}
E.-M. Mueller, F.~de~Bernardis, R.~Bean, and M.~D. Niemack,
\newblock Phys. Rev. {\bf D92}, 063501 (2015), 1412.0592.

\bibitem{GarciaBellido:2008gd}
J.~Garcia-Bellido and T.~Haugboelle,
\newblock JCAP {\bf 0809}, 016 (2008).

\bibitem{Caldwell:2007yu}
R.~R. Caldwell and A.~Stebbins,
\newblock Phys. Rev. Lett. {\bf 100}, 191302 (2008).

\bibitem{Clifton:2011sn}
T.~Clifton, C.~Clarkson, and P.~Bull,
\newblock Phys. Rev. Lett. {\bf 109}, 051303 (2012).

\bibitem{Bull:2011wi}
P.~Bull, T.~Clifton, and P.~G. Ferreira,
\newblock Phys. Rev. {\bf D85}, 024002 (2012).

\bibitem{Maartens2011}
R.~Maartens,
\newblock Philosophical Transactions of the Royal Society A: Mathematical,
  Physical and Engineering Sciences {\bf 369}, 5115 (2011).

\bibitem{2011CQGra..28p4005Z}
J.~P. {Zibin} and A.~{Moss},
\newblock Classical and Quantum Gravity {\bf 28}, 164005 (2011).

\bibitem{Yoo:2010ad}
C.-M. Yoo, K.-i. Nakao, and M.~Sasaki,
\newblock JCAP {\bf 1010}, 011 (2010).

\bibitem{Zhang10d}
P.~{Zhang},
\newblock MNRAS {\bf 407}, L36 (2010), 1004.0990.

\bibitem{Zhang11b}
P.~{Zhang} and A.~{Stebbins},
\newblock Physical Review Letters {\bf 107}, 041301 (2011).

\bibitem{Zhang:2015uta}
P.~Zhang and M.~C. Johnson,
\newblock JCAP {\bf 1506}, 046 (2015), 1501.00511.

\bibitem{Terrana2016}
A.~Terrana, M.-J. Harris, and M.~C. Johnson,
\newblock Journal of Cosmology and Astroparticle Physics {\bf 2017}, 040
  (2017), 1610.06919.

\bibitem{Hu:2001tn}
W.~Hu,
\newblock Astrophys. J. {\bf 557}, L79 (2001), astro-ph/0105424.

\bibitem{Hu:2001kj}
W.~Hu and T.~Okamoto,
\newblock Astrophys. J. {\bf 574}, 566 (2002), astro-ph/0111606.

\bibitem{Okamoto:2003zw}
T.~Okamoto and W.~Hu,
\newblock Phys. Rev. {\bf D67}, 083002 (2003), astro-ph/0301031.

\bibitem{Li:2014mja}
M.~Li, R.~E. Angulo, S.~D.~M. White, and J.~Jasche,
\newblock Mon.Not.Roy.Astron.Soc. {\bf 443}, 2311 (2014), 1404.0007.

\bibitem{Alonso:2016jpy}
D.~Alonso, T.~Louis, P.~Bull, and P.~G. Ferreira,
\newblock Phys. Rev. {\bf D94}, 043522 (2016), 1604.01382.

\bibitem{Deutsch:2017ybc}
A.-S. Deutsch, E.~Dimastrogiovanni, M.~C. Johnson, M.~Munchmeyer, and
  A.~Terrana,
\newblock (2017), 1707.08129.

\bibitem{Dore:2003ex}
O.~Dor\'e, J.~F. Hennawi, and D.~N. Spergel,
\newblock Astrophys. J. {\bf 606}, 46 (2004), astro-ph/0309337.

\bibitem{Ferraro:2016ymw}
S.~Ferraro, J.~C. Hill, N.~Battaglia, J.~Liu, and D.~N. Spergel,
\newblock Phys. Rev. {\bf D94}, 123526 (2016), 1605.02722.

\bibitem{Smith:2016lnt}
K.~M. Smith and S.~Ferraro,
\newblock Phys. Rev. Lett. {\bf 119}, 021301 (2017), 1607.01769.

\bibitem{Ferraro:2018izc}
S.~Ferraro and K.~M. Smith,
\newblock (2018), 1803.07036.

\bibitem{Battaglia:2016xbi}
N.~Battaglia,
\newblock JCAP {\bf 1608}, 058 (2016), 1607.02442.

\bibitem{Flender:2016cjy}
S.~Flender, D.~Nagai, and M.~McDonald,
\newblock Astrophys. J. {\bf 837}, 124 (2017), 1610.08029.

\bibitem{Louis:2017hoh}
T.~Louis, E.~F. Bunn, B.~Wandelt, and J.~Silk,
\newblock Phys. Rev. {\bf D96}, 123509 (2017), 1707.04102.

\bibitem{Soergel:2017ahb}
B.~Soergel, A.~Saro, T.~Giannantonio, G.~Efstathiou, and K.~Dolag,
\newblock Mon. Not. Roy. Astron. Soc. {\bf 478}, 5320 (2018), 1712.05714.

\bibitem{Keisler:2012eg}
R.~Keisler and F.~Schmidt,
\newblock Astrophys. J. {\bf 765}, L32 (2013), 1211.0668.

\bibitem{Sugiyama:2016rue}
N.~S. Sugiyama, T.~Okumura, and D.~N. Spergel,
\newblock (2016), 1606.06367.

\bibitem{Seljak:2008xr}
U.~Seljak,
\newblock Phys. Rev. Lett. {\bf 102}, 021302 (2009), 0807.1770.

\bibitem{Moritz}
M.~M\"unchmeyer, M.~S. Madhavacheril, S.~Ferraro, M.~C. Johnson, and K.~M.
  Smith,
\newblock To appear .

\bibitem{Skillman:2014qca}
S.~W. Skillman {\em et~al.},
\newblock (2014), 1407.2600.

\bibitem{Utkarsh}
U.~Giri {\em et~al.},
\newblock To appear .

\bibitem{Cayuso:2018lhv}
J.~I. Cayuso, M.~C. Johnson, and J.~B. Mertens,
\newblock Phys. Rev. {\bf D98}, 063502 (2018), 1806.01290.

\bibitem{LSSTSciBook}
{LSST Science Collaboration} {\em et~al.},
\newblock ArXiv e-prints  (2009), 0912.0201.

\bibitem{LSSTSRD}
{LSST Dark Energy Science Collaboration} {\em et~al.},
\newblock ArXiv e-prints  (2018), 1809.01669.

\bibitem{DESIReport}
{DESI Collaboration} {\em et~al.},
\newblock ArXiv e-prints  (2016), 1611.00036.

\bibitem{Dunkley2011}
J.~{Dunkley} {\em et~al.},
\newblock \apj {\bf 739}, 52 (2011), 1009.0866.

\bibitem{Ade:2018sbj}
{Simons Observatory Collaboration} {\em et~al.},
\newblock (2018), 1808.07445.

\bibitem{Calafut:2017mzp}
V.~Calafut, R.~Bean, and B.~Yu,
\newblock Phys. Rev. {\bf D96}, 123529 (2017), 1710.01755.

\bibitem{Cooray:2002dia}
A.~Cooray and R.~K. Sheth,
\newblock Phys. Rept. {\bf 372}, 1 (2002), astro-ph/0206508.

\bibitem{Sheth:1999su}
R.~K. Sheth, H.~J. Mo, and G.~Tormen,
\newblock Mon. Not. Roy. Astron. Soc. {\bf 323}, 1 (2001), astro-ph/9907024.

\bibitem{Duffy:2008pz}
A.~R. Duffy, J.~Schaye, S.~T. Kay, and C.~Dalla~Vecchia,
\newblock Mon. Not. Roy. Astron. Soc. {\bf 390}, L64 (2008), 0804.2486,
\newblock [Erratum: Mon. Not. Roy. Astron. Soc.415,L85(2011)].

\bibitem{Smith:2002dz}
R.~E. Smith {\em et~al.},
\newblock Mon. Not. Roy. Astron. Soc. {\bf 341}, 1311 (2003), astro-ph/0207664.

\bibitem{Komatsu:2001dn}
E.~Komatsu and U.~Seljak,
\newblock Mon. Not. Roy. Astron. Soc. {\bf 327}, 1353 (2001), astro-ph/0106151.

\bibitem{Shaw:2011sy}
L.~D. Shaw, D.~H. Rudd, and D.~Nagai,
\newblock Astrophys. J. {\bf 756}, 15 (2012), 1109.0553.

\bibitem{Berlind:2001xk}
A.~A. Berlind and D.~H. Weinberg,
\newblock Astrophys. J. {\bf 575}, 587 (2002), astro-ph/0109001.

\bibitem{Leauthaud:2011zt}
A.~Leauthaud, J.~Tinker, P.~S. Behroozi, M.~T. Busha, and R.~Wechsler,
\newblock Astrophys. J. {\bf 738}, 45 (2011), 1103.2077.

\bibitem{2012ApJ...744..159L}
A.~{Leauthaud} {\em et~al.},
\newblock \apj {\bf 744}, 159 (2012), 1104.0928.

\bibitem{Behroozi:2010rx}
P.~S. Behroozi, C.~Conroy, and R.~H. Wechsler,
\newblock Astrophys. J. {\bf 717}, 379 (2010), 1001.0015.

\bibitem{Hearin:2015jnf}
A.~P. Hearin, A.~R. Zentner, F.~C. van~den Bosch, D.~Campbell, and E.~Tollerud,
\newblock Mon. Not. Roy. Astron. Soc. {\bf 460}, 2552 (2016), 1512.03050.

\bibitem{Hu:1999vq}
W.~Hu,
\newblock Astrophys. J. {\bf 529}, 12 (2000), astro-ph/9907103.

\bibitem{Vishniac:1987wm}
E.~T. Vishniac,
\newblock Astrophys. J. {\bf 322}, 597 (1987).

\end{thebibliography}

\appendix

\section{Mode-counting integral}
\label{app:integrals}

The purpose of this appendix is to derive Eq.~(\ref{eq:F_scalar}) for the Fisher matrix $F_{BB'}$
as an integral over scalar wavenumbers.  We start from the definition of $F_{BB'}$:
\be
F_{BB'} = \frac{V}{2} \int \frac{d^3\k}{(2\pi)^3} \frac{d^3\k'}{(2\pi)^3} \frac{d^2\l}{(2\pi)^2}
     \frac{B(k,k',l,k_r)^* \, B'(k,k',l,k_r)}{P_{gg}^{\rm tot}(k) \, P_{gg}^{\rm tot}(k') \, C_l^{TT,\rm tot}} \,
     (2\pi)^3 \delta^3\left( \k + \k' + \frac{\l}{\chi_*} \right)
\ee
and insert the following expression in the integrand on the RHS:
\be
1 = \int dK \, dK' \, dL \, d\kappa \, \Big( \delta(|\k| - K) \, \delta(|\k'| - K') \, \delta(|\l|-L) \, \delta(k_r - \kappa) \Big)
\ee
to write $F_{BB'}$ in the form
\be
F_{BB'} = \frac{V}{2} \int dK \, dK' \, dL \, d\kappa \, I(K,K',L,\kappa) \frac{B(K,K',L,\kappa)^* \, B'(K,K',L,\kappa)}{P_{gg}^{\rm tot}(K) \, P_{gg}^{\rm tot}(K') \, C_L^{TT,\rm tot}}
\ee
where $I(K,K',L,\kappa)$ is the ``mode-counting integral''
\be
I(K,K',L,\kappa) = \int \frac{d^3\k}{(2\pi)^3} \frac{d^3\k'}{(2\pi)^3} \frac{d^2\l}{(2\pi)^2}
   (2\pi)^3 \delta^3\left(\k+\k'+\frac{\l}{\chi_*}\right)
   \delta(k-K) \delta(k'-K') \delta(l-L') \delta(k_r-\kappa)
\ee
which counts the number of closed triangles $\k+\k'+(\l/\chi_*)=0$
with lengths $(K,K',L)$ and radial wavenumber $\kappa$.

It remains to calculate $I$ explicitly.
First note that by rotational symmetry, the quantity
\be
J(K,K',\kappa,\l) = \int \frac{d^3\k}{(2\pi)^3} \frac{d^3\k'}{(2\pi)^3} 
   (2\pi)^3 \delta^3\left(\k+\k'+\frac{\l}{\chi_*}\right)
   \delta(k-K) \delta(k'-K') \delta(k_3-\kappa)
\ee
only depends on $\l$ through its length $l=|\l|$.  Therefore $I$ and $J$ are related by:
\be
I = \int \frac{d^2\l}{(2\pi)^2} J(\l) \delta(l-L) = \frac{L}{2\pi} J(L)  \label{eq:ij}
\ee
To compute $J$, we assume $\l$ points in the $x$-direction, and use the 3D delta function
to eliminate the $d^3\k'$ integral, obtaining:
\be
J = \int \frac{d^3\k}{(2\pi)^3}
   \delta(k-K) \delta(k'-K') \delta(k_3-\kappa)
\ee
where $k'$ is defined in the integrand by $k'{}^2 = (k_1 - l/\chi_*)^2 + k_2^2 + k_3^2$.
Since this a 3D integral with three delta functions, it is given by the inverse Jacobian
\be
J = 2 \frac{1}{(2\pi)^3} \left( \frac{\partial \{ k, k', k_3 \}}{\partial \{ k_1, k_2, k_3 \}} \right)^{-1}  \label{eq:jjac}
\ee
where the prefactor 2 is because the delta function constraints have two solutions.
A short calculation now gives the Jacobian:
\be
\frac{\partial \{ k, k', k_3 \}}{\partial \{ k_1, k_2, k_3 \}} 
   = \frac{l}{kk'\chi_*} \left[ \Gamma\left(k,k',\frac{l}{\chi_*}\right)^2 - k_3^2 \right]^{1/2}  \label{eq:jac}
\ee
where we have defined
\be
\Gamma(k,k',k'') = 
    \frac{\sqrt{ (k+k'+k'')(k+k'-k'')(k+k''-k')(k'+k''-k) }}{2 k''}  \label{eq:Gamma_def}
\ee
Note that all factors under the square root are positive if the wavenumbers $k,k',k''$
satisfy the inequalities needed for $k,k',k''$ to form a closed triangle.
By Heron's formula, $\Gamma(k,k',k'')$ can be interpreted as the component of $k$ (or $k'$)
perpendicular to $k''$, in a closed triangle $\k+\k'+\k''=0$.
Thus the inequality that $\kappa$ must satisfy to ensure that the delta function constraints
have solutions is simply $|\kappa| \le \Gamma(k,k',l/\chi)$.

Putting Eqs.~(\ref{eq:ij}),~(\ref{eq:jjac}),~(\ref{eq:jac}) together, we get our bottom-line formula for $I$:
\be
I(K,K',L,\kappa) = \frac{KK'\chi_*}{8\pi^4} \left[ \Gamma\left(K,K',\frac{L}{\chi_*}\right)^2 - \kappa^2 \right]^{-1/2}  \label{eq:I_done}
\ee
where the formula is understood to apply when $K,K',(L/\chi_*)$ form a closed triangle,
and $|\kappa| \le \Gamma(K,K',L/\chi_*)$.  Otherwise, $I=0$.


\section{Halo model}
\label{app:halo_model}

Throughout this paper, we use the halo model to compute nonlinear power spectra involving dark matter, electron, and galaxy fields.
In this appendix, we describe the details.

In the halo model, one makes the fundamental assumption that all the dark and baryonic matter is bound up in halos with varying mass and density profiles. The correlation function for density fluctuations then receives two contributions: a "two halo term" which arises from the clustering properties of distinct halos, and a "one halo term" which arises from the correlation in density between two points in the same halo. A review of the halo model can be found in Ref.~\cite{Cooray:2002dia}.

\subsection{Dark matter}

In Fourier space, the dark matter power spectrum is given by
\begin{eqnarray} \label{eq:Pmm}
P_{mm} (k,z) &=& P_{mm}^{1h}(k,z) + P_{mm}^{2h}(k,z) \\
P_{mm}^{1h}(k,z) &=& \int_{-\infty}^\infty d \ln m \ m n(m,z) \left( \frac{m}{\rho_m} \right)^2 |u(k|m,z)|^2 \\
P_{mm}^{2h}(k,z) &=& P^{\rm lin} (k,z) \left[ \int_{-\infty}^\infty d \ln m \ m n(m,z) \left( \frac{m}{\rho_m} \right) b_h(m,z) u(k|m,z) \right]^2
\end{eqnarray}
In these expressions, $m$ is the halo mass, $\rho_m$ is the present day cosmological matter density, $n(m,z)$ is the halo mass function (e.g. the differential number density of halos with respect to mass), $u(k|m,z)$ is the normalized fourier transform of the halo profile, $P^{\rm lin} (k)$ is the linear matter power spectrum, and $b_h(m,z)$ is the linear halo bias. 

The halo mass function is defined by
\begin{equation}
n(m,z) = \frac{\rho_m}{m^2} f(\sigma,z) \frac{d\ln \sigma (m,z)}{d\ln m},
\end{equation}
where $\sigma^2(m,z)$ is the rms variance of mass within a sphere of radius $R$ that contains mass $m=4\pi\rho_m R^3/3$, defined as
\begin{equation}
\sigma^2(m,z) = \frac{1}{2\pi^2} \int_0^\infty dk \ k^2 P^{\rm lin}(k,z) W^2(kR)
\end{equation}
Here, $R=R(m)$ and the window function in Fourier space is
\begin{equation}
W(kR) = \frac{3\left[ \sin(kR) -kR\cos(kR) \right]}{(kR)^3}
\end{equation}
We assume the Sheth-Tormen collapse fraction~\cite{Sheth:1999su}:
\begin{equation}
f(\sigma,z) = A \sqrt{\frac{2a}{\pi}} \left[ 1+ \left( \frac{\sigma^2}{a\delta_c^2}  \right)^p   \right]  \frac{\delta_c}{\sigma} \exp\left[ -\frac{a \delta_c^2}{2\sigma^2}\right]
\end{equation}
with $A=0.3222$, $a=0.75$, $p=0.3$, and $\delta_c=1.686$. The linear halo bias $b_h(m,z)$ accounts for the biasing of halos in the presence of variations in the density field, and is given by the response of the number density to variations in the collapse threshold $\delta_c$.  We use the Sheth-Tormen bias:
\ba
 b_h (m,z) 
   &=& 1 + \frac{1}{\delta_c} \frac{d\log f}{d\log\sigma} \nn \\
   &=& 1 + \frac{1}{\delta_c} \left( a\frac{\delta_c^2}{\sigma^2} - 1  \right) + \frac{2 p}{\delta_c} \left(  1+ \left(a\frac{\delta_c^2}{\sigma^2} \right)^p \right)^{-1}
\ea
Note that the halo bias satisfies a consistency relation:
\begin{equation}\label{eq:halobiasconsistency}
 \int_{-\infty}^\infty d \ln m \ m n(m,z) \left( \frac{m}{\rho_m(z)} \right) b_h(m,z) = 1.
\end{equation}
Finally, we need $u(k|m,z)$, the Fourier transform of the dark matter halo density profile, which for spherically symmetric profiles is defined as
\begin{equation}\label{eq:ukm}
u(k|m,z) = \int_0^{r_{\rm vir}} dr \ 4 \pi r^2 \frac{\sin (kr)}{kr} \frac{\rho(r|m,z)}{m}.
\end{equation}
We assume that halos are truncated at the virial radius, and have mass
\begin{equation}
m = \int_0^{r_{\rm vir}} dr \ 4 \pi r^2 \rho(r|m,z)
\end{equation}
Note that with this definition of mass, $u(k|m,z) \rightarrow 1$ as $k \rightarrow 0$. Returning to the two-halo term and using the consistency relation in Eq.~(\ref{eq:halobiasconsistency}), this property of $u(k|m,z)$ ensures that $P_{mm}^{2h}(k,z) \simeq P^{\rm lin} (k,z)$ in the limit where $k\rightarrow 0$, as it should.

We assume that dark matter halos follow an NFW profile:
\begin{equation}
\rho(r|m,z) = \frac{\rho_s}{(r/r_s) (1+r/r_s)^2}
\end{equation}
and relate the scale radius $r_s$ to the virial radius $r_{\rm vir}$ by the concentration parameter $c=r_{\rm vir}/r_s$. We model the concentration by the median power law fit of~\cite{Duffy:2008pz}, neglecting stochasticity:
\begin{equation}
c(m,z) = A \left( \frac{m}{2 \times 10^{12} h^{-1} M_\odot} \right)^\alpha (1+z)^\beta 
\end{equation}
with $A=7.85$, $\alpha=-0.081$, and $\beta=-0.71$. 

Including halos in the range $10^4 M_{\odot} < m < 10^{17} M_{\odot}$, our model reproduces the non-linear matter power spectrum using the commonly used 'halofit' model of Ref.~\cite{Smith:2002dz} at the $<10\%$ level over the range $10^{-5} \  {\rm Mpc^{-1}} < k <  20 \ {\rm Mpc^{-1}}$.

\subsection{Electrons}

The electron distribution in the halo model is modelled by assuming gas is bound within dark matter halos, having  density profiles $\rho_{\rm gas}(m,z)$ which we assume to be a function of the host halo mass and redshift only. The gas power spectrum is given by Eq.~\ref{eq:Pmm} with $u(k|m,z)$ calculated through Eq.~\ref{eq:ukm} by replacing $\rho(m,z)$ with $\rho_{\rm gas}(m,z)$ and computing a grid of templates. To estimate the systematic uncertainty associated with the distribution of free electrons within halos, we employ three models for the electron profile: the universal gas profile of Ref.~\cite{Komatsu:2001dn} and two fitting functions from Ref.~\cite{Battaglia:2016xbi} based on simulations with two different sub-grid feedback models ("AGN" and "SH"). Throughout we assume that electrons trace gas, and neglect the deficit in large scale power caused by collapse of gas into stars within halos~\footnote{This can cause a $\sim 30-50\%$ decrease in power on large scales~\cite{Shaw:2011sy}}.

The universal gas profile of Ref.~\cite{Komatsu:2001dn} is obtained by assuming that the gas has a polytropic equation of state $P \propto \rho^\gamma$ with unknown $\gamma$ and demanding hydrostatic equilibrium within the gravitational potential well of the dark matter halo (assumed NFW, as above). The two unknown parameters, $\gamma$ and an integration constant from the equation for hydrostatic equilibrium, are fixed by demanding that the slope of the gas profile matches that of the dark matter at twice the virial radius. Therefore, within this model, we explicitly require that gas traces dark matter on the largest scales.

The fitting function for the AGN and SH models of Ref.~\cite{Battaglia:2016xbi} is given by~\footnote{To be consistent with a universal NFW profile, where $\beta$ is the power law index at large $r$, one must correct Eq. A1 of Ref.~\cite{Battaglia:2016xbi} as we have done here.}
\begin{equation}
\rho_{\rm gas} = \frac{\Omega_b}{\Omega_m} \rho_c (z) \bar{\rho}_0(m,z) \left( \frac{r}{2 R_{200} (m,z)} \right)^\gamma \left[1 + \left( \frac{r}{2 R_{200}(m,z)} \right)^{\alpha(m,z)} \right]^{-(\beta(m,z)+\gamma)/\alpha(m,z)}
\end{equation}
where $\gamma=-0.2$, $R_{200} (m,z)$ is radius at which the dark matter halo reaches a density $200 \rho_c(z)$, and the parameters $\bar{\rho}_0(m,z)$, $\alpha(m,z)$, and $\beta(m,z)$ are fitted with a power law in halo mass and redshift:
\begin{equation}
A = A_0^x \left( \frac{M_{200}}{10^{14} M_\odot} \right)^{\alpha_m^x} (1+z)^{\alpha_z^x}
\end{equation}
with parameters in the AGN and SH model given from Table 2 of Ref.~\cite{Battaglia:2016xbi}. For the AGN model we have $\{A_0^{\rho_0},  \alpha_m^{\rho_0}, \alpha_z^{\rho_0}\} = \{4000,0.29,-0.66 \}$, $\{A_0^{\alpha},  \alpha_m^{\alpha}, \alpha_z^{\alpha}\} = \{0.88,-0.03,0.19 \}$, and $\{A_0^{\beta},  \alpha_m^{\beta}, \alpha_z^{\beta}\} = \{3.83,0.04,-0.025 \}$. For the SH model, we have $\{A_0^{\rho_0},  \alpha_m^{\rho_0}, \alpha_z^{\rho_0}\} = \{19000,0.09,-0.95 \}$, $\{A_0^{\alpha},  \alpha_m^{\alpha}, \alpha_z^{\alpha}\} = \{0.70,-0.017,0.27 \}$, and $\{A_0^{\beta},  \alpha_m^{\beta}, \alpha_z^{\beta}\} = \{4.43,0.005,0.037 \}$. 
 
\begin{figure}[tbh]
  \includegraphics[width=0.5\textwidth]{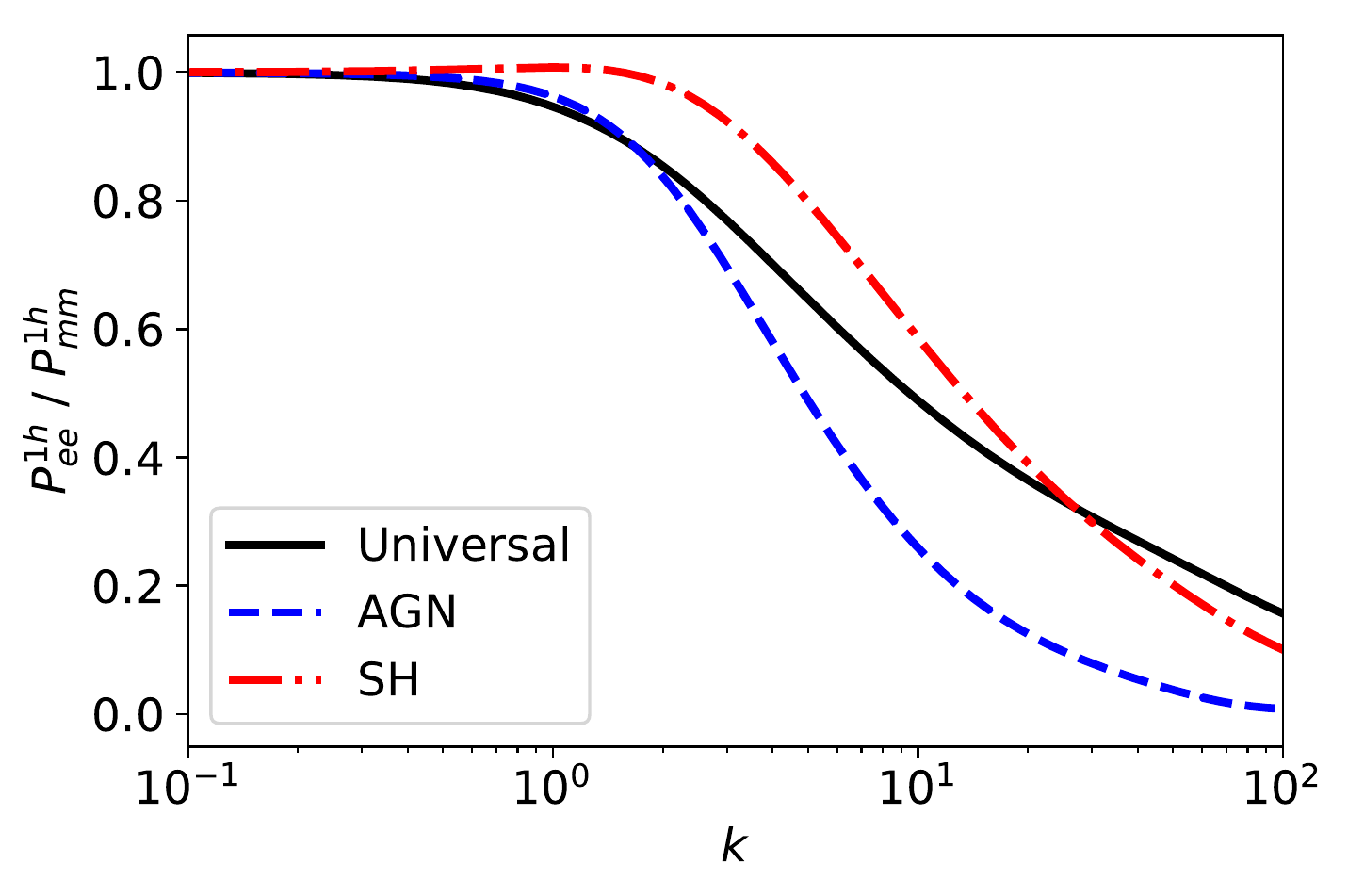}
  \caption{The ratio of the one halo gas power spectrum $P_{ee}^{1h}$ and the one halo dark matter power spectrum $P_{mm}^{1h}$ for three models of the gas profile. }
  \label{fig:onehalogas}
\end{figure}

In Fig.~\ref{fig:onehalogas} we compare the 1-halo terms in the power spectrum for the three gas models to the 1-halo term for dark matter for halos in the range $10^{10} M_{\odot} < m < 10^{17} M_{\odot}$. The one halo term is the dominant contribution to the power spectrum over the plotted range. On scales $k \alt .5 \ {\rm Mpc}^{-1}$, one can approximate the gas power spectrum by the dark matter power spectrum. At higher $k$, the difference between the gas profiles and dark matter and among the various gas models becomes apparent, with the three models giving different predictions at the $\sim 50\%$ level. This is indicative of the 'theory' error bar on the electron power spectrum, which depends in detail on how the various feedback processes are modelled.

\subsection{Galaxies}

We model the distribution of galaxies within dark matter halos by the Halo Occupation Distribution (HOD) model~\cite{Berlind:2001xk} of Refs.~\cite{Leauthaud:2011zt,2012ApJ...744..159L}. This model has been calibrated using measurements of the galaxy-galaxy power spectrum, galaxy-galaxy weak lensing, and the stellar mass function~\cite{2012ApJ...744..159L} to a redshift of $z=1$. We extrapolate the applicable redshift range to $z \sim 4$ using fits for the stellar mass-halo mass relation in Ref.~\cite{Behroozi:2010rx}. This is the 'baseline HOD model' of Ref.~\cite{Hearin:2015jnf}.

Briefly, the ingredients going into the HOD model are as follows. 
First, we assume separate distributions for central and satellite galaxies.
The number of central galaxies in a halo is always 0 or 1, and centrals are at exact halo centers.
The mean number of centrals $\bar N_c(m)$ in a halo of mass $m$ is fixed by the amount of stellar mass in each dark matter halo and given by:
 \begin{equation}
\bar N_c(m) = \frac{1}{2} - \frac{1}{2} {\rm erf} \left(  \frac{\log_{10} (m_*^{\rm thresh}) - \log_{10} \left[ m_*(m) \right]}{\sqrt{2} \sigma_{\log m_*}}  \right)
\end{equation}
where $m_*(m)$ is the stellar mass in a halo of mass $m$.  We specify the galaxy sample by imposing a threshold $m_*^{\rm thresh}$ in stellar mass of observable galaxies, and we assume a log-normal distribution for the stellar mass at fixed halo mass with constant redshift-independent scatter $\sigma_{\log m_*}=0.2$ (consistent with ~\cite{2012ApJ...744..159L}). We employ the model developed in Ref.~\cite{Behroozi:2010rx} for $m_*(m)$, which we refer the reader to for more details. A fiducial threshold is $m_*^{\rm thresh} = 10^{10.5} \ M_{\odot}$, which corresponds to a halo mass of $m \simeq 10^{12} \ M_\odot$ at $z=0$. In the body of the text, we match the number densities for various surveys by adjusting $m_*^{\rm thresh}$.

\begin{figure}[tbh]
\includegraphics[width=0.5\textwidth]{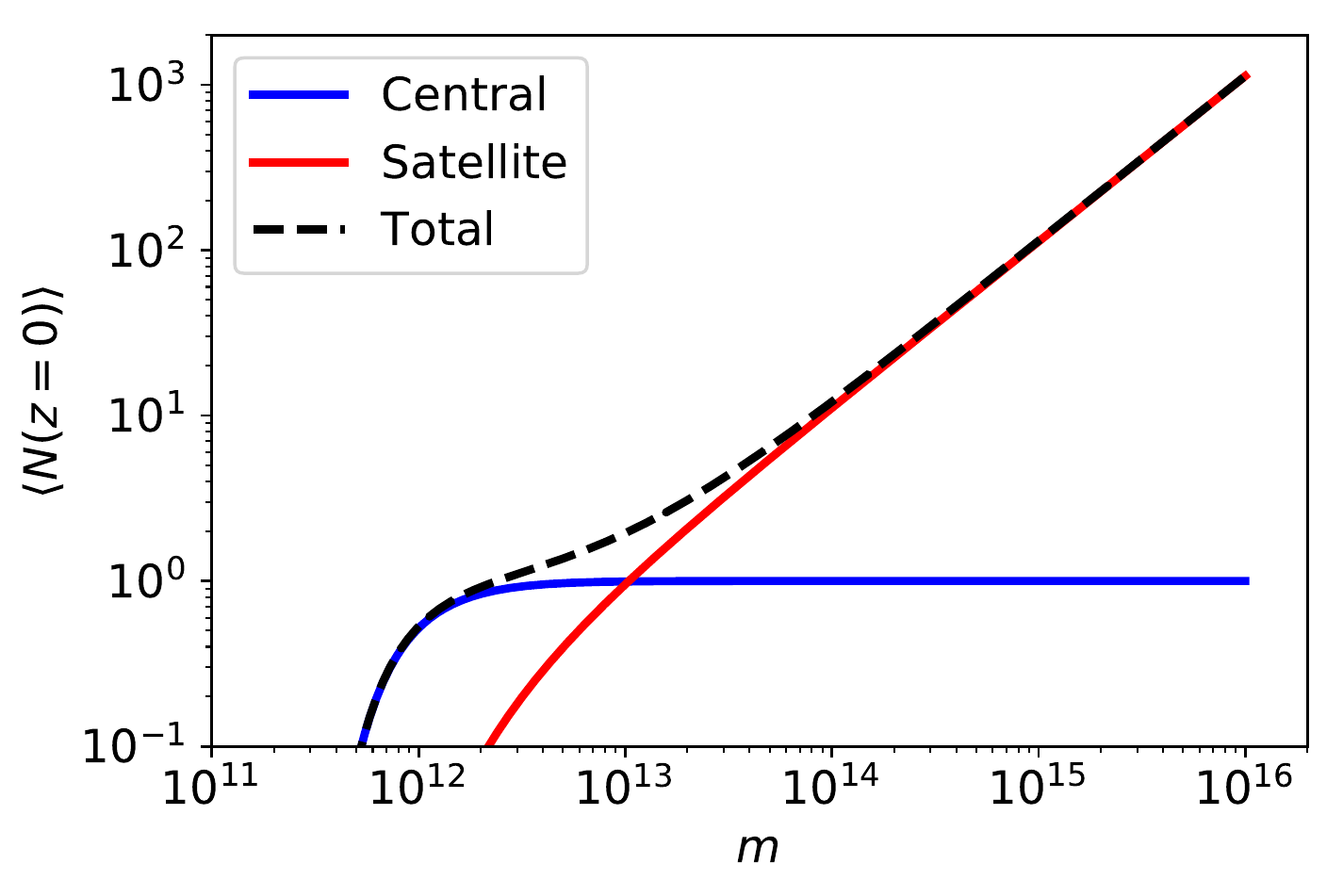}
  \caption{The number of central and satellite galaxies as a function of halo mass using the assumed HOD at $z=0$.}
  \label{fig:galnumber}
\end{figure}

For the satellite galaxies, we assume that the spatial profile is NFW, and
the mean number of satellites $\bar N_s(m)$ in a halo of mass $m$ is given by:
\begin{equation}
\bar N_s(m)  = \bar N_c(m) \left( \frac{m}{m_{\rm sat}} \right)^{\alpha_{\rm sat}} e^{-m_{\rm cut}/m}
\end{equation}
We choose values for the free parameters $m_{\rm sat}$, $\alpha_{\rm sat}$, and $m_{\rm cut}$ (which depend on the choice of $m_*^{\rm thresh}$) consistent with the 'SIG\_MOD1' model of Ref.~\cite{2012ApJ...744..159L} (from the median redshift bin). 
We show $\bar N_c$ and $\bar N_s$ at $z=0$ for our choice of parameters in Fig.~\ref{fig:galnumber}.

We define the one and two halo contributions to the galaxy-galaxy power spectrum as (see e.g.~\cite{Berlind:2001xk}):
\begin{eqnarray}
P_{gg} (k,z) &=& P_{gg}^{1h}(k,z) + P_{gg}^{2h}(k,z) \\
P_{gg}^{1h}(k,z) &=& \int_{-\infty}^\infty d \ln m \ \frac{m n(m,z)}{n_g^2} 
   \Big( 2 \big\langle N_c(m) N_s(m) \big\rangle u_c(k) u_s(k|m,z) 
         + \big\langle N_s(m) (N_s(m)-1) \big\rangle u_s(k|m,z)^2 \Big) \label{eq:pgg_1h} \\
P_{gg}^{2h}(k,z) &=& P^{\rm lin} (k,z) \left[ \int_{-\infty}^\infty d \ln m \ m n(m,z)  b_h(m,z) \frac{\bar N_c(m) u_c(k) + \bar N_s(m) u_s(k|m,z)}{n_g} \right]^2 \label{eq:pgg_2h}
\end{eqnarray}
Here, $n_g$ is the mean number of galaxies as a function of halo mass and redshift:
\begin{equation}
n_g =  \int_{-\infty}^\infty d \ln m \ m n(m,z) \left( \bar N_c(m) + \bar N_s(m) \right).
\end{equation}
and $u_c(k), u_s(k|m,z)$ denote the Fourier-space profiles of the centrals and satellites.
Since we are assuming that centrals are at exact halo centers, and satellites are NFW-distributed,
we have $u_c(k)=1$, and $u_s(k|m,z)$ is given by the Fourier-space NFW profile.\footnote{In
 Eqs.~(\ref{eq:pgg_1h}),~(\ref{eq:pgg_2h}), we have denoted the profile $u_c(k)$ explicitly,
 rather than setting it to 1.  This is to clarify a technical point which arises in~\S\ref{sec:photoz_rsd}
 when modeling photometric redshift errors.  As explained there, photo-$z$ errors
 modify galaxy profiles as $u(k) \rightarrow W(k_r) u(k)$, where $W(k_r)$ is the Fourier-space
 photo-$z$ error distribution.  This convolution is applied to both profiles $u_c(k), u_s(k)$.
 By Eqs.~(\ref{eq:pgg_1h}),~(\ref{eq:pgg_2h}) it follows that both $P_{gg}^{1h}(k)$ and $P_{gg}^{2h}(k)$
 are multiplied by factors of $W(k_r)^2$, as claimed in the body of the paper (Eq.~(\ref{eq:photoz_pgg})).}

The expectation values $\langle N_s(m) (N_s(m)-1) \rangle$ and $\langle N_c N_s \rangle$
appearing in Eq.~(\ref{eq:pgg_1h}) depend on the assumed correlation between centrals and satellites.
We consider two extremes: (1) centrals and satellites are totally uncorrelated, and (2) a central is required for a satellite, and therefore 
centrals and satellites are maximally correlated.
In these cases, and assuming that the number of satellites is Poisson distributed, a short calculation shows:
\ba
\langle N_s(m) (N_s(m)-1) \rangle &=& \left\{ \begin{array}{cl}
  \bar N_s(m)^2 & \mbox{if centrals and satellites are uncorrelated} \\
  \bar N_s(m)^2 / \bar N_c(m) & \mbox{if centrals and satellites are maximally correlated} \\
\end{array} \right. \\
\langle N_c(m) N_s(m) \rangle &=&\left\{ \begin{array}{cl}
  \bar N_c(m) \bar N_s(m) & \mbox{if centrals and satellites are uncorrelated} \\
  \bar N_s(m) & \mbox{if centrals and satellites are maximally correlated} \\
\end{array} \right.
\ea
When deriving this, note that in the maximally-correlated model, the number of satellites in a halo which contains
a central (i.e.~the conditional PDF $P(N_s|N_c=1)$) is a Poisson random variable with mean $\bar N_s(m) / \bar N_c(m)$ 
(not mean $\bar N_s(m)$).

As our fiducial choice in the following we use the maximally-correlated model.
At the level of the galaxy galaxy power spectrum, the difference between these two models is minimal (at the $\sim 5\%$ level for $k<10^2 \ {\rm Mpc}^{-1}$).

Examining the two-halo term, and using the property that $u_g(k|m,z) \rightarrow 1$ as $k \rightarrow 0$, we see that the linear galaxy bias is given by
\begin{equation}
b_g (z) = \int_{-\infty}^\infty d \ln m \ m n(m,z)  b_h(m,z) \frac{\langle N_c(m) \rangle + \langle N_s(m) \rangle}{n_g}
\end{equation}
yielding $P_{gg}^{2h}(k,z) \simeq b_g(z)^2  P^{\rm lin} (k,z)$ on large scales.

\subsection{Cross-power}

The one and two halo contributions to the cross-power between galaxies and gas (or matter) is given by (see e.g.~\cite{Berlind:2001xk})
\begin{eqnarray}
P_{ge} (k,z) &=& P_{ge}^{1h}(k,z) + P_{ge}^{2h}(k,z) \\
P_{ge}^{1h}(k,z) &=& \int_{-\infty}^\infty d \ln m \ m n(m,z) \frac{m}{\rho_m} u_e(k|m,z) \frac{\langle N_c(m) \rangle u_c(k) + \langle N_s(m) \rangle u_s(k|m,z)}{n_g} \label{eq:pge_1h} \\
P_{ge}^{2h}(k,z) &=& P^{\rm lin} (k) \left[ \int_{-\infty}^\infty d \ln m \ m n(m,z)  b_h(m,z) \frac{\langle N_c(m) \rangle u_c(k) + \langle N_s(m) \rangle u_s(k|m,z)}{n_g} \right] \nonumber \\ 
&& \times \left[ \int_{-\infty}^\infty d \ln m \ m n(m,z) \left( \frac{m}{\rho_m} \right) b_h(m,z) u_e(k|m,z) \right]  \label{eq:pge_2h}
\end{eqnarray}
where notation has been introduced above.

\begin{figure}[tbh]
\includegraphics[width=0.45\textwidth]{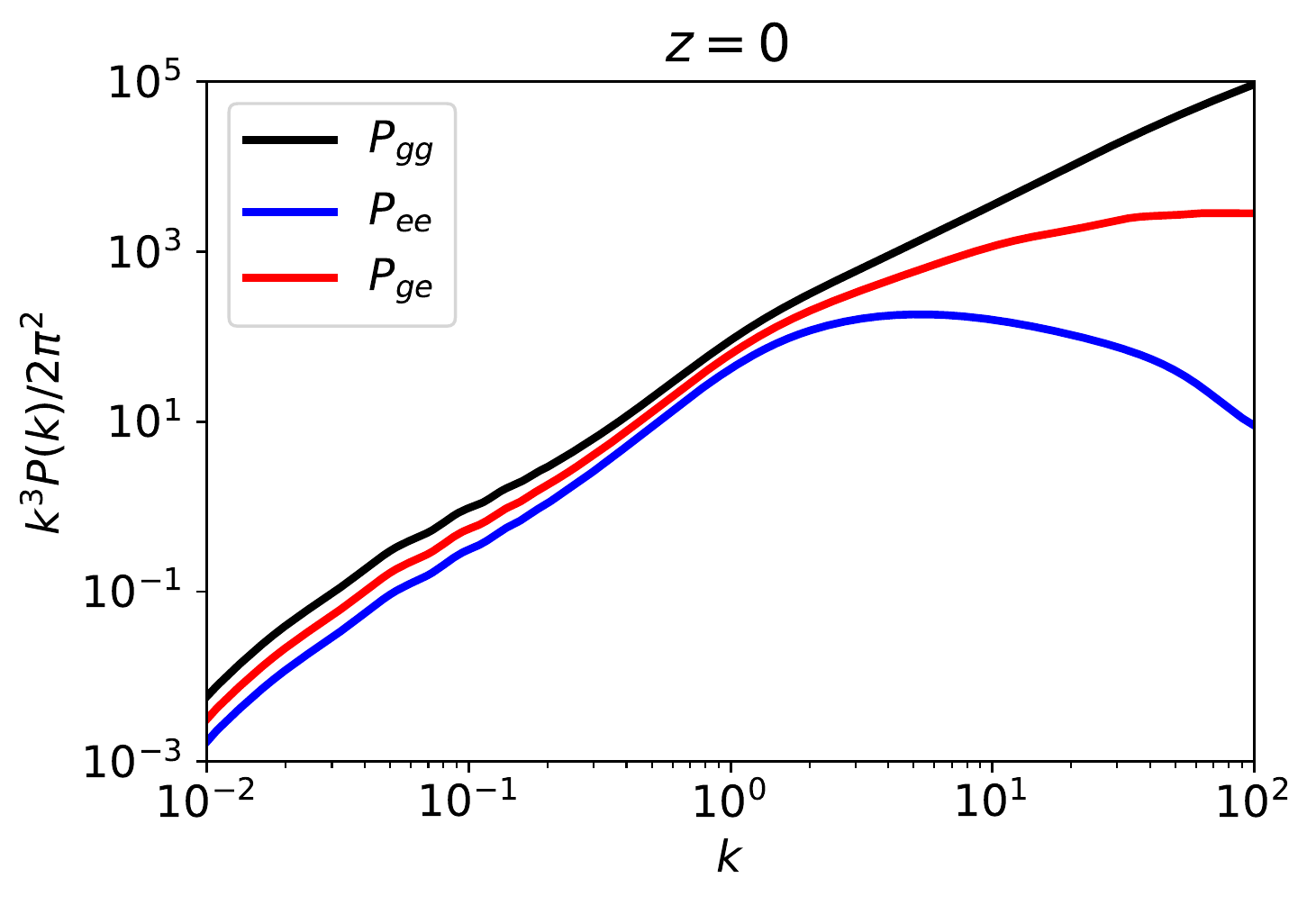}
\includegraphics[width=0.45\textwidth]{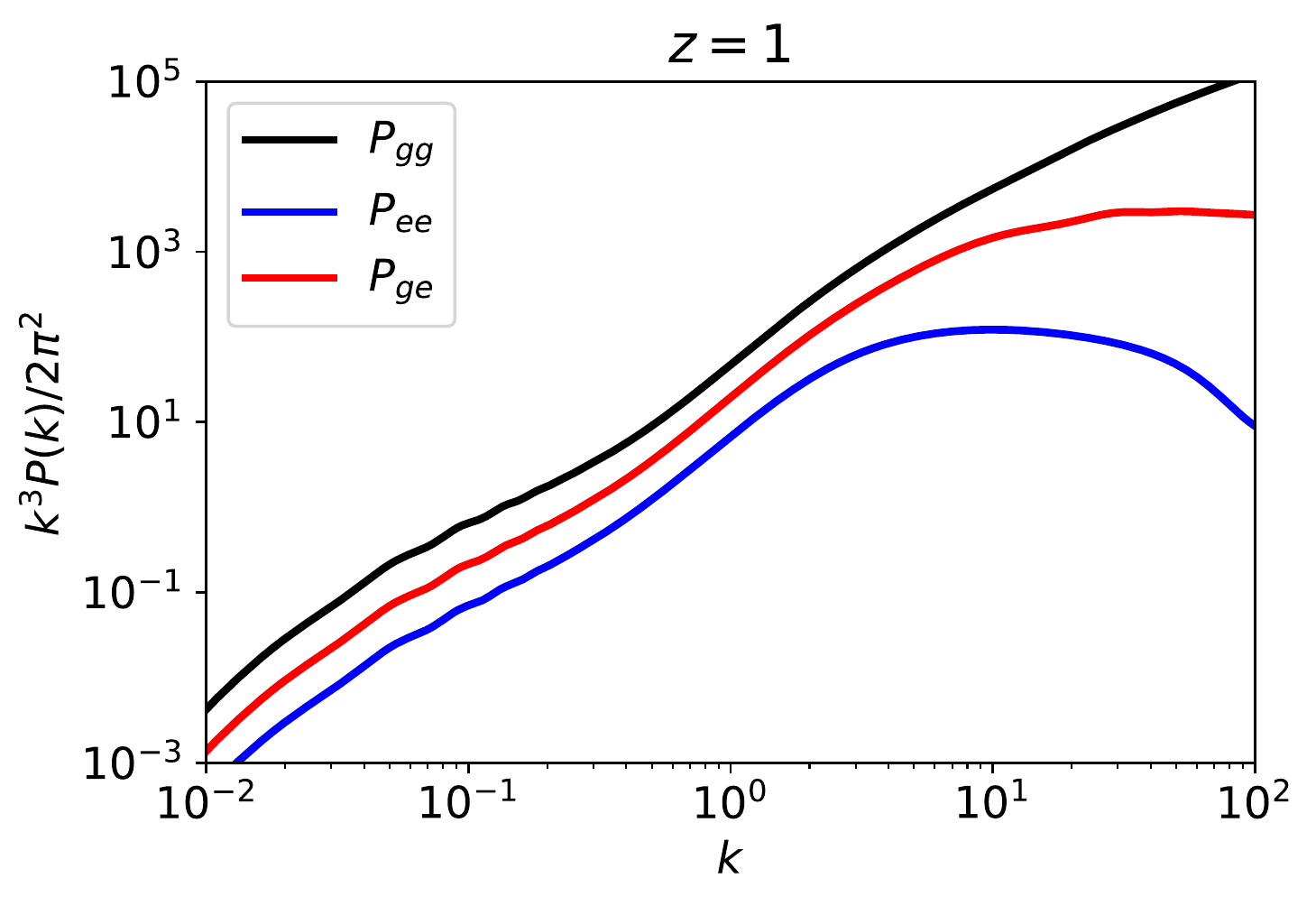}
  \caption{Auto and cross power spectra in our halo model assuming the 'AGN' model for the gas profile at $z=0$ (left) and $z=1$ (right) including halo masses in the range $10^{10} M_{\odot} < m < 10^{17} M_{\odot}$. }
  \label{fig:pspecs}
\end{figure}

In Fig.~\ref{fig:pspecs}, we compare the auto and cross power for galaxies at redshifts $z=0$ and $z=1$ including halo masses in the range $10^{10} M_{\odot} < m < 10^{17} M_{\odot}$ assuming the 'AGN' model for the gas profile.

\subsection{kSZ from the halo model}
\label{sec:kSZ}

\begin{figure}[tbh]
  \includegraphics[width=0.5\textwidth]{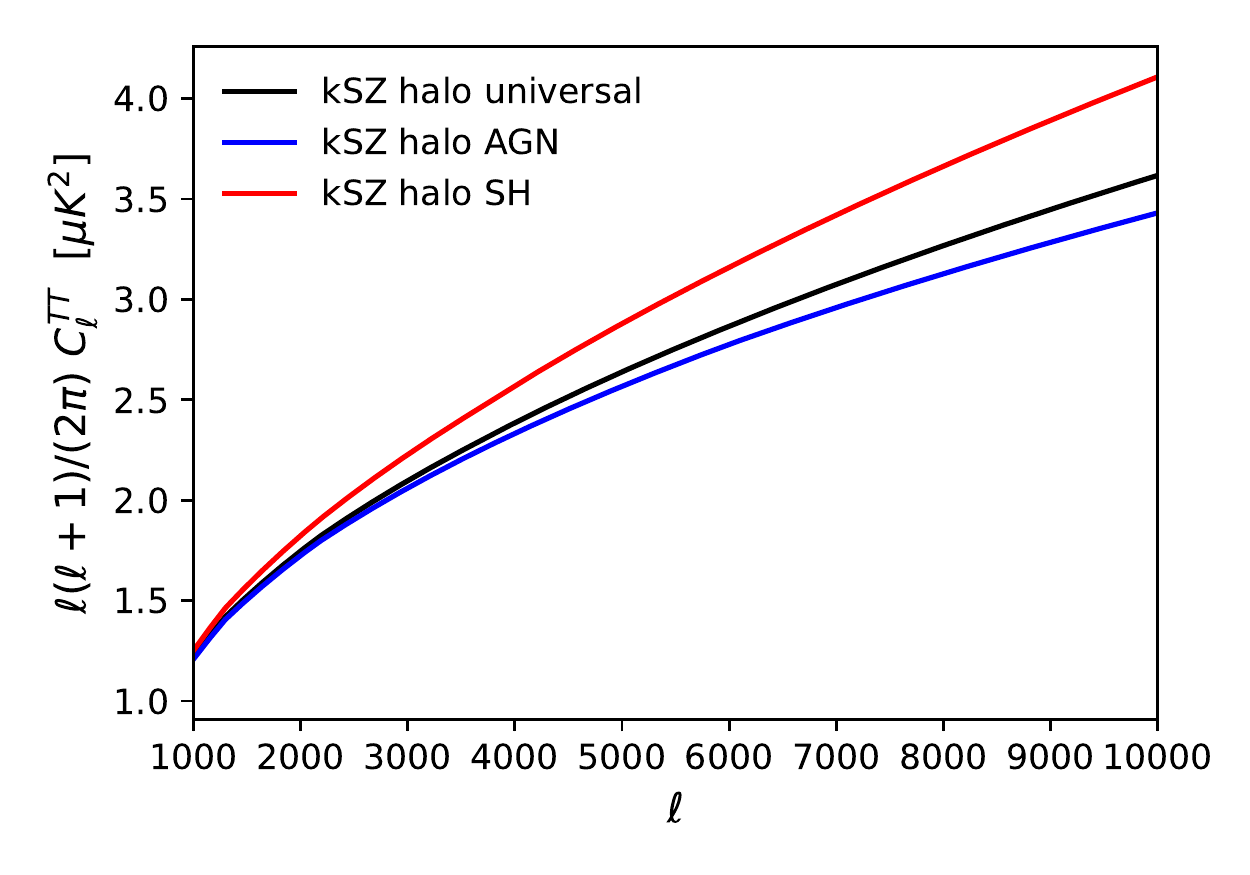}  
  \caption{CMB power spectrum from kSZ from redshifts $0<z<6$ calculated in the halo model using different electron distribution profiles.} 
\label{fig:clkszprofiles}
\end{figure}

We will also need a model for the kSZ contribution to the CMB power spectrum $C_l^{TT}$. We model this as the sum of two terms, from late times (i.e.~after reionization) and reionization.
We use the model from~\cite{Park:2013mv} for the reionization contribution to $C_l^{TT}$. We calculate the late-time kSZ contribution in the well known non-linear approximation from~\cite{Hu:1999vq}. The kSZ angular power spectrum at large multipoles is dominated by the power spectrum of the transverse momentum field, $P_{q_{\perp}}(k)$, and is given by~\cite{Vishniac:1987wm}
\begin{equation}
C_\ell = \frac12
\left(\frac{\sigma_T \bar{n}_{e,0}}{c}  \right)^2
\int\frac{d\chi}{\chi^2a^4} e^{-2\tau}
P_{q_{\perp}}\left(k=\frac{l}{\chi},\chi\right).
\end{equation}
The power spectrum of the transverse momentum field can be approximated as~\cite{Hu:1999vq}
\begin{eqnarray} \label{NLOV}
\nonumber 
P^{\rm{S}}_{q_{\perp}}(k,z) &=& \dot{a}^2f^2\int \frac{d^3k^\prime}{(2\pi)^3} P_{ee}^{nl} (|\bold{k} - \bold{k^\prime}|,z) P^{lin}_{\delta\delta}(k^\prime,z) \frac{k(k - 2k^\prime\mu^\prime)(1-{\mu^\prime}^2)}{{k^\prime}^2(k^2 + {k^\prime}^2-2kk^\prime\mu^\prime)}
\end{eqnarray}
where $P_{ee}^{nl}$ is the non-linear power spectrum of the electron distribution, which we calculated in the halo model. We show the resulting kSZ power spectra for different halo profiles in Fig.~\ref{fig:clkszprofiles}. The differences between these profiles at the $\ell$-range of interest in this paper is only of the order of $10\%$. However the true uncertainty on the kSZ signal size is likely larger than that (compare for example the simulations in~\cite{Shaw:2011sy}). Nevertheless we are using consistent assumptions in this paper by calculating $P_{gg}$, $P_{ge}$ and $P_{ee}$ from the same model.

\end{document}